# THE HYPERBOLIC THEORY OF SPECIAL RELATIVITY

*J.F. BARRETT*

--------------------------

**J.F. BARRETT**

# *THE HYPERBOLIC THEORY OF SPECIAL RELATIVITY*

-------------------------

## The interpretation of the Special Theory in hyperbolic space.

*'The principle of relativity corresponds to the hypothesis that the kinematic space is a space of constant negative curvature the space of Lobachevski and Bolyai The value of the radius of curvature is the speed of light.'*

**Borel 1913**

# Preface

This monograph can be considered as the outcome of an early interest in the theory of relativity when I felt uneasy at its presentation, particularly in its need to use an imaginary fourth dimensional time coordinate.There seemed to be something basically wrong so that, many years later, when I saw by chance the work of Varićak expressing the theory in terms of Bolyai-Lobachevski geometry (or 'hyperbolic geometry'), it came as a revelation. Being convinced that this was without doubt the correct approach, I started to work on it on my retirement publishing preliminary results at the conferences "Physical Interpretations of Relativity Theory " (PIRT) held biannually in London. The present monograpsh collects together ideas described there expanded with additional material. It is intended to give an introductory systematic account of this theory intended for the reader acquainted with the standard theory of Special Relativity.

Most of the mathematics in this work is elementary and known. The novelty lies in the arrangement of the material and showing inter-relationships. But there are new formulations and much use is made of historical aspects which the author believes essential for a correct perspective.  It is hoped to show that,keeping close to the historical development, advances can be made even in such a well established field as the Special Theory of Relativity.

The work was completed while Visitor to ISVR, University of Southampton

## The present version (2019)

The present version has many changes to that of 2011. The introductory comments of the first chapter have been omitted as elementary and unnecessary.The chapter on the hyperbolic theory has been considerably modified from the previous version. Also there have been many corrections and minor improvements throughout the work.

*Contact address:*

Institute of Sound and Vibration,
University of Southampton,
Southampton UK

jfb@soton.ac.uk

# On the Hyperbolic Interpretation of Special Relativity

The Special Theory of Relativity, which received its initial formulation by Poincaré and Einstein in 1905, gained general acceptance in 1908 about the same time as Minkowski's interpretation in terms of the 4 dimensional world   Soon after, in the years 1910-1914, the Yugoslav mathematician Vladimir Varićak showed that this theory finds a natural interpretation in hyperbolic (or Bolyai-Lobachevski) geometry,  an idea also put forward in less detail by a few other writers about the same time, notably Robb (1910, etc.) and Borel (1913).  Despite its apparently fundamental nature, this hyperbolic interpretation remains little known and has not yet found its way into standard texts on relativity theory, even after nearly a century.  This lack of interest has historical roots since hyperbolic geometry had from early times gained a reputation as an imaginary geometry of interest only in pure mathematics and its possible application to physical science therefore to most scientists not seriously considered.

The hyperbolic theory as put forward by Varićak in 1910 arose in connexion with the velocity composition law of Einstein.  Sommerfeld in 1909 had shown how, using Minkowski's ideas, this law may be reinterpreted in an intuitively clear way in terms of spherical rotations in space time. But his interpretation relied essentially on Minkowski's imaginary complex coordinate ict and was not without its difficulties. Varićak reinterpreted Sommerfeld's theory in hyperbolic space and avoided the need for complex representation. His basic result is that the relativistic law of combination of velocities can be interpreted as the triangle of velocities in hyperbolic space and so the kinematic space of Special Relativity is hyperbolic.  This view leads to the redefinition of velocity as a corresponding hyperbolic velocity more appropriate to relativity. It is needed for the correct definition of relative velocity which is fundamental to the theory.

Although Varićak's theory attracted some interest when it was first proposed, it soon became overshadowed by the appearance of the General Theory of Relativity which was of course also an interpretation using non-Euclidean geometry though in its Riemannian form. Afterwards the hyperbolic theory was only mentioned rarely and the mainstream exposition of Special Relativity followed lines laid down by Einstein and Minkowski. Hyperbolic trigoonometry   was used as a notation by just a very few authors e.g. Karapetoff (1944) and Whittaker (1953). Hyperbolic geometry itself was used in cosmology by Milne (1934) and Fock (1955) and in the period after the Second World War, was found of use in particle physics principally by Smorodinsky.

Recently the hyperbolic theory has shown signs of a revival. Attention was drawn to it bythe extensive historical review by Scott-Walter (1999) although this emphasized only the use of trigonometrical notation  Also, the many articles and books of Ungar in the period from the 1990s have drawn attention to the hyperbolic interpretation. The theory however has yet to become generally known by the majority of physicists and accepted as a correct intepretation, if not the correct interpetation, of the theory of Special Relativity.

# CHAPTER 1 - Product of Lorentz matrices

## 1. The Standard Form of a Lorentz Matrix

The standard 4-dimensional Lorentz matrix (the so-called 'boost matrix') relates time and space coordinates of a reference frame S' moving relative to a similarly oriented frame S The equation giving differential coordinate changes of S' in terms of those for S is

$$\begin{bmatrix} cdt' \\ dx' \\ dy' \\ dz' \end{bmatrix} = \begin{bmatrix} \gamma & -\gamma v_1/c & -\gamma v_2/c & -\gamma v_3/c \\ -\gamma v_1/c & 1+(\gamma-1)n_1^2 & (\gamma-1)n_1n_2 & (\gamma-1)n_1n_3 \\ -\gamma v_2/c & (\gamma-1)n_2n_1 & 1+(\gamma-1)n_2^2 & (\gamma-1)n_2n_3 \\ -\gamma v_3/c & (\gamma-1)n_3n_1 & (\gamma-1)n_3n_2 & 1+(\gamma-1)n_3^2 \end{bmatrix} \begin{bmatrix} cdt \\ dx \\ dy \\ dz \end{bmatrix} \tag{1}$$

Here velocity is $(v_1, v_2, v_3)$ in the direction of the unit vector $(n_1, n_2, n_3)$**.** The coefficient matrix is symmetric and its inverse mapping S' to S, is found by changing the sign of the velocity With three frames S, S' and S" in sequence multiplication of matrices $L_1$, $L_2$ corresponds to mappings $L_1$: S' ← S followed by $L_2$ :S" ← S' resulting in $L_2L_1$: S" ← S. Composition of the matrices consequently takes place in reversed order which is somewhat awkward but standard practice.

Equation (1) can be written concisely as in (2) by partitioned matrices using bold letters for vectors written as column matrices. This applies equally well to 1, 2 or 3 space dimensions.

$$\begin{bmatrix} cdt' \\ \mathbf{dr'} \end{bmatrix} = \begin{bmatrix} \gamma & -\gamma \mathbf{v}^T/c \\ -\gamma \mathbf{v}/c & I+(\gamma-1)\mathbf{nn}^T \end{bmatrix} \begin{bmatrix} cdt \\ \mathbf{dr} \end{bmatrix} \tag{2}$$

Here $\beta = v/c$ may also be used giving $\beta\mathbf{n}$ instead of $\mathbf{v}/c$.

*Terminology*: The Lorentz matrix and transformation (1) was, not long ago, renamed 'boost', who by is not clear. Minkowski had named it a Special Lorentz transformation, arguably a more appropriate name for:the matrix which has been so fundamental in Special Relativity and which surely has the right to be named after Lorentz
The correct geometrical interpretation of the Lorentz matrix depends on hyperbolic geometry (footnote). In the present account 'boost' will not be used so (1) will be called just the Lorentz matrix denoted by the letter L and products of Lorentz matrices (members of the Restricted Lorentz Group) will be referred to as general Lorentz matrices and denoted by Λ.

---

*Footnote*: According to Varićak the transformation is a Galilean type translation in hyperbolic space. See his 1910 paper "Theory of Relativity and Lobachevskian geometry", §3 "Lorentz-Einstein transformation as translation"

## 2. The product of Lorentz matrices.

Explicit representation of the product of two Lorentz matrices has been an enduring problem of Special Relativity. As is now well known, it is a Lorentz matrix followed or preceded by a spatial rotation (originally shown by Silberstein 1914 ) So the product of $L_1$ with $L_2$ may be written using left and right Lorentz matrices $L_l$, $L_r$ and a rotation matrix R as

$$L_2 L_1 = R L_l = L_r R \tag{1}$$

Lorentz matrices being symmetric and transposition of a rotation matrix giving its inverse, the relations above imply on transposing that

$$L_1 L_2 = L_l R^{-1} = R^{-1} L_r \tag{2}$$

here $L_l$ and $L_r$ are rotated forms of each other so $L_l = R\, L_r\, R^{-1}$ and $L_r = R\, L_l R^{-1}$

*The relation between right and left representations* can be shown in detail as follows. Consider the product written using a spatial rotation matrix Ω as

$$\begin{bmatrix} 1 & 0 \\ 0 & \Omega \end{bmatrix}\begin{bmatrix} \gamma & -\beta\gamma\mathbf{n}^T \\ -\beta\gamma\mathbf{n} & I+(\gamma-1)\mathbf{n}\mathbf{n}^T \end{bmatrix} \tag{3}$$

It is

$$\begin{bmatrix} \gamma & -\beta\gamma\mathbf{n}^T \\ -\beta\gamma\Omega\mathbf{n} & \Omega+(\gamma-1)(\Omega\mathbf{n})\mathbf{n}^T \end{bmatrix} \tag{4}$$

Using the unit vector $\mathbf{n}' = \Omega\,\mathbf{n}$ (so conversely $\mathbf{n} = \Omega^{-1}\mathbf{n}' = \Omega^T\mathbf{n}'$) it becomes symmetrical

$$\begin{bmatrix} \gamma & -\beta\gamma\mathbf{n}^T \\ -\beta\gamma\mathbf{n}' & \Omega+(\gamma-1)\mathbf{n}'\mathbf{n}^T \end{bmatrix} \tag{5}$$

From this it can be transformed with the same rotation into the alternative form

$$\begin{bmatrix} \gamma & -\beta\gamma\,\mathbf{n}'^T \\ -\beta\gamma\mathbf{n}' & I+(\gamma-1)\mathbf{n}'\mathbf{n}'^T \end{bmatrix}\begin{bmatrix} 1 & 0 \\ 0 & \Omega \end{bmatrix} \tag{6}$$

Here (5) gives a standard form for the product of Lorentz matrices and represents the general element of the Restricted Lorentz Group generated by Lorentz matrix products and inverses, here called general Lorentz transformations and denoted by Λ.The characterisation (5) has not been used in the literature as far as writer knows.

## 3. The Explicit Product

In terms of the symmetrical form described in the last section the product $L_2 L_1$ is

$$\begin{bmatrix} \gamma_2 & -\beta_2\gamma_2\mathbf{n}_2^{T} \\ -\beta_2\gamma_2\mathbf{n}_2 & I+(\gamma_2-1)\mathbf{n}_2\mathbf{n}_2^{T} \end{bmatrix} \begin{bmatrix} \gamma_1 & -\beta_1\gamma_1\mathbf{n}_1^{T} \\ -\beta_1\gamma_1\mathbf{n}_1 & I+(\gamma_1-1)\mathbf{n}_1\mathbf{n}_1^{T} \end{bmatrix}$$

$$= \begin{bmatrix} \gamma & -\beta\gamma\,\mathbf{n}^{T} \\ -\beta\gamma\,\mathbf{n'} & \Omega+(\gamma-1)\,\mathbf{n'}\mathbf{n}^{T} \end{bmatrix} \tag{1}$$

There is found, after adjustments for sign, transposition, etc

$$\begin{aligned} \gamma &= \gamma_1\,\gamma_2\,\{1+\beta_1\beta_2\,\mathbf{n}_2^{T}\mathbf{n}_1\} \\ \beta\gamma\,\mathbf{n} &= \gamma_2\,\{(I+(\gamma_1-1)\,\mathbf{n}_1\mathbf{n}_1^{T})\,\beta_2\,\mathbf{n}_2+\gamma_1\,\beta_1\,\mathbf{n}_1\} \\ \beta\gamma\,\mathbf{n'} &= \gamma_1\,\{(I+(\gamma_2-1)\,\mathbf{n}_2\mathbf{n}_2^{T})\,\beta_1\,\mathbf{n}_1+\gamma_2\,\beta_2\,\mathbf{n}_2\} \\ \Omega+(\gamma-1)\,\mathbf{n'}\,\mathbf{n}^{T} &= \beta_2\,\beta_1\,\gamma_2\,\gamma_1\mathbf{n}_2\mathbf{n}_1^{T}+(I+(\gamma_2-1)\,\mathbf{n}_2\mathbf{n}_2^{T})(I+(\gamma_1-1)\,\mathbf{n}_1\mathbf{n}_1^{T}) \end{aligned} \tag{2}$$

From the 1st and 2nd of these equations follows

$$\beta\,\mathbf{n} = (I+(\gamma_1-1)\,\mathbf{n}_1\,\mathbf{n}_1^{T})\,\beta_2\,\mathbf{n}_2+\gamma_1\beta_1.\mathbf{n}_1/\gamma_1\{1+\beta_1\beta_2\,\mathbf{n}_1^{T}\mathbf{n}_2\} \tag{3}$$

$$= \{\gamma_1^{-1}\,(I-\mathbf{n}_1\,\mathbf{n}_1^{T})+\mathbf{n}_1\,\mathbf{n}_1^{T}\}\,\beta_2\,\mathbf{n}_2+\beta_1.\mathbf{n}_1/\{1+\beta_1\beta_2\,\mathbf{n}_1^{T}\mathbf{n}_2\} \tag{4}$$

Here $\mathbf{n}_1^{T}\mathbf{n}_2 = \cos\theta$ where $\theta$ is the angle between $\mathbf{n}_1$ and $\mathbf{n}_2$. Now $\mathbf{n}_2 = \mathbf{n}_1\cos\theta+\mathbf{n}_1^{\perp}\sin\theta$ resolves $\mathbf{n}_2$ parallel and perpendicular to $\mathbf{n}_1$ So that $(I-\mathbf{n}_1\,\mathbf{n}_1^{T})\,\mathbf{n}_2 = \mathbf{n}_2-\cos\theta\,\mathbf{n}_1 = \mathbf{n}_1^{\perp}\sin\theta$. The equation now becomes

$$\beta\,\mathbf{n} = \frac{\sqrt{(1-\beta_1^2)}\,\beta_2\sin\theta\,\mathbf{n}_1^{\perp}+\{\beta_2\cos\theta+\beta_1\}\mathbf{n}_1}{\{1+\beta_1\beta_2\cos\theta\}} \tag{5}$$

Similarly

$$\beta\,\mathbf{n}' = \frac{\sqrt{(1-\beta_2^2)}\,\beta_1\sin\theta\,\mathbf{n}_2^{\perp}+\{\beta_1\cos\theta+\beta_2\}\mathbf{n}_2}{\{1+\beta_1\beta_2\cos\theta\}} \tag{6}$$

The spatial rotation matrix $\mathbf{\Omega}$ is now, in principle, determined by substituting the values of **n, n'** into the final equation in (2).This general form of matrix $\mathbf{\Omega}$ clearly is very complicated. Since the rotation axis of $\mathbf{\Omega}$ is orthogonal to both **n** and **n'** a simpler characterization of $\mathbf{\Omega}$ by its rotation angle and axis direction follows on taking scalar and vector products of **n** and **n'** If $\mathbf{\Omega}$ is known, left and right forms for the Lorentz matrix L follow from the previous section.

The calculation may be simplified by taking the plane determined by **n, n'** as the xy plane so reducing the dimensionality of the problem as in the following example

---

*Remarks*: (a) The method of calculation given here is a modification of that of Feinstein (1914)
(b) Taking the magnitude squared of $\beta$ **n** and $\beta$ **n'** gives Einstein's composition formula

# 4. Composition of Matrices for Orthogonal Motions

The composition formulae are here simplified by taking the plane of the two velocities as the xy plane so that the z-axis becomes redundant. The Lorentz matrices can then be written conveniently as 3×3 matrices transforming only the variables ct, x, y. The spatial rotation matrix $\Omega$ becomes a 2x2 matrix through a rotation angle $\Psi$ and from it the 3x3 rotation matrix R constructed as.

$$\begin{bmatrix} 1 & 0 & 0 \\ 0 & \cos\Psi & -\sin\Psi \\ 0 & \sin\Psi & \cos\Psi \end{bmatrix} \tag{1}$$

The velocities may be taken along the x and y axes and introducing $\beta_1$, $\beta_2$, $\beta$, for the ratios $v_1/c$, $v_2/c$, $v/c$ the Lorentz matrices $L_1$, $L_2$ for translation along x and y axes will be

$$\begin{bmatrix} \gamma_1 & -\gamma_1\beta_1 & 0 \\ -\gamma_1\beta_1 & \gamma_1 & 0 \\ 0 & 0 & 1 \end{bmatrix} \qquad \begin{bmatrix} \gamma_2 & 0 & -\gamma_2\beta_2 \\ 0 & 1 & 0 \\ -\gamma_2\beta_2 & 0 & \gamma_2 \end{bmatrix} \tag{2}$$

The required reversed product $L_2\, L_1$ is equal to

$$\begin{bmatrix} \gamma_1\gamma_2 & -\gamma_1\gamma_2\beta_1 & -\gamma_2\beta_2 \\ -\gamma_1\beta_1 & \gamma_1 & 0 \\ -\gamma_1\gamma_2\beta_2 & \gamma_1\gamma_2\beta_1\beta_2 & \gamma_2 \end{bmatrix} \tag{3}$$

So

$$\begin{aligned} \gamma &= \gamma_1\,\gamma_2 \\ \beta\,\mathbf{n}^T &= [\gamma_1\,\gamma_2\,\beta_1,\ \gamma_2\,\beta_2] \\ \gamma\,\beta\,\mathbf{n}'^T &= [\gamma_1\,\beta_1,\ \gamma_1\gamma_2\,\beta_2] \end{aligned} \tag{4}$$

where the value of $\beta$ is given by

$$\begin{aligned} \beta^2 &= \beta_1^{\,2} + \beta_2^{\,2}/\gamma_1^{\,2} = \beta_1^{\,2}/\gamma_2^{\,2} + \beta_2^{\,2} \\ &= \beta_1^{\,2} + \beta_2^{\,2} - \beta_1^{\,2}\,\beta_2^{\,2} = 1 - (1 - \beta_1^{\,2})(1 - \beta_2^{\,2}) \end{aligned} \tag{5}$$

The unit vectors **n**, **n**' for the two velocity compositions are

$$\begin{aligned} \mathbf{n}^T &= [\beta_1,\ \beta_2\,/\gamma_1]^T\,\beta^{-1} = [\beta_1,\ \beta_2\,\sqrt{(1 - \beta_1^{\,2})}]^T\,\beta^{-1} \\ \mathbf{n}'^T &= [\beta_1/\gamma_2,\ \beta_2]^T\,\beta^{-1} = [\beta_1\sqrt{(1 - \beta_2^{\,2})},\ \beta_2]^T\,\beta^{-1} \end{aligned} \tag{6}$$

From these cos Ψ and sin Ψ are found from scalar and vector products as

$$\cos\psi = \frac{\gamma_1 + \gamma_2}{1 + \gamma_1 \gamma_2} \qquad \sin\psi = \frac{\beta_1 \beta_2 \gamma_1 \gamma_2}{1 + \gamma_1 \gamma_2} \tag{7}$$

The right hand Lorentz matrix can now be constructed using **n** and the parameters $\beta$, $\beta_1$, $\beta_2$, $\gamma_1$ and the product $L_2.L_1$ written

$$\begin{bmatrix} 1 & 0 & 0 \\ 0 & \cos\psi & -\sin\psi \\ 0 & \sin\psi & \cos\psi \end{bmatrix} \begin{bmatrix} \gamma & -\gamma\beta_1 & -\gamma(\beta_2/\gamma_1) \\ -\gamma\beta_1 & 1+(\gamma-1)(\beta_1/\beta)^2 & (\gamma-1)(\beta_1/\beta)(\beta_2/\beta\gamma_1) \\ -\gamma(\beta_2/\gamma_1) & (\gamma-1)(\beta_2/\beta\gamma_1)(\beta_1/\beta) & 1+(\gamma-1)(\beta_2/\beta\gamma_1)^2 \end{bmatrix} \tag{8}$$

Similarly the left-hand Lorentz matrix can be constructed using **n'** and the parameters $\beta$, $\beta_1$, $\beta_2$, $\gamma_2$ and the product $L_2.L_1$ written

$$\begin{bmatrix} \gamma & -\gamma(\beta_1/\gamma_2) & -\gamma\beta_2 \\ -\gamma(\beta_1/\gamma_2) & 1+(\gamma-1)(\beta_1/\beta\gamma_2)^2 & (\gamma-1)(\beta_1/\beta\gamma_2)(\beta_2/\beta) \\ -\gamma\beta_2 & (\gamma-1)(\beta_2/\beta)(\beta_1/\beta\gamma_2) & 1+(\gamma-1)(\beta_2/\beta)^2 \end{bmatrix} \begin{bmatrix} 1 & 0 & 0 \\ 0 & \cos\psi & -\sin\psi \\ 0 & \sin\psi & \cos\psi \end{bmatrix} \tag{9}$$

The extended **n** and **n'** matrices here are $[1 \;\; \beta_1, \beta_2/\gamma_1]^T \beta^{-1}$ and $[1 \;\; \beta_1/\gamma_2 \;\; \beta_2]^T \beta^{-1}$
That the rotation matrix R interchanges these is seen as follows. Consider the product

$$\begin{bmatrix} 1 & 0 & 0 \\ 0 & \cos\Psi & -\sin\Psi \\ 0 & \sin\Psi & \cos\Psi \end{bmatrix} \begin{bmatrix} 1 \\ \beta_1 \\ \beta_2/\gamma_1 \end{bmatrix} \tag{10}$$

Use of the formulae (7) gives

$$\begin{aligned} \beta_1 \cos\Psi - \beta_2/\gamma_1 \sin\Psi &= \{\beta_1 (\gamma_1 + \gamma_2) - \beta_1\beta_2^2 \gamma_2\}/(1 + \gamma_1\gamma_2) \\ &= \{\beta_1 (1 - \beta_2^2)\gamma_2 + \beta_1\gamma_1\}/(1 + \gamma_1\gamma_2) \\ &= \{\beta_1/\gamma_2 + \beta_1\gamma_1\}/(1 + \gamma_1\gamma_2) \\ &= \beta_1/\gamma_2 \end{aligned} \tag{11}$$

$$\begin{aligned} \beta_1 \sin\Psi - \beta_2/\gamma_1 \cos\Psi &= \{\beta_1 (\beta_1\gamma_1\beta_2\gamma_2) + (\beta_2/\gamma_1)(\gamma_1 + \gamma_2)\}/(1 + \gamma_1\gamma_2) \\ &= \{\beta_1^2\gamma_1^2 \beta_2\gamma_2 + \beta_2(\gamma_1 + \gamma_1)\}/(1 + \gamma_1\gamma_2)\gamma_1 \\ &= \{\beta_2\gamma_2 (1 + \beta_1^2\gamma_1^2) + \beta_2\gamma_1\}/(1 + \gamma_1\gamma_2)\gamma_1 \\ &= \{\beta_2\gamma_1\gamma_1 + \beta_2\}/(1 + \gamma_1\gamma_2) \\ &= \beta_2 \end{aligned} \tag{12}$$

So the product gives $[1 \;\; \beta_1/\gamma_2 \;\; \beta_2]^T \beta^{-1}$ as required. Left multiplication by $R^{-1}$ on both sides makes the reverse change.

## 5. Rotation Angle for Orthogonal Motions

The rotation angle Ψ may be determined by forming scalar and vector products of **n** and **n**' which give cos Ψ and sin Ψ. In the case of composition of orthogonal motions there is found in this way, using values of **n** and **n**' found in the previous section,

$$\cos\Psi = \left(\sqrt{(1 - \beta_2^2)}\ \beta_1^2 + \sqrt{(1 - \beta_1^2)}\ \beta_2^2\right)/\ \beta^2 = \left(\gamma_2^{-1}\ \beta_1^2 + \gamma_1^{-1}\ \beta_2^2\right)/\ \beta^2$$
$$\sin\Psi = \beta_1\beta_2 \left(1 - \sqrt{(1 - \beta_1^2)(1 - \beta_2^2)}\right)/\ \beta^2 = \beta_1\beta_2 \left(1 - \gamma_1^{-1}\gamma_2^{-1}\right)/\ \beta^2$$
$$\tan\Psi = \beta_1\beta_2 \left(\gamma_1\gamma_2 - 1\right)/\ \left(\gamma_1\ \beta_1^2 + \gamma_2\ \beta_2^2\right) \qquad (1)$$

These may be expressed in various ways by algebraic transformation e.g.

$$\cos\Psi = (\gamma_1 + \gamma_2)(1 - \gamma_1\gamma_2)/\ \beta^2 = (\gamma_1 + \gamma_2)/\ (1 + \gamma_1\gamma_2)$$
$$\sin\Psi = \beta_1\beta_2/\ (1 + \gamma_1^{-1}\gamma_2^{-1}) = \beta_1\gamma_1\beta_2\gamma_2/\ (1 + \gamma_1\gamma_2)$$
$$\tan\Psi = \beta_1\gamma_1\beta_2\gamma_2\ /\ (\gamma_1 + \gamma_2) \qquad .(2)$$

Corresponding formulae may be given in terms of the velocities $v_1$, $v_2$, e.g.

$$\cos\Psi = \frac{v_1^2\sqrt{(1 - v_2^2/c^2)} + v_2^2\sqrt{(1 - v_1^2/c^2)}}{v_1^2 + v_2^2 - (v_1 v_2/c)^2} \qquad (3)$$

$$\sin\Psi = \frac{v_1 v_2 \{1 - \sqrt{(1 - v_1^2/c^2)}\sqrt{(1 - v_2^2/c^2)}\}}{v_1^2 + v_2^2 - (v_1 v_2/c)}$$
$$= \frac{v_1 v_2}{c^2 \{1 + \sqrt{(1 - v_1^2/c^2)}\sqrt{(1 - v_2^2/c^2)}\}} \qquad (4)$$

$$\tan\Psi = \frac{v_1 v_2 \{1 - \sqrt{(1 - v_1^2/c^2)}\sqrt{(1 - v_2^2/c^2)}\}}{v_1^2\sqrt{(1 - v_2^2/c^2)} + v_2^2\sqrt{(1 - v_1^2/c^2)}} \qquad (5)$$

* *Half-angle formulae*: The half-angle formulae for rotation angle Ψ take a simpler form. They have importance in the theory.They may be found from:

$$\cos^2\Psi/2 = \frac{1 + \cos\Psi}{2} = \frac{(\gamma_1 + 1)(\gamma_2 + 1)}{2(1 + \gamma_1\gamma_2)}$$
$$\sin^2\Psi/2 = \frac{1 - \cos\Psi}{2} = \frac{(\gamma_1 - 1)(\gamma_2 - 1)}{2(1 + \gamma_1\gamma_2)}$$
$$\tan^2\Psi/2 = \frac{\sin^2\Psi/2}{\cos^2\Psi/2} = \frac{(\gamma_1 - 1)(\gamma_2 - 1)}{(\gamma_1 + 1)(\gamma_2 + 1)} \qquad (6)$$

From the last follows the formula of Liebmann (Varićak 1912, 1924) which can be written

$$\cot\Psi/2 = \frac{\sqrt{(\gamma_1 + 1)}\sqrt{(\gamma_2 + 1)}}{\sqrt{(\gamma_1 - 1)}\sqrt{(\gamma_2 - 1)}} \qquad (7)$$

## 6. The Thomas Precession

The Thomas precession occurs when acceleration $\alpha$ is in a different direction to velocity so that velocities v, v + $\delta$v at successive instants t, t + $\delta$t do not have the same direction. If vector velocities **v** and **$\delta$v** are inclined at an angle $\theta$, the increment **$\delta$v** has components $\delta v \cos\theta$, $\delta v \sin\theta$ parallel and orthogonal to **v**. Being infinitesimal, their effects may be superimposed and then the resulting rotation $\delta\Psi$ arises only from the orthogonal component The rotation angle can be calculated from the previously given formula

$$\sin\Psi = \frac{v_1 v_2}{c^2 \{1 + \sqrt{(1- v_1^2/c^2)}\,\sqrt{(1 - v_2^2/c^2)}\}} \tag{1}$$

Setting

$$v_1 = v, \; v_2 = \delta v \sin\theta = \alpha\, \delta t \sin\theta, \; \Psi = \delta\Psi \tag{2}$$

the following approximations will hold to first order:

$$\sqrt{(1 - v_1^2/c^2)} \approx \sqrt{(1 - v^2/c^2)} = \gamma^{-1}, \quad \sqrt{(1 - v_2^2/c^2)} \approx 1 \tag{3}$$

The angular velocity relative to the observer is

$$\frac{d\Psi}{dt} = \frac{v\,\alpha \sin\theta}{c^2 \{1 + \sqrt{(1- v^2/c^2)}\}} = \frac{\{1 - \sqrt{(1- v^2/c^2)}\}\, v\,\alpha \sin\theta}{v^2} \tag{4}$$

Relative to the moving point it is

$$\frac{d\Psi}{d\tau} = \frac{\{1 - \sqrt{(1- v^2/c^2)}\}\, v\,\alpha \sin\theta}{v^2\sqrt{(1- v^2/c^2)}} = \{\gamma - 1\}\,\frac{v\,\alpha \sin\theta}{v^2} \tag{5}$$

The rotation is about an axis perpendicular to **v** and **$\alpha$** and so can be written as

$$\frac{d\boldsymbol{\Psi}}{d\tau} = (\gamma - 1)\,\frac{\mathbf{v} \times \boldsymbol{\alpha}}{v^2} \tag{6}$$

If the velocity v is small compared with that of light, the formula simplifies as

$$\frac{d\boldsymbol{\Psi}}{d\tau} = \frac{\mathbf{v} \times \boldsymbol{\alpha}}{2c^2} \tag{7}$$

---

*Historical note*: The Thomas rotation occurred with the accelerated motion of an electron in an electric field when the formula for spin of Goudschmidt and Uhlenbeck was corrected by Thomas (1926, 1927). Thomas' first paper gave the approximate formula, the more accurate formula being given in his second paper. It is interesting that Borel had in 1913 deduced on purely mathematical grounds from the hyperbolic representation that such an effect would occur.
*References*: Thomas: *Nature* 1926 117 p.514; *Phil. Mag.* 7 1927 1-23; Borel: *C.R. Acad. Sci.* Paris 1913; Borel's observation is described by Scott-Walker 1999.

# CHAPTER 2 – Sommerfeld's Spherical Theory

## 1. The Lorentz Transformation as Rotation

In his fundamental 1909 paper 'On the composition of velocities', Sommerfeld aimed to show how the space-time view of Minkowski, then recently introduced, could be of real use to physicists. His spherical interpretation of Einstein's composition formula led on directly to the hyperbolic theory.

As previously described, Minkowski had, In his investigation of Maxwell's equations, somewhat incidentally, represented the Lorentz transformation as a Euclidean rotation

$$\begin{aligned} x' &= x \cos\varphi + ict \sin\varphi \\ ict' &= -x \sin\varphi + ict \cos\varphi \end{aligned} \qquad (1)$$

The purely imaginary angle $\varphi$ is defined by

$$\tan\varphi = i\, v/c \qquad (2)$$

implying

$$\cos\varphi = \frac{1}{\sqrt{(1 - v^2/c^2)}} = \gamma \qquad \sin\varphi = \frac{i\,(v/c)}{\sqrt{(1 - v^2/c^2)}} = i\,\beta\gamma \qquad (3)$$

Sommerfeld pointed out that this representation simplifies the composition rule for rectilinear motion since the result of two successive Lorentz transformations of angles $\varphi_1$, $\varphi_2$ is the Lorentz transformation with angle $\varphi_1 + \varphi_2$ given by

$$iv/c = \tan(\varphi_1 + \varphi_2) = \frac{\tan\varphi_1 + \tan\varphi_2}{1 - \tan\varphi_1 \tan\varphi_2} = \frac{iv_1/c + iv_2/c}{1 - iv_1/c\; iv_2/c} \qquad (4)$$

There follows the composition rule

$$v = \frac{v_1 + v_2}{1 + v_1 v_2/c^2}. \qquad (5)$$

This new method of deriving the composition rule soon came into general use and was, for example, used by Einstein in his 1921 Princeton lectures replacing his earlier method. It is now well known. Not so well known however, is the generalization of this idea to the non-rectilinear case introduced by Sommerfeld in his 1909 paper and subsequent writings.

---

*References*: Sommerfeld A:'Über die Zusammensetzung der Geschwindigkeiten' *Phys.Z.* 1909 826-829; Einstein's 1921 Princeton lectures were published as a book *The Meaning of Relativity* which went through many editions the last being in 1945 (Methuen)

## 2. Non-commutativity of Velocity Addition

Einstein's 1905 derivation of the composition rule gave the magnitude of the resultant of two velocities but had said nothing about its direction. Sommerfeld attempted to clarify this situation by combining two orthogonal velocities

$$\mathbf{v}_1 = (v_1, 0), \quad \mathbf{v}_2 = (0, v_2) \qquad (1)$$

This can be done in two possible ways - $\mathbf{v}_1$ followed by $\mathbf{v}_2$ and $\mathbf{v}_2$ followed by $\mathbf{v}_1$.In agreement with Einstein's combination law (cf chapter 1) he found, for the resultant velocities corresponding to these two ways the values

$$(v_1, v_2\sqrt{(1 - v_1^2/c^2)}), \quad (v_1\sqrt{(1 - v_2^2/c^2)}, v_2) \qquad (2)$$

These have the same magnitude squared

$$v^2 = v_1^2 + v_2^2 - (v_1 v_2/c)^2 \qquad (3)$$

which can be written in either of two ways corresponding to the two resultants:

$$v^2 = v_1^2 + v_2^2 (1 - v_1^2/c^2) = v_1^2 (1 - v_2^2/c^2) + v_2^2 \qquad (4)$$

From these are found Pythagoras formulae corresponding to the figure below. The resultant velocities are represented by lines AC and C'A' of equal length.

Multiplication of the transverse velocity components by square root contraction factors results in failure of the rectangular figure to close giving rise to what became known as non-commutativity of velocity addition. The angle Ψ between the resultants, is easily also found from this diagram by taking scalar and vector products of the vectors (2).

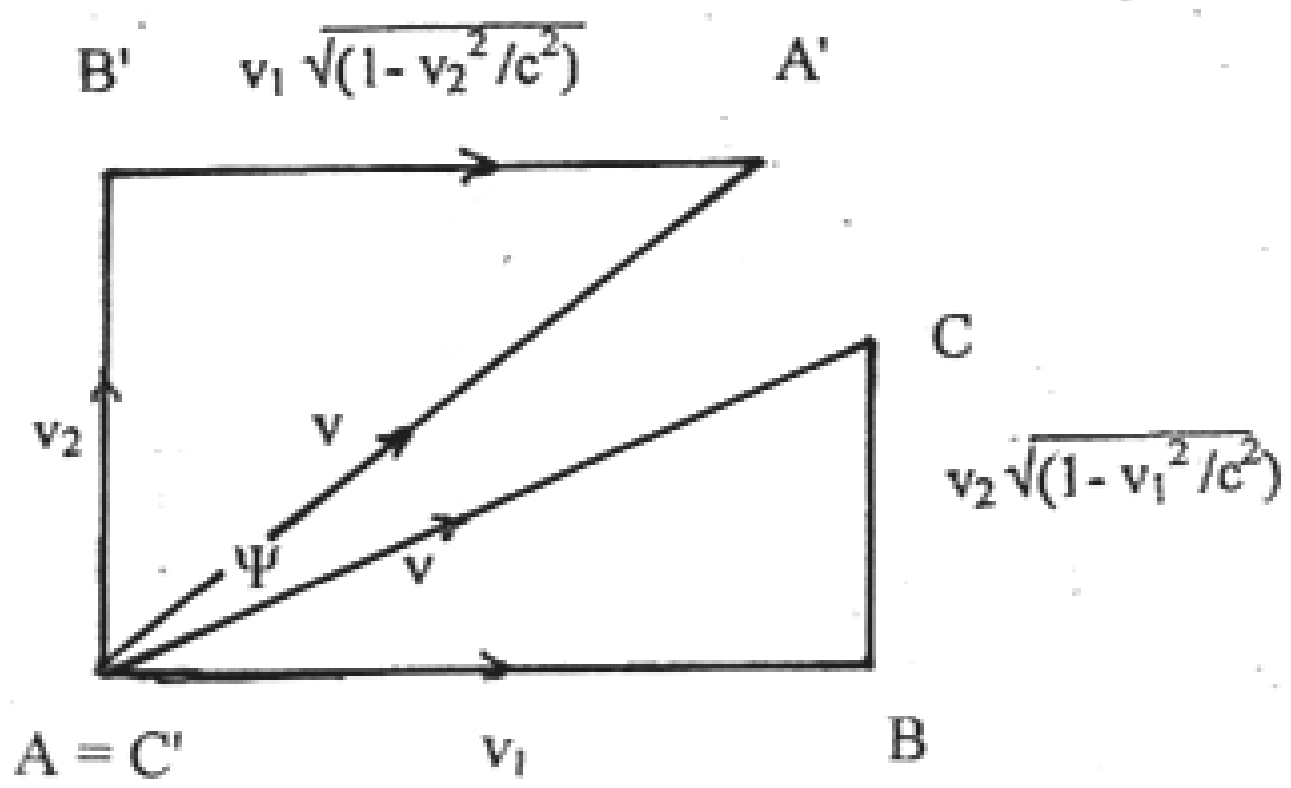

*Fig*: Non-commutativity of velocity composition

As seen in the last chapter, composition of the Lorentz matrices results in a rotation through an angle, Ψ in the present case. This has become known as the *Thomas Rotation* takimg the name previously used for the effect with electrons. The angle Ψ will here be called the *rotation angle*. The curious consequence of this would be to make an interchange of directions so that $\mathbf{v}_1$ followed by $\mathbf{v}_2$ results, not in vector AC, but in C'A' It is similar in reverse for $\mathbf{v}_2$ followed by $\mathbf{v}_1$.

The rules govening the application of this rotation are not clear to the writer, but the rotation does not apply to the present discussion of pure velocity addition followimg Sommerfeld.

# 3. The Spherical Representation

Sommerfeld explained non-commutativity of velocity addition as arising from addition of displacements on a sphere determined by the Minkowski angles.
In the case of addition of orthogonal velocities, the relation between the velocities $v_1$, $v_2$ and their resultant v gives the equations

$$(1 - v^2/c^2) = (1 - v_1^2/c^2)\,(1 - v_2^2/c^2) \qquad (1)$$

$$\frac{1}{\sqrt{(1 - v^2/c^2)}} = \frac{1}{\sqrt{(1 - v_1^2/c^2)}}\;\frac{1}{\sqrt{(1 - v_2^2/c^2)}}\,. \qquad (2)$$

By using Minkowski angles $\varphi$, $\varphi_1$, $\varphi_2$, corresponding to v, $v_1$, $v_2$, this becomes the Pythagoras formula for the right angled spherical triangle the Minkowski angles project on the surface of a sphere.

$$\cos\varphi = \cos\varphi_1 \cos\varphi_2 \qquad (3)$$

That Minkowski angles are purely imaginary is seen from (2) since all factors exceed unity

The figure for the addition of velocities at right now takes the following form (called spherical Lambert quadrilateral). There are two right-angled spherical triangles ABC, A'B'C' with sides AB=A'B'= then BC=B'C'= and the hypotenuse AC=A'C'=

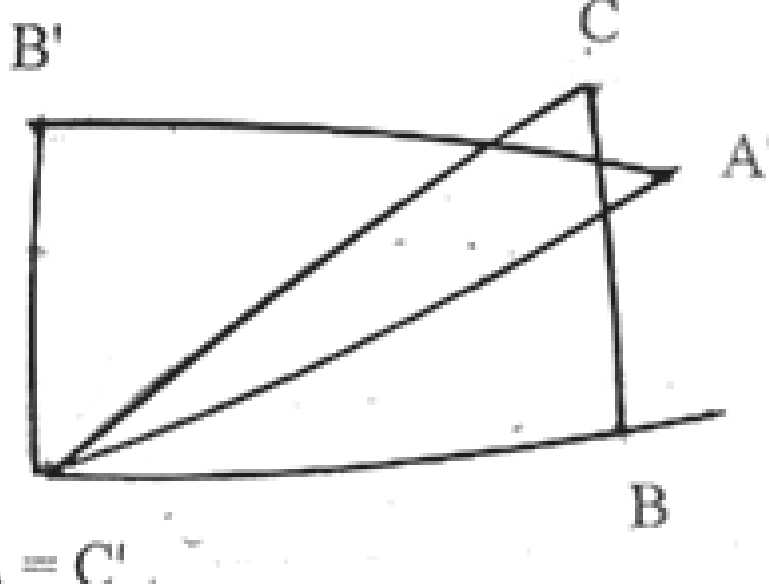

*Fig*: Non-commutativity of velocity composition (spherical form)

Unlike the Cartesian case non-commutativity arises naturally here without any contraction factors. Spherical geometry thus gives a natural explanation of noncommutativity
But the diagram shows non-commutativity differently to the previous Cartesian case because of the switch-over of directions of the resultants AC, A'C'. However Sommerfeld noted that this would not happen in the imaginary case.This is because angle A'AC represents the spherical excess of either triangle ABC or A'B'C' (i.e the angle by which the sum of angles of either triangle exceeds $\pi$ radians) When the radius is imaginary as here, the excess becomes negative and there is no swichover, the figure taking the same form as in the Cartesian case though without contraction factors.Sommerfeld concluded:

> *"For the combination of velocities in relativity theory there are valid, not formulae of the plane, but instead of spherical trigonometry (with imaginary sides.)"*

## 4. Spherical Form of the Einstein Composition Formula:

Extending these ideas, Sommerfeld derived the general composition law for inclined velocities. The Minkowski angle $\varphi$ of the resultant velocity v is given by vector addition of the arcs on a sphere of the Minkowski angles $\varphi_1$, $\varphi_2$ for $v_1$,$v_2$ so that by the spherical cosine rule where $\theta$ is the angle of inclination of velocities $v_1$, $v_2$:

$$\cos\varphi = \cos\varphi_1 \cos\varphi_2 + \sin\varphi_1 \sin\varphi_2 \cos(\pi - \theta)$$
$$= \cos\varphi_1 \cos\varphi_2 - \sin\varphi_1 \sin\varphi_2 \cos\theta \qquad (1)$$

Changing from Minkowski angles to velocities gives

$$\frac{1}{\sqrt{(1 - v^2/c^2)}} = \frac{1}{\sqrt{(1 - v_1^2/c^2)}\sqrt{(1 - v_2^2/c^2)}} - \frac{(iv_1/c)}{\sqrt{(1 - v_1^2/c^2)}}\,\frac{(iv_2/c)}{\sqrt{(1 - v_2^2/c^2)}}\cos\theta \qquad (2)$$

$$= \frac{(1 + v_1 v_2/c^2 \cos\theta)}{\sqrt{(1 - v_1^2/c^2)}\sqrt{(1 - v_2^2/c^2)}}\;. \qquad (3)$$

Inversion and squaring results in

$$(1 - v^2/c^2) = \frac{(1 - v_1^2/c^2)\,(1 - v_2^2/c^2)}{(1 + v_1 v_2/c^2 \cos\theta)^2} \qquad (4)$$

Then solving for v gives Einstein's composition law

$$v = \frac{\sqrt{\{v_1^2 + v_2^2 + 2\,v_1 v_2 \cos\theta - (v_1 v_2/c \sin\theta)^2\}}}{1 + (v_1 v_2/c^2)\cos\theta}\;. \qquad (5)$$

The algebra here is reversible so, starting from the composition formula, the spherical cosine rule can be deduced.

# 5. Spherical Excess and Rotation Angle

As later described by Sommerfeld (1931), the spherical representation gives a geometrical method for determination of the rotation angle Ψ and from it the formula for. Thomas rotation. Consider his figure illustrated here.

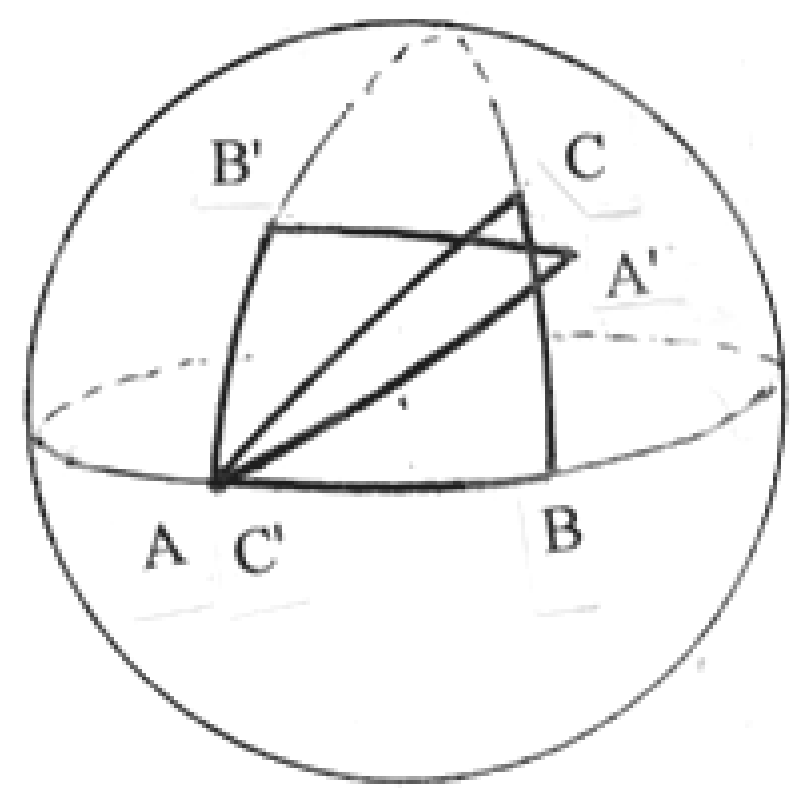

*Fig*: Rotation angle (spherical form)

The rotation angle is the angle between AC and C'A' seen to be

$$\Psi = (A + B + C) - \pi \tag{1}$$

Ψ is the spherical excess E of either triangle ABC or A'B'C' and equation (1) is not dependent on triangles being right angled. Using its spherical excess, inter--pretation the determination of the rotation angle is reduced to the purely trigonometrical problem of finding the spherical excess of a triangle.

Sommerfeld (1931) used this method to find the rotation angle for velocities at right angles for application to the Thomas rotation. To get the spherical excess E, he used a formula quoted from spherical trigonometry and interpreting angles as the Minkowski angles corresponding to two velocities $v_1$, $v_2$.he found the formula already derived in chapter 1:by a different method:

$$\sin\Psi = \frac{v_1 v_2}{c^2} \frac{1}{\{1 + \sqrt{(1 - v_1^2/c^2)}\sqrt{(1 - v_2^2/c^2)}\}} \approx \frac{v_1 v_2}{2\,c^2} \tag{2}$$

From this follows, as before, the formula for Thomas rotation.

The spherical representation used here consequently gives a nice explanation of the Thomas Rotation and its magnitude. Sommerfeld appears to assume that this explanation will also be valid in the case of imaginary radius i.e. the hyperbolic case. But it seems that this is not so as will be seen in the next chapter.and so the explanation of the rotational effect in this case remains unclear.The formula for magnitude of the rotation (i.e. the angle between the two non-commutating resultants) however remains the same.

---

# CHAPTER 3 – The Hyperbolic Theory

## 1. Rapidity

The basic quantity of the hyperbolic theory is the *rapidity* w defined in terms of velocity by

$$\mathrm{th}\, w = v/c \tag{1}$$

the principal value of the inverse hyperbolic tangent being used for the determination of w from this equation. Corresponding to any value of v less in magnitude to c this equation determines a value of w lying in the range $-\infty < w < \infty$. Transforming the Minkowski rotational representation and its inverse

$$\begin{aligned} x' &= x \cos\varphi + ict \sin\varphi \qquad & x &= x' \cos\varphi - ict' \sin\varphi \\ ict' &= -x \sin\varphi + ict \cos\varphi \qquad & ict &= x' \sin\varphi + ict' \cos\varphi \end{aligned} \tag{2}$$

by setting φ to be iw and using the identities

$$\begin{aligned} \cos\varphi &= \cos iw = \mathrm{ch}\, w \\ \sin\varphi &= \sin iw = i\, \mathrm{sh}\, w \end{aligned} \tag{3}$$

there arises the symmetric transformation and its inverse

$$\begin{aligned} ct' &= ct\, \mathrm{ch}\, w - x\, \mathrm{sh}\, w \qquad & ct &= ct'\, \mathrm{ch}\, w + x'\, \mathrm{sh}\, w \\ x' &= -ct\, \mathrm{sh}\, w + x\, \mathrm{ch}\, w \qquad & x &= ct'\, \mathrm{sh}\, w + x'\, \mathrm{ch}\, w \end{aligned} \tag{4}$$

which is the representation in terms of rapidity. Note that from (1) follows

$$\mathrm{ch}\, w = \frac{1}{\sqrt{(1 - v^2/c^2)}} \qquad \mathrm{sh}\, w = \frac{(v/c)}{\sqrt{(1 - v^2/c^2)}}. \tag{5}$$

* *Additivity*: The characteristic property of rapidity is its additivity for velocity composition. This additivity is obvious from the relation with Minkowski's imaginary Euclidean form but can also be seen directly from the identity

$$\mathrm{th}\,(w_1 + w_2) = \frac{\mathrm{th}\, w_1 + \mathrm{th}\, w_2}{1 + \mathrm{th}\, w_1\, \mathrm{th}\, w_2} \tag{6}$$

resulting in the composition law for the velocity v corresponding to rapidity $w_1 + w_2$

$$v = \frac{v_1 + v_2}{1 + v_1 v_2/c^2} \tag{7}$$

So that for the rapidities $w_1, w_2, w$ corresponding to $v_1, v_2, v$ it is true that

$$w = w_1 + w_2 \tag{8}$$

* *Matrix representation*: Using rapidity the Lorentz matrix takes a symmetric form (9) having inverse (10)

$$L(w) = \begin{bmatrix} \operatorname{ch} w & -\operatorname{sh} w \\ -\operatorname{sh} w & \operatorname{ch} w \end{bmatrix} \tag{9}$$

$$L(-w) = \begin{bmatrix} \operatorname{ch} w & \operatorname{sh} w \\ \operatorname{sh} w & \operatorname{ch} w \end{bmatrix} \tag{10}$$

The multiplication law is expressed by the equations

$$\begin{bmatrix} \operatorname{ch}(w_1 + w_2) & -\operatorname{sh}(w_1 + w_2) \\ -\operatorname{sh}(w_1 + w_2) & \operatorname{ch}(w_1 + w_2) \end{bmatrix} = \begin{bmatrix} \operatorname{ch} w_2 & -\operatorname{sh} w_2 \\ -\operatorname{sh} w_2 & \operatorname{ch} w_2 \end{bmatrix} \begin{bmatrix} \operatorname{ch} w_1 & -\operatorname{sh} w_1 \\ -\operatorname{sh} w_1 & \operatorname{ch} w_1 \end{bmatrix} \tag{11}$$

$$L(w_1+w_2) = L(w_1)\, L(w_2) \tag{12}$$

This semi-group property makes possible the exponential representation

$$L(w) = \exp Kw = I + Kw + \frac{(Kw)^2}{2!} + \frac{(Kw)^3}{3!} + \ldots \tag{13}$$

where K is the infinitesimal generator

$$K = \begin{bmatrix} 0 & -1 \\ -1 & 0 \end{bmatrix} \quad \ldots\ldots\ldots\ldots\ldots\ldots\ldots\ldots\ldots\ldots \tag{14}$$

---------------------------------------------------------------------------------------------------------

* *Historical note*: As previously seen, Minkowski (1908) defined a quantity ψ satisfying th ψ = v/c in modern notation. Here ψ is the rapidity but Minkowski only used it to define his imaginary angle φ as equal to iψ and then made no further use of it. Varićak (1910 etc) used the notation w and described its properties. Then it was independently defined by Robb in 1911 who gave it the name rapidity. It has since been quite commonly used for representation of the Lorentz transformation even though the Lorentz transformation loses the intuitive appeal of the imaginary rotational form when written in this way. Notably, Whittaker, in his well known historical book defined rapidity in the first edition (1910) and used it consistently in later editions. Various notations have been used instead of w including Minkowski's ψ. It has been commonly regarded as an arbitrary mathematical parameter without any special physical significance

## 2. Hyperbolic Velocity

Instead of the non-dimensional rapidity w it is more natural in physical applications to use the corresponding dimensional quantity

$$V = c\,w = c\,\mathrm{th}^{[-1]}(v/c) \qquad (1)$$

This was used by Varićak (1910) though without a name.He regarded it as the true velocity from which the usual velocity v is found as a Euclidean projection. This is here accepted as a correct view but for the sake of conforming with customary usage of the word 'velocity' the term *hyperbolic velocity* will be used to denote velocity as defined by equation (1).Since v and V have the same physical dimensions the relation between them can be shown as below where the scales of v and V are the same

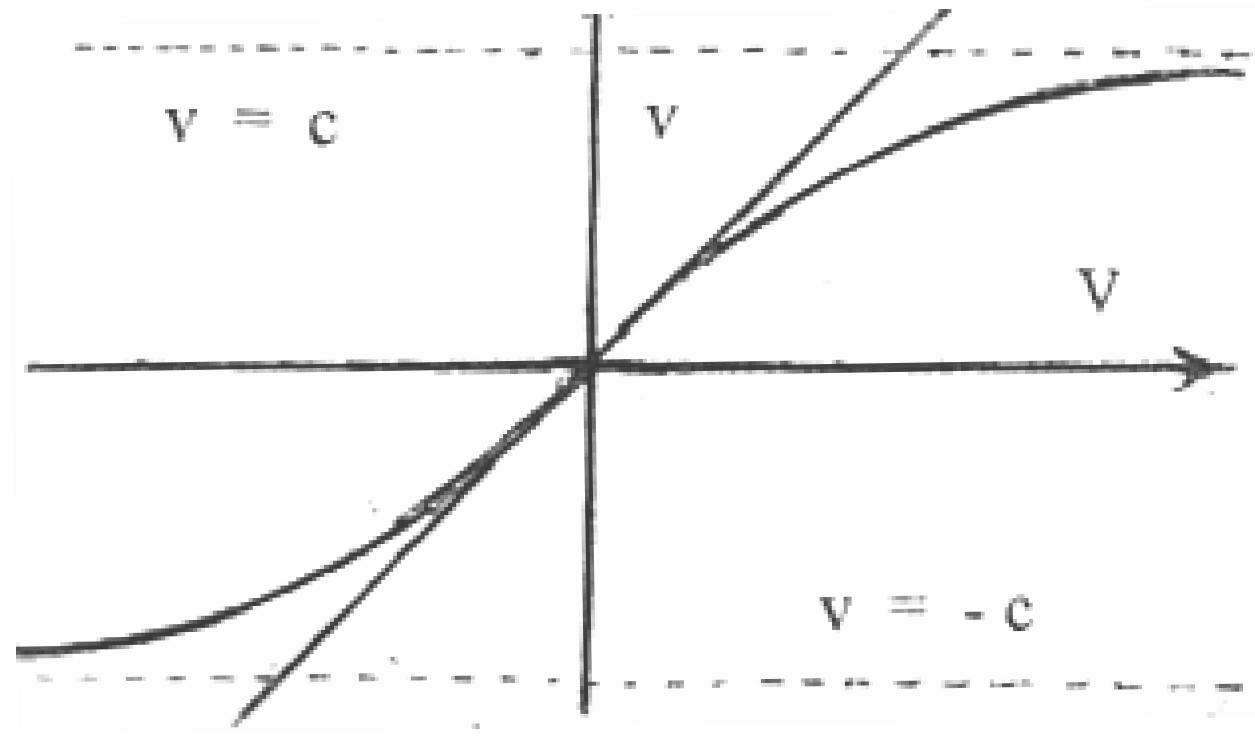

.... ......... ......... *Fig*: The relation between velocity and hyperbolic velocity

* *Properties of hyperbolic velocity*
(a) Like rapidity, it can take any value from $-\infty$ to $+\infty$, the hyperbolic velocity of light being infinite. When $v \rightarrow c$ correspondingly $V \rightarrow \infty$
(b) At low velocities ($v << c$) V approximates v or more precisely,

$$V = v\{\, 1 + 1/3\,(v/c)^2 + 1/5\,(v/c)^4 + \dots \} \qquad (2)$$

For rectilinear motion, hyperbolic velocities V combine by the same rules of addition as do the proportional rapidities w. So if $w = w_1 + v_2$ then

$$V = V_1 + V_2 \qquad (3)$$

* *The space of hyperbolic velocities*: hyperbolic velocity vectors **V** are defined by their magnitude V and direction. The space of such vectors forms a hyperbolic space of radius of curvature equal to c and defines the kinematic space in Special Relativity. This provides the constant c with a natural meaning and leads to a very satisfactory view of the principle of relativity well expressed by Borel (1913) by the quotation in the title page of this book The kinematic space approximates the classical velocity space locally for velocities small compared with the speed of light.

# 3. The Hyperbolic Triangle Law of Velocity Addition.

Since the early days of hyperbolic geometry it has been realized that the trigonometric formulae can be obtained from those of spherical trigonometry by using imaginary angles or equivalently by using a sphere of imaginary radius as in the last chapter. So the change from spherical to hyperbolic form was mathematically a very small step but was of great importance in setting the theory on its correct path: establishing the physical reality of hyperbolic geometry, which was often disbelieved, and its central significance for the accurate statement of the Principle of Relativity. The change was made by Varićak in 1910 soon after Sommerfeld's 1909 paper followed by independent work by a few others also arriving at hyperbolic representation. (see footnote)

Substituting iw for angles φ in the spherical cosine formula leads to the cosine rule in hyperbolic space:

$$\text{ch } w = \text{ch } w_1 \text{ ch } w_2 - \text{sh } w_1 \text{ sh } w_2 \cos(\pi - \theta)$$

$$\text{ch } \frac{V}{c} = \text{ch } \frac{V_1}{c} \text{ ch } \frac{V_2}{c} - \text{sh } \frac{V_1}{c} \text{ sh } \frac{V_2}{c} \cos(\pi - \theta) \qquad (1)$$

This equation is illustrated by the triangle ABC with inwardly curving sides suggesting negative curvature.

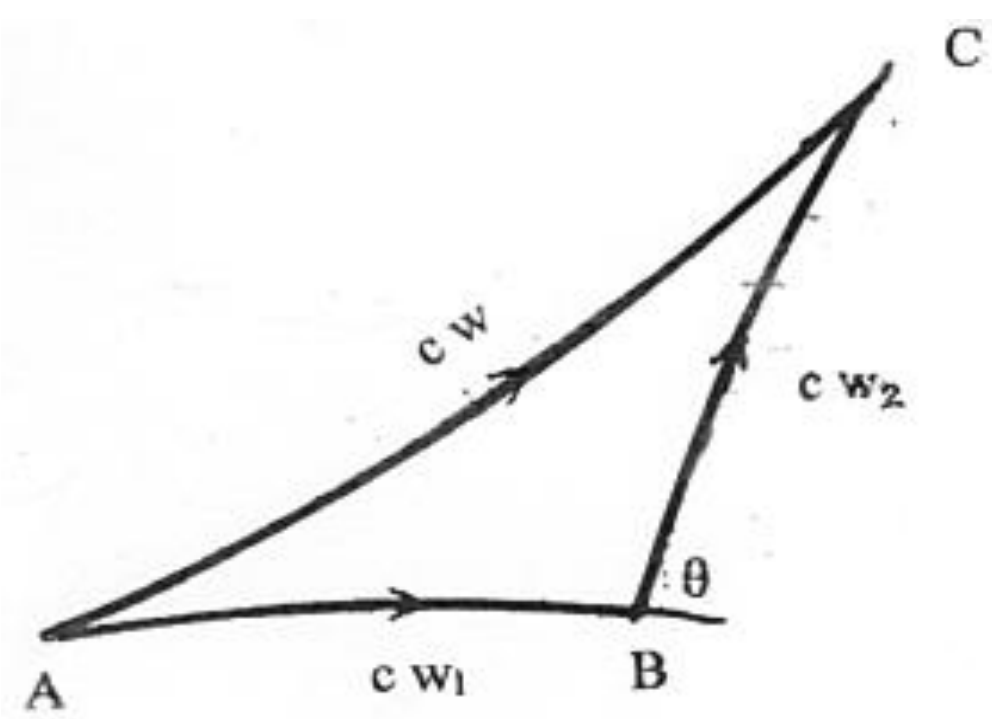

*Fig*. A triangle of rapidities

The direct derivation of formula (1) starting from Einstein's composition rule is:

$$v^2 = \left\{\frac{v_1^2 + v_2^2 + 2\, v_1 v_2 \cos\theta - (v_1 v_2/c \sin\theta)^2}{1 + (v_1.v_2/c^2)\cos\theta}\right\} \qquad (2)$$

$$(1 - v^2/c^2) = \frac{(1 - v^2/c^2)\,(1 - v^2/c^2)}{(1 + v_1 v_2/c^2 \cos\theta)^2} \qquad (3)$$

$$\frac{1}{\sqrt{(1 - v^2/c^2)}} = \frac{1}{\sqrt{(1 - v^2/c^2)}}\,\frac{1}{\sqrt{(1 - v^2/c^2)}} + \frac{(v_1/c)}{\sqrt{(1 - v^2/c^2)}}\,\frac{(v_2/c)}{\sqrt{(1 - v^2/c^2)}}\cos\theta \qquad (4)$$

Formula (1) results on substitution of hyperbolic functions.

---

* *Historical note on Varicak*:etc Vladimir Varićak (1865-1942) was professor of mathematics at Zagreb University. His biography and detailed list of all his publications was given by Kurepa (1965). Further contributions were made by Lewis & Tolman (1909), LeRoux(1922), Milne(1934), Karapetoff (1936, 1944), Patria (1956), Smorodinski (1963,1964). An extensive review has been given by Scott-Walker (1996, 1999)

## 4 Addition of rapidities at right angles.

The interpretation of non-commutativity in hyperbolic geometry for velocities at right angles corresponding to that of Sommerfeld for the spherical case is shown in the figure If rapidities are used in the Sommerfeld representation, there is no reduction in length of two sides by a square root factor which are equal in length to the opposite sides. This makes it necessary to use the different form of representation shown - the rectangular hyperbolic Lambert quadrilateral.

The appearance of the square root factor in the velocity representation comes naturally from the geometry since, by hyperbolic trigonometry equation (1) holds implying (2) and then (3) showing that an extra root factor occurs in velocity representation

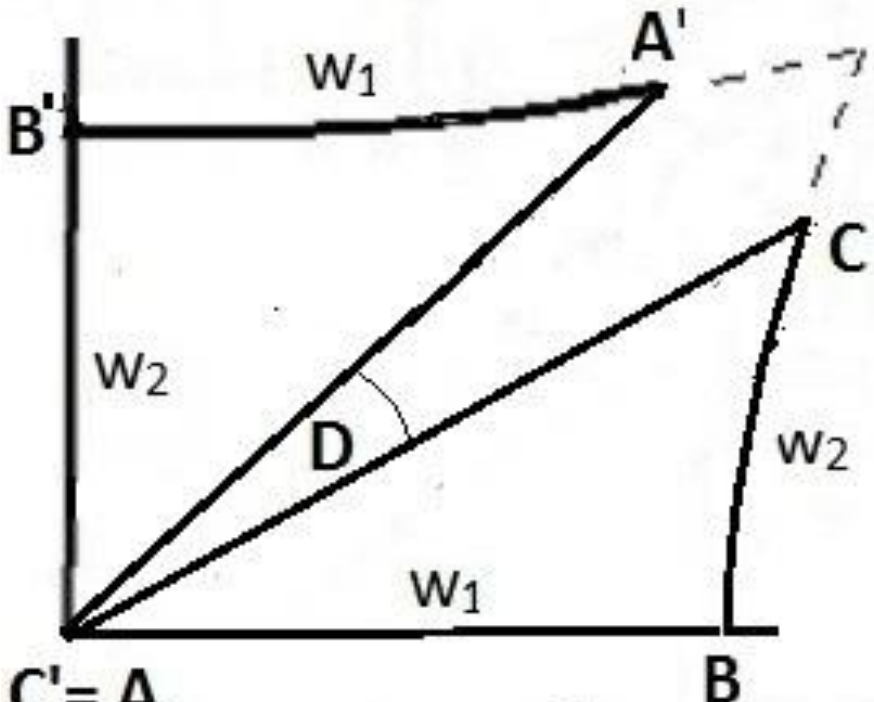

*Fig*. Non-commutativity of addition of rapidities

$$\tan A \,\mathrm{sh}\, w_1 = \mathrm{th}\, w_2 \qquad (1)$$
$$\mathrm{th}\, w_1 \tan A = \mathrm{th}\, w_2 \,\mathrm{sech}\, w_1 \qquad (2)$$
$$v_1 \tan A = v_2 \sqrt{(1 - v_1^2/c^2)} \qquad (3)$$

*Determination of defect (or rotation angle)*
This follows the same method as in the corresponding spherical representation. But the congruent triangles ABC, A'B'C' no longer overlap as in the spherical case From the congruence of triangles ABC and A'B'C' it can be seen that the defect D is equal to the defect D of either triangle because e.g. using triangle ABC

$$D = \pi/2 - (A+C') = \pi - (A+B+C) \qquad (4)$$

The rotation angle Ψ equals the defect D as in the figure. This equality is not dependent on the triangle being right-angled

*Liebmann's half-angle formula for defect*: In its original form it was

$$\cot D/2 = \frac{\sqrt{(\gamma_1 + 1)(\gamma_2 + 1)}}{\sqrt{(\gamma_1 - 1)(\gamma_2 - 1)}} \qquad (5)$$

Now (6) holds for any rapidity w so Liebmann's formula may be written as (7)

$$\frac{(\gamma + 1)}{(\gamma - 1)} = \frac{\mathrm{ch}\, w + 1}{\mathrm{ch}\, w - 1} = \frac{\mathrm{ch}^2(w/2)}{\mathrm{sh}^2(w/2)} = \coth^2 (w/2) \qquad (6)$$

$$\cot D/2 = \coth (w_1/2) \coth (w_2/2) \qquad (7)$$

-----------------------------------------------------------------------------------------------------------

*References* The formula is from Liebmann *Hyperbolic Geometry* (1905) p.149 quoted by Varićak (1912), (1924) It was proved previously in chapter 1.

*Deficit for two velocities inclined at any angle*: The deficit is given by the hyperbolic form of Lagrange's formula

$$\cot (D/2) = \frac{\text{ch}\,(w_1/2)\,\text{ch}\,(w_2/2) + \text{sh}\,(w_1/2)\,\text{sh}\,(w_2/2)\cos\theta}{\text{sh}\,(w_1/2)\,\text{sh}\,(w_2/2)\sin\theta} \qquad (8)$$

$\theta$ being angle of inclination of velocities On division of numerator and denominator by the sinh half angles, there results a formula which may be written

$$\cot D/2 = \frac{C + \cos\theta}{\sin\theta} \qquad (9)$$

C being the right hand side of Liebmann's formula (7) which is the value for orthogonal velocities when $\cos\theta = 0$, $\sin\theta = 1$

The derivation of (7) is immediate from transcribing a formula of Lagrange for spherical excess (see the mathematical appendix). It may be further transformed to the following symmetric form (see mathematical appendix)

$$\cot D/2 = \frac{\sqrt{(1 - \text{ch}^2 w_1 - \text{ch}^2 w_2 - \text{ch}^2 w_3 + 2\,\text{ch}\,w_1\,\text{ch}\,w_2\,\text{ch}\,w_3)}}{1 + \text{ch}\,w_1 + \text{ch}\,w_2 + \text{ch}\,w_3} \qquad (10)$$

From there it transforms to symmetrical formulae which have been used in the particle physics literature: These may all be deduced by the method given in Lagrange's paper.

$$\sin \frac{D}{2} = \frac{1 + \text{ch}\,w_1 + \text{ch}\,w_2 + \text{ch}\,w_3}{(\text{ch}\,w_1/2)\,(\text{ch}\,w_2/2)\,(\text{ch}\,w_3/2)} \qquad (11)$$

$$\cos \frac{D}{2} = \frac{\sqrt{(1 - \text{ch}^2 w_1 - \text{ch}^2 w_2 - \text{ch}^2 w_3 + 2\,\text{ch}\,w_1\,\text{ch}\,w_2\,\text{ch}\,w_3)}}{4\,(\text{ch}\,w_1/2)\,(\text{ch}\,w_2/2)\,(\text{ch}\,w_3/2)} \qquad (12)$$

---

* *Note*: The value of D given by Liebmann's formula was proved by spinors, by van Wyk (1984) and also used by Ben-Menahem (1985). Subsequently many papers were devoted to the derivation of related formulae. The present derivation by the hyperbolic form of Lagrange's (1799) formula, giving also the extension to the case of non-orthogonal velocities, appears simplest and clearest It is due to the writer (PIRT 2000) and has apparently not appeared elsewhere although Smorodinski (1962 etc.) used related formulae. The symmetrical formulae (9), (10) were used in the physics literature from 1962. See Wick (1962), Smorodinski (1963) etc. On their derivation see also Hestenes *Space-time Algebra* 1966

# 5 Hyperbolic geometry on a hyperboloid

Hyperbolic geometry is the geometry of a hyperbolic surface just as spherical geometry is the geometry on a sphere. But whereas spherical representation is in Euclidean space, hyperbolic representation must be made in Minkowski space. For easy visualization the three-dimensional version will be described. With z as ct, consider the surface in 3 dimensional Minkowski space.

$$-x^2 - y^2 + z^2 = R^2 \tag{1}$$

It is a hyperboloid in two sheets, the upper half-sheet having parameterization

$$\begin{aligned} x &= R \, \mathrm{sh}\, u \cos\theta \\ y &= R \, \mathrm{sh}\, u \sin\theta \\ z &= R \, \mathrm{ch}\, u \end{aligned} \tag{2}$$

The parameter u here takes positive values, the vertex being at $u = 0$. A differential displacement (dx dy dz) on the surface of the hyperboloid will satisfy

$$-x\,dx - y\,dy + z\,dz = 0 \tag{3}$$

This equation shows (x y z) and (dx dy dz) are orthogonal (normal) relative to the semi definite quadratic form (1) of the Minkowski space. Since from equation (1) (x y z) is time-like, (dx dy dz) must be space-like since no two time-like vectors can be orthogonal. Consequently a surface line-element ds may be defined by

$$ds^2 = dx^2 + dy^2 - dz^2 \tag{4}$$

Substituting parametric values we find the 2-dimensional hyperbolic metric

$$ds^2 = R^2 (du^2 + \mathrm{sh}^2 u \, d\theta^2) \tag{5}$$

On integrating along a curve $\theta$ = const., there is found the distance $\rho$ from vertex at $u = 0$ to a point on the surface having parameter u as:

$$\rho = Ru \tag{6}$$

So that the parameter u has the interpretation

$$u = \rho/R \tag{7}$$

---

*Note*: For some history of this mathematical representation see. Wikipedia 'The hyperboloid model'. Its Special Relativity application was explicitly described by Wick CERN report 1973.
For further comments on representation in Minkowski space see the writer's 2015 paper 'Minkowski Space-time and hyperbolic geometry'.

## 6. The Beltrami projection

A point P on the surface of the hyperboloid can be projected from the origin O on to the plane tangent to the vertex V at z = R of the hyperboloid (diagram).

The whole upper surface is mapped inside a circle, *the Beltrami-Klein disc*. A general point on the hyperboloid at (R sh u cos θ, R sh u sin θ, R ch u) is mapped on to the point with coordinates

$$x = R \,\mathrm{th}\, u \cos\theta$$
$$y = R \,\mathrm{th}\, u \sin\theta$$
$$z = R \qquad (1)$$

whose radial distance to the vertex is given by

$$\sqrt{(x^2+ y^2)} = R \,\mathrm{th}\, u \qquad (2)$$

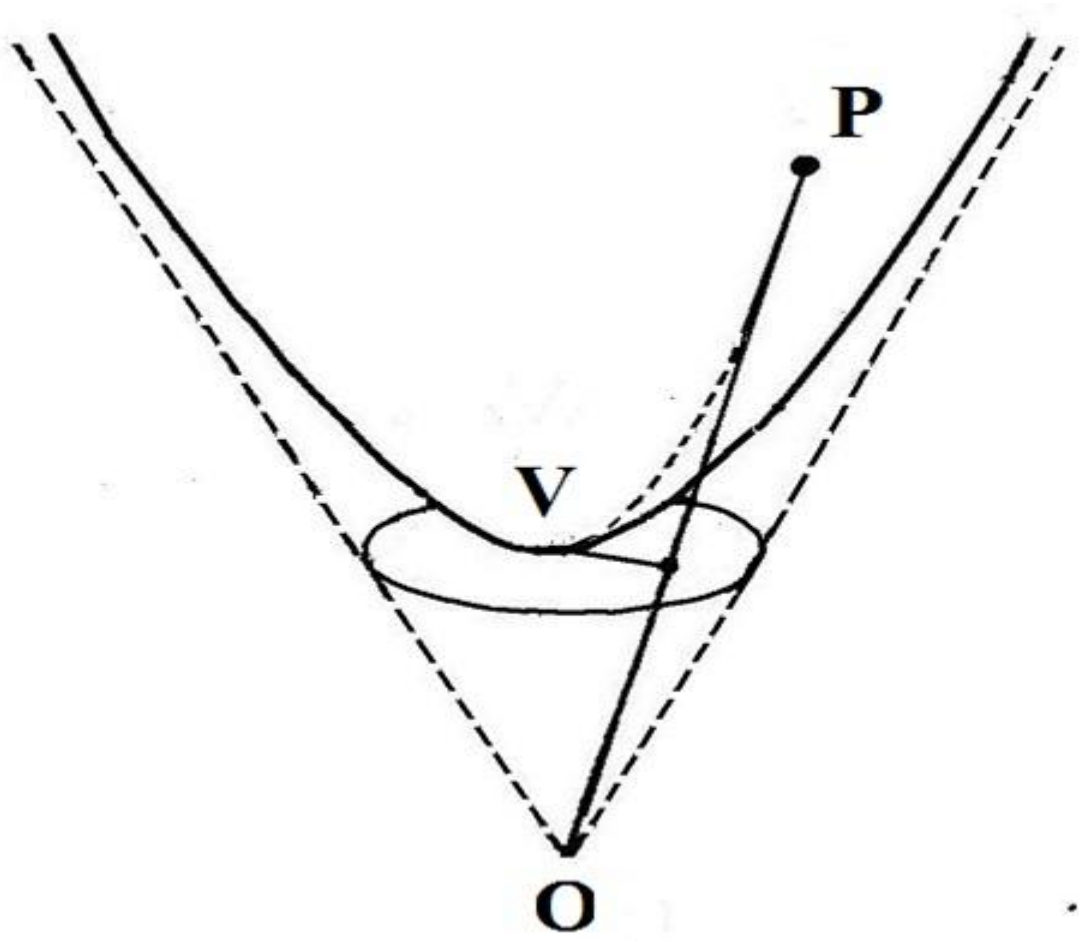

*Fig*: Beltrami projection

The representation can be considered either from the viewpoint of differential geometry (Beltrami) or from the viewpoint of projective geometry (Klein).

The metric of the disc due to Beltrami is (see mathematical appendix)

$$ds^2 = \frac{R^2\{R^2 (dx^2 + dy^2) + (y\, dx - x\, dy)^2\}}{(R^2 - x^2 - y^2)^2} \qquad (3)$$

On using the transformation of variables in (1) from x and y to u and θ (3) becomes identical with the metric element on the hyperboloid (5) of the last section. Thus lengths are preserved in the hyperbolic metric during projection from the hyperboloid on to the Klein disc

From the point of view of projective geometry, the plane determined by two points on the hyperboloid and the centre O cuts the hyperboloid in the geodesic joining the two points. So geodesics on the hyperboloid are projected into straight lines on the disc. In particular the geodesic from the vertex V to a point P on the hyperboloid projects into a radial segment on the disc from V as in the diagram.

The 3-dimensional form given here for illustration immediately extends to the 4-dimensional case needed in relativity, its principal use being the representation of velocity space.In this case 4-dimensional relativistic velocity represented by Minkowski's 4-vector, is projected into a 3 dimensional velocity vector in the ball representing velocities less than light.

Further details on the Beltrami projection are given in the mathematical appendix.

---

*Note*: Original references are Beltrami's two papers of 1868 (in Italian with English translations in Stillwell 1996) and Klein's Non-Euclidean geometry (in German). See also Coxeter Non-Euclidean Geometry.(with a different notation)

# CHAPTER 4 – Relative Velocity

## 1. Relative Velocities in Rectilinear Motion

In Special Relativity all velocities are relative but the standard literature pays little attention to relative velocities. It is necessary to give a general definition of relative velocities and the rules for combining them in a way independent of origin. This will be discussed in this chapter. with the limitation to relative velocity of two points moving in one plane.

*One dimensional* motion: The usual derivation of the addition of velocities uses the two frames S and S' of the Einstein formulation together with their origins O, O'. Origin O' of S' moves away from origin O of S with a velocity say $v_1$ and a point P moves relative to origin O' with velocity u. The velocity $v_2$ of point P relative to origin O is,by the Composition Rule

$$v_2 = \frac{v_1 + u}{1 + v_1 u / c^2} \tag{1}$$

From this formula follows

$$u = \frac{v_2 - v_1}{1 - v_2 v_1 / c^2} \tag{2}$$

This gives the relative velocity of two points (here O' and P) moving with velocities $v_1$ and $v_2$ relative to the one origin O. However, even though formula (2) has been deduced using the origin O, it is actually independent of any origin. This can be verified algebraically by substituting for $v_1$ and $v_2$ the expressions corresponding to another choice of origin using a further arbitrary relative velocity u'.

$$\frac{v_1 + u'}{1 + v_1 u' / c^2}, \qquad \frac{v_2 + u'}{1 + v_2 u' / c^2} \tag{3}$$

It is then found that the expression (2) for relative velocity remains unchanged. In this respect formula (2) differs from the origin–dependent composition formula (1)

Using this fact the composition formula can be restated in an origin-independent way suggested by Prokhovnik (1967) – see next section.

---

*Reference*: Prokhovnik :*The Logic of Special Relativity:*1967

*Re-derivation of the composition rule:* Consider the situation in the figure where 3 points $P_1$, $P_2$, $P_3$ move relative to origin O with velocities $v_1$, $v_2$ , $v_3$.

```
                                   O          P1          P2        P3
Fig. Composition of relative       o----------o-----------o---------o
      rectilinear velocities                  v1 →        v2 →      v3 →
```

The relative velocities $u_{2/1}$ of $P_2$ to $P_1$ and $u_{3/2}$ of $P_3$ to $P_2$, are

$$u_{2/1} = \frac{v_2 - v_1}{1 - v_2 v_1/c^2} \qquad u_{3/2} = \frac{v_3 - v}{1 - v_3 v_2/c^2} \tag{4}$$

and the relative velocity $u_{3/1}$ of $P_3$ relative to $P_1$ is

$$u_{3/1} = \frac{v_3 - v_1}{1 - v_3 v_1/c^2} \tag{5}$$

These relations are valid with any origin. After some algebra it is found the validity of the composition rule (6) which now involves only origin-independent relative velocities

$$u_{3/1} = \frac{u_{3/2} + u_{2/1}}{1 + u_{3/2} u_{2/1}/c^2} \tag{6}$$

*Use of hyperbolic velocities*: These relations simplify and become more transparent on using rapidities or the equivalent hyperbolic velocities. Suppose, as in the previous diagram, there are three moving points $P_1$, $P_2$, $P_3$ referred to an origin O. Then there are found, using an obvious subscript notation, the following relations between their rapidities w or hyperbolic velocities V

$$w_{2/1} = w_2 - w_1 \qquad V_{2/1} = V_2 - V_1 \tag{7}$$

$$w_{3/2} = w_3 - w_2 \qquad V_{3/2} = V_3 - V_2 \tag{8}$$

By addition,

$$w_{3/2} + w_{2/1} = w_3 - w_1 = w_{3/1} \qquad V_{3/2} + V_{2/1} = V_3 - V_1 = V_{3/1} \tag{9}$$

* *Galilean and Lorentzian translational invariance*: Use of rapidity and hyperbolic velocity makes clearer the distinction between two forms of translation invariance. The two forms are

$$v \rightarrow v + u \qquad \text{(Galilean)} \tag{10}$$

$$V \rightarrow V + U \qquad \text{(Lorentzian)} \tag{11}$$

## 2. Fock’s definition of relative velocity

As suggested by the rectilinear case, the natural definition of the relative velocity of two points having velocities $\mathbf{v}_1$ and $\mathbf{v}_2$ would be found by writing the Einstein composition rule for velocity difference $-\mathbf{v}_1 + \mathbf{v}_2$. If θ is the angle between the velocities. this would lead to

$$\mathbf{v}_{2/1} = \frac{\{v_2 \cos\theta - v_1\}\mathbf{n}_1 + \sqrt{(1 - v_1^2/c^2)}\, v_2 \sin\theta\, \mathbf{n}_1^{\perp}}{\{1 - v_1 v_2/c^2 \cos\theta\}} \qquad (1)$$

The equation is illustrated in the figure showing classical mechanics resolution of $\mathbf{v}_2 - \mathbf{v}_1$ into components parallel and orthogonal to $\mathbf{v}_1$. The characteristic feature of formula (3) is the multiplication of the orthogonal component by the root factor $\sqrt{(1 - v_1^2/c^2)}$.

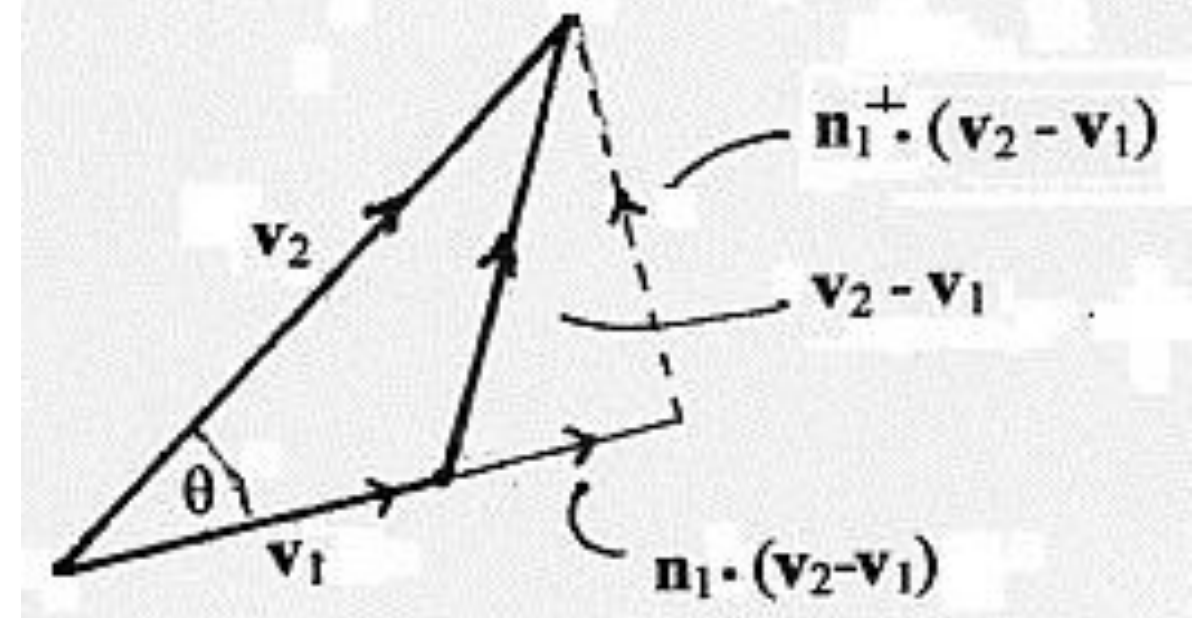

*Fig*. Components of classical relative velocity

*Fock's derivation*: This expression (1) was derived by Fock (1955) by transferring the origin from O to $P_1$ so considering $P_1$ to be at rest. As in the classical case, the velocity of $P_2$ with respect to this new origin is defined as the velocity of $P_2$ relative to $P_1$.

By the same definition the reverse relative velocity is

$$\mathbf{v}_{1/2} = \frac{\{v_1 \cos\theta - v_2\}\mathbf{n}_2 + \sqrt{(1 - v_2^2/c^2)}\, v_1 \sin\theta\, \mathbf{n}_2^{\perp}}{\{1 - v_1 v_2/c^2 \cos\theta\}} \qquad (2)$$

The two relative velocities are not negatives of one another but have the same magnitude given by the difference form of Einstein’s composition rule:

$$v_{2/1}^2 = v_{1/2}^2 = \frac{(v_1^2 - 2\, v_1 v_2 \cos\theta + v_2^2) - (v_1 v_2/c)^2 \sin^2\theta}{\{1 - v_1 v_2/c^2 \cos\theta\}^2} \qquad (3)$$

$$= \frac{\sqrt{\{|\mathbf{v}_1 - \mathbf{v}_2|^2 - |\mathbf{v}_1 \times \mathbf{v}_2|^2/c^2\}}}{\{1 - \mathbf{v}_1.\mathbf{v}_2/c^2\}^2} \qquad (4)$$

The two relative velocities consequently differ only in direction. The rotational effect is taken into account in the matrix definition given in the next section.

---

## 3. Relative velocity and hyperbolic geometry

Since the magnitude of the relative velocity is given by the Einstein composition formula for the difference of velocities, the corresponding calculation for velocity addition given previously gives the relation

$$1-\frac{v^2}{c^2} = \frac{(1-v_1^2/c^2)(1-v_2^2/c^2)}{(1-v_1 v_2 \cos\theta/c^2)^2} \qquad (1)$$

between the velocities $v_1$, $v_2$, the angle θ between them, and the relative velocity v. Taking square roots and inverting results in

$$\frac{1}{\sqrt{(1-v^2/c^2)}} = \frac{(1-v_1 v_2 \cos\theta/c^2)}{(1-v_1^2/c^2)(1-v_2^2/c^2)} = \frac{c^2 - \mathbf{v_1 v_2}}{\sqrt{[(c^2-v_1^2)(c^2-v_2^2)]}} \qquad (2)$$

Using corresponding rapidities, this gives for the relative rapidity w

$$\mathrm{ch}\, w = \mathrm{ch}\, w_1\, \mathrm{ch}\, w_2 - \mathrm{sh}\, w_1\, \mathrm{sh}\, w_2 \cos\theta \qquad (3)$$

or equally in terms of hyperbolic velocities V, $V_1$, $V_2$

$$\mathrm{ch}\, V/c = \mathrm{ch}\, V_1/c\, \mathrm{ch}\, V_2/c - \mathrm{sh}\, V_1/c\, \mathrm{sh}\, V_2/c \cos\theta \qquad (4)$$

Here w, V may be denoted by $w_{2/1}$, $V_{2/1}$ The relationships between $V_{2/1}$, $V_1$, $V_2$ may be represented by the triangle in hyperbolic space as shown.

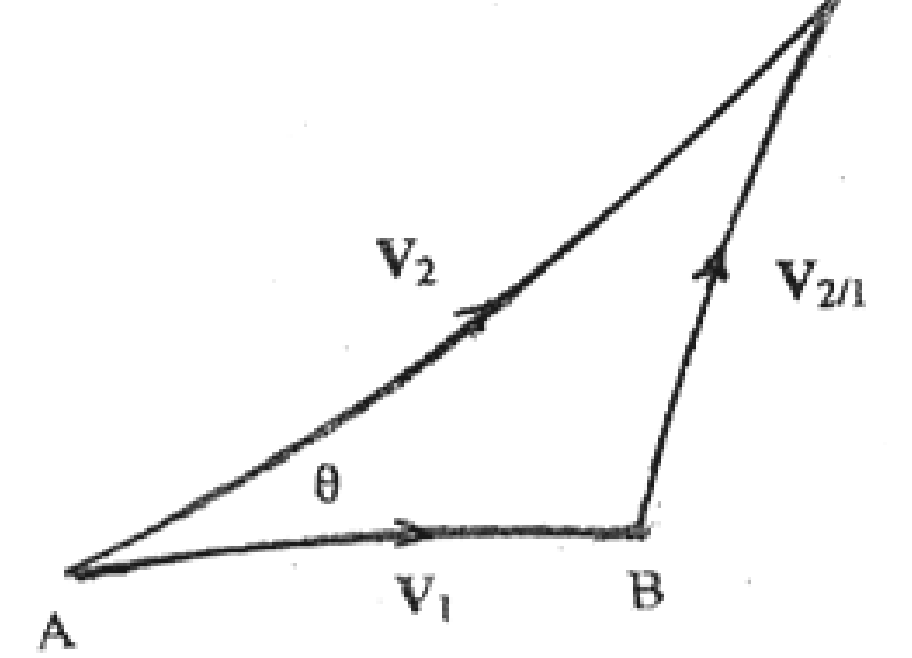

*Fig*: Relative hyperbolic velocity

An alternative hyperbolic representation is the formula for relative velocity in the Beltrami-Klein disc which is essentially the same as Fock's definition already described which includes the square root factor for the relative velocity. The (non-dimensional) metric in this space is given by the magnitude of the relative rapidities. $w_{1/2}$, $w_{2/1}$ satisfying the above equation

$$c\, \mathrm{ch}\, w_{1/2} = c\, \mathrm{ch}\, w_{2/1} = \frac{c^2 - \mathbf{v}_1.\mathbf{v}_2.}{\sqrt{(c^2-\mathbf{v}_1.\mathbf{v}_1)(c^2-\mathbf{v}_2.\mathbf{v}_2)}} \qquad (5)$$

---

*Historical note* :The relation between velocity and rapidity formulae was apparently first given by Weyl: Space-Time Matter (1922).p.182, chapter iii. Later Fock (1955) realized the importance of hyperbolic space for the correct representation of relative velocity in Special Relativity and introduced the appropriate Beltrami-Klein representation

# 4. Matrix Representation of Relative Velocity

Relative velocity may be defined by a similar idea to that used previously for the one dimensional case where all motions are referred to an observer O. For two moving points $P_1$, $P_2$ the relations between coordinate changes referred to an observer's frame at O are, using appropriate suffices

$$\begin{bmatrix} cdt_1 \\ \mathbf{dr}_1 \end{bmatrix} = \left[L(\mathbf{v}_1)\right] \begin{bmatrix} cdt_0 \\ \mathbf{dr}_0 \end{bmatrix}, \qquad \begin{bmatrix} cdt_2 \\ \mathbf{dr}_2 \end{bmatrix} = \left[L(\mathbf{v}_2)\right] \begin{bmatrix} cdt_0 \\ \mathbf{dr}_0 \end{bmatrix} \tag{1}$$

The matrices here are Lorentz matrices.From this follows

$$\begin{bmatrix} cdt_2 \\ \mathbf{dr}_2 \end{bmatrix} = \left[L(\mathbf{v}_2)\right]\left[L(\mathbf{v}_1)\right]^{-1} \begin{bmatrix} cdt_1 \\ \mathbf{dr}_1 \end{bmatrix} \tag{2}$$

Although matrices $L(\mathbf{v}_2)$, $L(\mathbf{v}_1)$ are defined relative to O, their product $L(\mathbf{v}_2)\, L(\mathbf{v}_1)^{-1}$ is independent of O. For supposing that the relation between coordinate changes between two different observers O, O' is

$$\begin{bmatrix} cdt_0 \\ \mathbf{dr}_0 \end{bmatrix} = \left[L(\mathbf{v}_0)\right] \begin{bmatrix} cdt'_0 \\ \mathbf{dr}'_0 \end{bmatrix} \tag{3}$$

the matrices $L(\mathbf{v}_2)$, $L(\mathbf{v}_1)$ will transform by

$$L(\mathbf{v}_2) \rightarrow L(\mathbf{v}_2)\, \Lambda(\mathbf{v_0}), \quad L(\mathbf{v}_1) \rightarrow L(\mathbf{v}_1)\, \Lambda(\mathbf{v_0}) \tag{4}$$

leaving the product $L(\mathbf{v}_2)\, L(\mathbf{v}_1)^{-1}$ unchanged. The relative velocity matrix is consequently well defined independently of observer as

$$\Lambda_{2/1} = L(\mathbf{v}_2)\, L(\mathbf{v}_1)^{-1} \tag{5}$$

From this there follows for the inverse

$$\Lambda_{1/2} = L(\mathbf{v}_1)\, L(\mathbf{v}_2)^{-1} = (\Lambda_{2/1})^{-1} \tag{6}$$

Note that a relative velocity of two moving points is represented by a pure Lorentz matrix when and only when the origin is taken at one of the moving points. This was in fact what Fock had done in his derivation which could give the incorrect impression that a relative velocity in general can be so represented.

# 5 Re-derivation of Fock's Expression for Relative Velocity

The product $L(\mathbf{v}_2)\, L(\mathbf{v}_1)^{-1}$ written as $L(\mathbf{v}_2)\, L(-\mathbf{v}_1)$ may be written in the form $R\, L(\mathbf{v})$ where R is a spatial rotation and $L(\mathbf{v})$ a Lorentz matrix. More explicitly it will be

$$\begin{bmatrix} 1 & 0 \\ 0 & \Omega \end{bmatrix} \begin{bmatrix} \gamma & -\gamma \mathbf{v}^T / c \\ -\gamma \mathbf{v} / c & I + (\gamma - 1)\mathbf{n}\mathbf{n}^T \end{bmatrix} \tag{1}$$

Here $\Omega$ is a 3x3 spatial rotation matrix. On forming the product there is found

$$\begin{aligned} \gamma \quad &= \gamma_1 \gamma_2 \{1 - v_2^T v_1/c^2\} \\ \gamma\, \mathbf{v} \quad &= \gamma_2 \{(I + (\gamma_1 - 1)\, \mathbf{n}_1 \mathbf{n}_1^T)\, \mathbf{v}_2 - \gamma_1 \mathbf{v}_1\} \\ \gamma\, (\Omega\, \mathbf{v}) &= \gamma_1 \{(I + (\gamma_2 - 1)\, \mathbf{n}_2 \mathbf{n}_2^T)(-\mathbf{v}_1) + \gamma_2 \mathbf{v}_2\} \end{aligned} \tag{2}$$

From the first and second of these equations is found (3) and slight rearrangement gives Fock's expression (4) and (5) so identifying $\mathbf{v}$ as $\mathbf{v}_{2/1}$

$$\mathbf{v} = \frac{\{I + (\gamma_1 - 1)\, \mathbf{n}_1 \mathbf{n}_1^T\}\mathbf{v}_2 - \gamma_1 \mathbf{v}_1}{\gamma_1 \{1 - \mathbf{v}_1^T \mathbf{v}_2/c^2\}} \tag{3}$$

$$= \frac{\mathbf{v}_2 - \mathbf{v}_1 + (\gamma_1 - 1)\mathbf{n}_1 \{(\mathbf{n}_1^T\, \mathbf{v}_2) - v_1\}}{\gamma_1\, (1 - \mathbf{v}_1^T\, \mathbf{v}_2/c^2)} \tag{4}$$

$$= \frac{\mathbf{n}_1 \mathbf{n}_1^T (\mathbf{v}_2 - \mathbf{v}_1) + \sqrt{(1 - v_1^2/c^2)}\, (I - \mathbf{n}_1 \mathbf{n}_1^T)\, (\mathbf{v}_2 - \mathbf{v}_1\}}{(1 - \mathbf{v}_1^T\, \mathbf{v}_2/c^2)} \tag{5}$$

Interchange of suffixes gives the reverse relative velocity:

$$\mathbf{v}_{1/2} = \frac{\mathbf{n}_2 \mathbf{n}_2^T (\mathbf{v}_1 - \mathbf{v}_2) + \sqrt{(1 - v_2^2/c^2)}\, (I - \mathbf{n}_2 \mathbf{n}_2^T)\, (\mathbf{v}_1 - \mathbf{v}_2)}{(1 - \mathbf{v}_1^T\, \mathbf{v}_2/c^2)} \tag{6}$$

From (2) above it is seen that the two relative velocities arerelated by

$$\Omega\, \mathbf{v}_{2/1} = \mathbf{v}_{1/2} \tag{7}$$

The matrix (1) may be written symmetrically in canonical form as

$$\begin{bmatrix} \gamma & -\beta\gamma\, \mathbf{n}_{2/1}^T \\ -\beta\gamma\, \mathbf{n}_{1/2} & \Omega + (\gamma - 1)\, \mathbf{n}_{1/2} \mathbf{n}_{2/1}^T \end{bmatrix} \tag{8}$$

## 6. Composition of Relative Velocity Matrices

Assume three points $P_1$, $P_2$, $P_3$ move with velocities $\mathbf{v}_1, \mathbf{v}_2, \mathbf{v}_3$ relative to an observer O. The relative velocity matrices are, with L() denoting Lorentz transformations and Λ() general Lorentz transformations,

$$\begin{aligned} \Lambda_{2/1} &= L(\mathbf{v}_2)\, L(\mathbf{v}_1)^{-1} \\ \Lambda_{3/2} &= L(\mathbf{v}_3)\, L(\mathbf{v}_2)^{-1} \\ \Lambda_{3/1} &= L(\mathbf{v}_3)\, L(\mathbf{v}_1)^{-1} \end{aligned} \qquad (1)$$

from which follows the composition rule:

$$\Lambda_{3/1} = \Lambda_{3/2}\, \Lambda_{2/1} \qquad (2)$$

showing the transitivity of matrix multiplication for connected relative velocities. This equation can obviously be extended to any number of consecutive stages.

* *Moving frames of reference*: In the standard situation of systems S and S' with S' moving with velocity **v** relative to S, consider the motion of a point P moving with velocities **u**, **u'** relative to S and S'. As seen by an observer at rest at the origin O of the S frame, the velocity of P relative to the origin O' in the S' frame is

$$\Lambda(\mathbf{u'}) = L(\mathbf{u})\, L(\mathbf{v})^{-1} \qquad (3)$$

where L(.) denotes a Lorentz transformation matrix. Now the velocity **u** will be defined by the equation

$$L(\mathbf{u'})\, L(\mathbf{v}) = R\, L(\mathbf{u}) \qquad (4)$$

R being here a rotation matrix So the observed relative velocity of P is

$$\Lambda(\mathbf{u'}) = R^{-1}\, L(\mathbf{u'}) \qquad (5)$$

The transformation law then takes the form

$$L(\mathbf{u}) = \{R^{-1}\, L(\mathbf{u'})\} L(\mathbf{v}) = \Lambda(\mathbf{u'})\, L(\mathbf{v}) \qquad (6)$$

which appears to imply that it is not possible to find L(**u**) by just multiplying Lorentz matrices L(**v**) and L(**u'**) as in, for example, the 1909 Sommerfeld paper in chapter 2

# CHAPTER 5 – Applications to Optics

## 1. Aberration and Oblique Doppler Effect

Following the original idea of Bradley on stellar aberration, the observer is thought of as moving while the source of light is considered to be stationary. So frame S will represent the stationary source frame for emitted light and frame S' the moving frame of the observer.

* *The principle of invariance of the phase*: If the incoming wave has direction cosines (l, m, n) relative to the source frame S, the phase relative to S is

$$\Phi = 2\pi f\{t - (lx + my + nz)/c\} \qquad (1)$$

f being frequency in Herz. Similarly if the wave has direction cosines (l', m', n') relative to the observer frame S' the phase relative to S' is

$$\Phi = 2\pi f'\{t' - (l'x' + m'y' + n'z')/c\} \qquad (2)$$

The basis of the calculation is that these two expressions must be the same. This is the 'Principle of invariance of the phase' used by Lorentz (1886,1895) with classical transformation formulae from which he found approximately correct formulae for both Doppler effect and aberration. Einstein (1905) used the same method with the corrected transformation and so found the relativistic formulae which have since become very much used. Varićak rewrote them in terms of rapidity

If the incoming wave is inclined at an angle φ to the direction of motion of the observer relative to the source along the x-axis then the phase of the incoming wave relative to the two frames S can be written in the simpler form

$$\Phi = 2\pi f\{t - (x\cos\varphi + y\sin\varphi)/c\}$$
$$\Phi = 2\pi f'\{t' - (x'\cos\varphi' + y'\sin\varphi')/c\} \qquad (3)$$

so that the principle of invariance gives

$$f(ct - x\cos\varphi - y\sin\varphi) = f'(ct' - x'\cos\varphi' - y'\sin\varphi') \qquad (4)$$

*(a) True values in terms of observed values*: from

$$ct' = \gamma(ct - xv/c)$$
$$x' = \gamma(x - vt)$$
$$y' = y \qquad (5)$$

---

* *Reference*: For a description of Lorentz's work on aberration and Doppler Effect see Miller (1998)

there are found the equations

$$f = f' \frac{(1 + v/c \cos \varphi')}{\sqrt{(1-v^2/c^2)}} \tag{6}$$

$$\cos \varphi = \frac{\cos \varphi' + v/c}{1 + v/c \cos \varphi'} \tag{7}$$

$$\sin \varphi = \frac{\sqrt{(1-v^2/c^2)} \sin \varphi'}{1 + v/c \cos \varphi'} \tag{8}$$

*(b) Observed values in terms of true values*: Similarly, starting from

$$\begin{aligned} ct &= \gamma\,(ct' + v\,x'/c) \\ x &= \gamma\,(x' + vt') \end{aligned} \tag{9}$$

there are found similar formulae with change in sign for v

$$f' = f \frac{(1 - v/c \cos \varphi)}{\sqrt{(1 - v^2/c^2)}} \tag{10}$$

$$\cos \varphi' = \frac{\cos \varphi - v/c}{1 - v/c \cos \varphi} \tag{11}$$

$$\sin \varphi' = \frac{\sqrt{(1-v^2/c^2)} \sin \varphi}{1 - v/c \cos \varphi} \tag{12}$$

* *Transverse Doppler Effect*: The transverse Doppler Effect occurs when a light source is observed at right angles to the direction of motion. This requires solving for $f'$ when $\varphi'$ is $\pi/2$. Putting $\cos \varphi$ in (11) in terms of $\cos \varphi'$ using (7), the observed frequency $f'$ in terms of $\varphi'$ is found as (14) giving transverse Doppler Effect (15) when $\varphi'$ is $\pi/2$.

$$f' = f \frac{\sqrt{(1-v^2/c^2)}}{(1 + v/c.\cos \varphi')} \tag{14}$$

$$f' = f \sqrt{(1-v^2/c^2)} \tag{15}$$

-----------------------------------------------------------------------------------------------------------

* *Reference*: The books of Stevenson & Kilminster (1958) and Prokhovnik (1969) have interesting comments on the transverse Doppler Effect and its history

# 2 The Radial Doppler Formula

For observation in the direction of motion, $\varphi = 0$ so that from the previous formula (10)

$$f' = f \sqrt{\frac{1 - v/c}{1 + v/c}} \tag{1}$$

This is easily found directly from the simplified form (2) taken by the phase and making a Lorentz transformation along the x-axis to new coordinate x', t' (Einstein 1907)

$$\Phi \ = \ 2\,\pi\, f\,(t - x/c) \tag{2}$$

As frequency f and wave length $\lambda$ satisfy $f\,\lambda \ = \ c$. formula (1) may equally be written

$$\lambda' = \lambda \sqrt{\frac{1 + v/c}{1 - v/c}} \tag{3}$$

*Transitivity of relativistic Doppler shift*: The relativistic formula has the *transitivity* property showing its dependence only on relative velocity. Suppose $P_0$, $P_1$, $P_2$ are three collinear points on the path of the wave having velocities $v_0$, $v_1$, $v_2$ and relative velocities v' (= $v_{1/0}$), v" (= $v_{2/1}$), v (= $v_{2/0}$) as in the figure below.

|……………. .v→……………  ...|
|……v'→... .....|……...v"→……...|

*Fig*. Doppler Effect with collinear moving points

---o----------------o---------------------o----

$P_0$ $P_1$ $P_2$

$v_0$ $v_1$ $v_2$

Then v can be obtained from v', v" by the velocity composition formula which can be written

$$\frac{1 + v/c}{1 - v/c} \ = \ \frac{1 + v'/c}{1 - v'/c}\ \frac{1 + v''/c}{1 - v''/c} \tag{4}$$

Taking square roots this implies consistency of the formulae (5) with (6) below.

$$\lambda_1 = \lambda_0 \sqrt{\frac{1 + v'/c}{1 - v'/c}}\ , \quad \lambda_2 = \lambda_1 \sqrt{\frac{1 + v''/c}{1 - v''/c}} \tag{5}$$

$$\lambda_2 = \lambda_0 \sqrt{\frac{1 + v/c}{1 - v/c}} \tag{6}$$

* *A modified linear redshift formula*: The transitivity property is important as it ensures the objectivity of measurements based on Doppler Effect which otherwise may depend on the source of the light. However, the linear Doppler formula used at present in cosmology only has the transitivity property for small (infinitesimal) redshifts. This formula is

$$\lambda ' = \lambda (1 + v/c) \tag{7}$$

It can be derived from binomial theorem approximation to the wave length relativistic formula (3). It is usually written as (8) in terms of z the *redshift* (9)

$$z = v/c \tag{8}$$

$$z = (\lambda ' - \lambda)/ \lambda \tag{9}$$

Although the linearized Doppler formula (8) does not have the transitivity property it can easily be modified so that it does. This is done by using the *logarithmic redshift* which becomes equal to z at small redshifts

$$Z = \ln (\lambda '/ \lambda) = \ln (1+z) \tag{10}$$

On taking logarithms in equation (3) there follows

$$\ln \sqrt{\frac{1+v/c}{1-v/c}} = \text{th}^{[-1]}(v/c) = \tag{11}$$

The right hand side is V/c so from this comes the modified Doppler redshift formula

$$Z = V/c \tag{12}$$

Z and V are both additive over subintervals as in the figure. below. So the transitive property is restored to the linear Doppler redshift formula

|……………...V→……………...|
|……V' →..…|..……...V''→……..|
$P_0$ $P_1$ $P_2$
----o--- ------------o-----------------------o--

< ------Z'----- >< -----------Z''------- >
< ----------- Z = Z' + Z'' ----------- >

*Fig*. The additive relation between hyperbolic velocity and logarithmic Doppler shift

---

*Note*: Transitivity of Einstein's radial Doppler formula was noted by Prokhovnik (1969),and Jánossy (1971). The classical Doppler formula had transitivity but it became lost in the linearized version

** Properties of logarithmic Doppler shift:* The quantity Z defined here gives an alternative and improved measure of Doppler shift. It is a strictly increasing function of the wave-length ratio $\lambda'/\lambda$. It reduces to normal redshift for low relative velocities since, when $\lambda' - \lambda$ is small,

$$Z = \ln \{1 + (\lambda' - \lambda)/\lambda\} \approx (\lambda' - \lambda)/\lambda = \Delta\lambda/\lambda = z \qquad (13)$$

The following properties of logarithmic Doppler shift Z contrast favourably with those of normal Doppler redshift z for which they fail to hold exactly:

(a) Change from wave-length to frequency or vice versa merely involves change of sign

$$\ln(\lambda'/\lambda) = -\ln(f'/f) \qquad (14)$$

(b) Z has an anti-symmetry between emitter and receiver undergoing a simple sign change when these are reversed:

$$\ln(\lambda'/\lambda) = -\ln(\lambda/\lambda'), \qquad \ln(f'/f) = -\ln(f/f') \qquad (15)$$

(c) Z is additive (transitive) as expressed by the equations

$$\ln(\lambda'/\lambda) + \ln(\lambda''/\lambda') = \ln(\lambda''/\lambda),$$
$$\ln(f'/f) + \ln(f''/f') = \ln(f''/f) \qquad (16)$$

---

** Notes*:
The logarithmic Doppler shift measure can also be used in General Relativity. Using the known formulae for cosmological redshift and time-varying Hubble parameter H(t):

$$\frac{\lambda_2}{\lambda_1} = \frac{R(t_2)}{R(t_1)}, \qquad H(t) = \frac{1}{R}\frac{dR(t)}{dt}$$

there is found the generalized logarithmic redshift law

$$Z = \ln(\lambda_2/\lambda_1) = \ln\{R(t_2)/R(t_1)\} = \ln R(t_2) - \ln R(t_1) = \int H(t)\,dt$$

Further details are given in the writer’s 1992 PIRT paper.(available on Researchgate)

## 3. The Hyperbolic Interpretation of Relativistic Aberration

The wave theory interpretation of aberration loses contact with the simple picture of the triangular velocity addition of the classical Bradley theory. This picture is partially restored by using the velocity composition formulae. Consider an incoming photon moving with velocity components relative to the source frame S of

$$u_x = -c\cos\varphi, \quad u_y = -c\sin\varphi \tag{1}$$

Using the velocity composition formula the corresponding components relative to the observer frame S ' are

$$u_x' = -\frac{(c\cos\varphi + v)}{(1 + v/c.\cos\varphi)} \tag{2}$$

$$u_y' = -\frac{c\sin\varphi\sqrt{(1 - v^2/c^2)}}{(1 + v/c.\cos\varphi)} \tag{3}$$

Here

$$u_x'^{\,2} + u_y'^{\,2} = c^2 \tag{4}$$

so that it is permissible to put

$$u_x' = -c\cos\varphi', \quad u_y' = -c\sin\varphi' \tag{5}$$

when substitution leads directly to the aberration formulae. There is however a difficulty. Since both $(u_x, u_y)$ and $(u_x', u_y')$ have magnitude c it is not possible to form a Euclidean triangle of velocities. This difficulty is overcome in the non-Euclidean representation.

The aberration formula resulting from the substitution (5) in (2) may be written

$$c.\cos\varphi' = \frac{c\cos\varphi + v}{1 + \cos\varphi.\ v/c} \tag{6}$$

It compounds the forward component of the light velocity $c\cos\varphi$ with v by the composition rule for rectilinear velocities. So introducing rapidity components w, w ', W by

$$\cos\varphi = \text{th}\, w, \quad \cos\varphi' = \text{th}\, w', \quad v = c\,\text{th}\, W \tag{7}$$

(6) becomes

$$c\cos\varphi' = \frac{c\,\text{th}\, w + c\,\text{th}\, W}{1 + \text{th}\, w.\ \text{th}\, W} = c\,\text{th}\,(w + W) \tag{8}$$

The simple addition

$$w' = w + W \tag{9}$$

of the x-components together with the fact that the y component of the light rapidity remains infinite makes it possible to reconstruct the triangle of rapidities as shown in the figure.

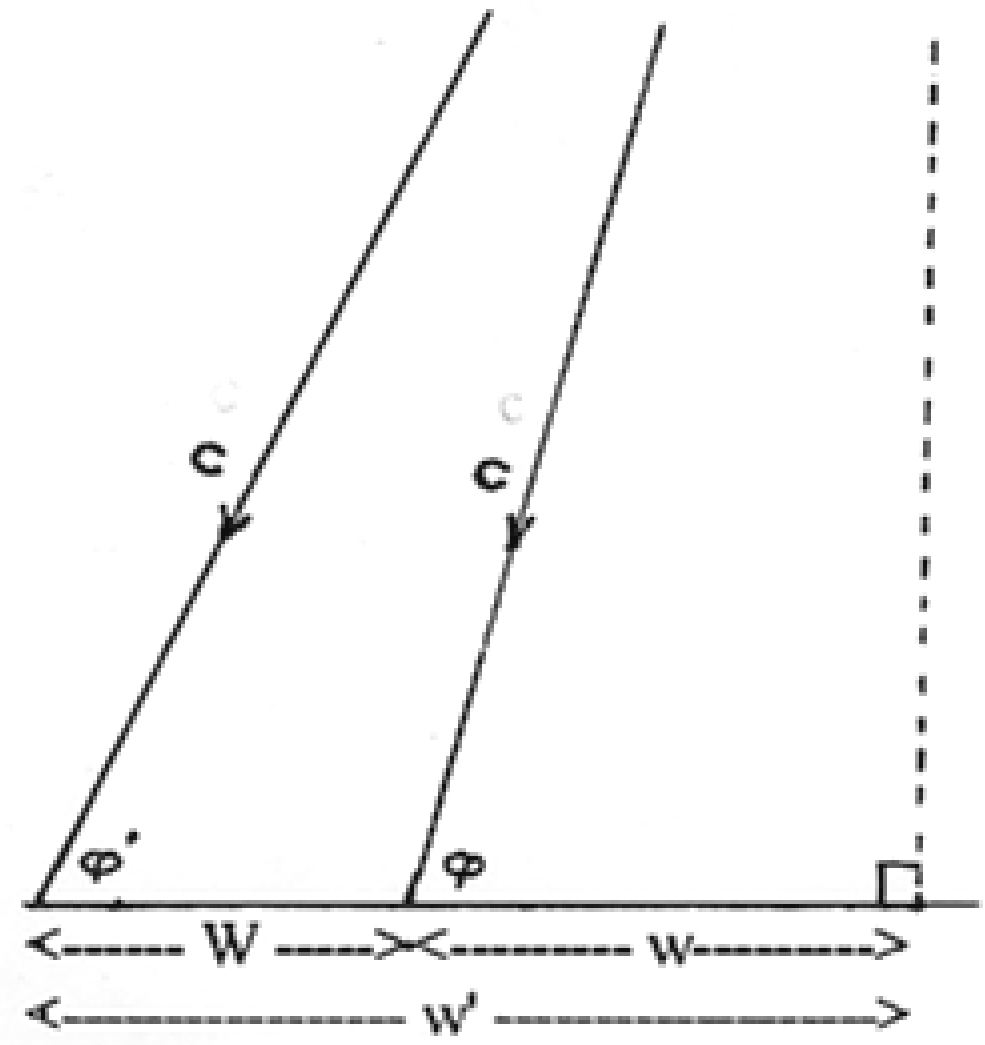

*Fig*: The triangle of rapidities in aberration.

Here the two sides representing velocity of light have become infinite, their components in the direction of motion remaining finite. The light paths are both Lobachevski parallels to the vertical line representing the transverse component of light velocity. Angles φ', φ are Lobachevski angles of parallelism as shown by the equations (5). In this way the triangle addition of Bradley's theory is restored.

---

*Note*: The use of Lobachevski parallels in light propagation was described by Varićak (1910a) and mentioned by Silberstein in his 1914 book but it did not appear in later literature.

# CHAPTER 6 – Applications to Dynamics

## 1. Hyperbolic Acceleration

Hyperbolic acceleration will be here defined as rate of change $dV/d\tau$ of hyperbolic velocity $V$ with respect to proper time $\tau$, the time observed relative to the moving body. In one dimension it coincides with the known expression in equation (1) for rectilinear acceleration given e.g. by Rindler 1991:

$$\alpha = \frac{1}{(1-(v/c)^2)^{3/2}} \frac{dv}{dt} \tag{1}$$

$$= \frac{1}{(1- v^2/c^2)^{1/2}} \frac{1}{1 - v^2/c^2} \frac{dv}{dt} = \frac{1}{(1- v^2/c^2)^{1/2}} \frac{d}{dt}\{th^{[-1]}\, v/c\} = \frac{dV}{d\tau} \tag{2}$$

Relative to the moving body v is zero and then from (2) hyperbolic acceleration is represent-ed by the normal derivative expression. Hyperbolic acceleration $dV/d\tau$ has a more familiar form than the expression (1) with its multiplying factor. It may also be defined more intuitively as follows.

Consider the situation in the figure below representing an accelerating point P.

```
              v →        v +δv →
 0 |-----------o P --------o P'--------------------------→
               x          x +δx                x -axis
```

*Fig*: On the definition of rectilinear acceleration

In any time increment $\delta t$, the increase in velocity $\delta v$ of an accelerating body does not occur at a fixed value of $x$ relative to an origin O but takes place over the interval $x$ to $x+\delta x$. The increase of velocity relative the origin O should consequently be calculated by the relative velocity formula as

$$\frac{(v+\delta v)-v}{1 - (v+\delta v).v/c^2} = \frac{\delta v}{1 - (v+\delta v).v/c^2}. \tag{3}$$

which gives, as $\delta t \rightarrow 0$ and $\delta v \rightarrow 0$, a first order increment of

$$\frac{\delta v}{1 - v^2/c^2} = c\, \delta\{ th^{[-1]}\, v/c \} = \delta V \tag{4}$$

Taking account of the time change from the origin to the frame of the moving particle, the acceleration is found as the hyperbolic acceleration as defined above.

## 2. Motion under Constant Acceleration

The relativistic motion of a particle under constant acceleration was analysed in detail by Born (1909) who called it 'hyperbolic motion' since the equation of the trajectory in the x, t plane is a hyperbola instead of the parabola of classical physics. To find the trajectory it is necessary to integrate the equation

$$\frac{dv}{(1-v^2/c^2)^{3/2}} = \alpha\ dt \tag{1}$$

This gives, if v = 0 when t=0,

$$\frac{v}{\sqrt{(1-v^2/c^2)}} = \alpha t \tag{2}$$

On solving for v and again integrating, there is found

$$x - x_0 = \int_0^t v\,dt = \int_0^t \frac{\alpha t}{\sqrt{(1+(\alpha t/c)^2)}}\,dt \tag{3}$$

which gives

$$x = x_0 + (c^2/\alpha)\left[\sqrt{(1+(\alpha t/c)^2)} - 1\right] \ = \ x_0 + \alpha t^2/2 - \ldots \tag{4}$$

from which the trajectory is found as the hyperbola

$$(x - x_0 + b)^2 - c^2 t^2 = b^2 \tag{5}$$

where b is the constant $c^2/\alpha$.

It is convenient to parameterize the trajectory as

$$x = x_0 - b + b\ \mathrm{ch}\ u, \quad ct = b\ \mathrm{sh}\ u \tag{6}$$

so that

$$dx = b\ \mathrm{sh}\ u\ du, \quad dt = (b/c)\ \mathrm{ch}\ u\ du \tag{7}$$

from which u can be identified with rapidity from

$$dx/dt = c\ \mathrm{th}\ u \tag{8}$$

---

Born 'Die Theorie des starren Elektrons ...' *Ann. Phys*. 1909, On hyperbolic motion see e.g. Pauli: *Theory of Relativity,* The integration method here follows Prokhovnik: *Theory of Relativity* 1967.

Further,

$$d\tau = \sqrt{\{dt^2 - (dx.c)^2\}} = b\, du \qquad (9)$$

again leading to the value of hyperbolic acceleration of

$$dV/d\tau = c\, du/d\tau = c^2/ b = \alpha \qquad (10)$$

Integration gives an analogue of Galileo's law for velocity increase as in elementary mechanics:

$$V = V_0 + \alpha\tau \qquad (11)$$

* *The Minkowski interpretation*: The current standard interpretation of the motion uses the ideas of Minkowski. The hyperbola is centralized by choosing $x_0$ equal to b in which case.

$$x^2 - c^2 t^2 = b^2 \qquad (12)$$

and given the complex representation

$$x^2 + (ict)^2 = b^2 \qquad (13)$$

In polar coordinates using an imaginary angle φ this is

$$\begin{aligned} x &= b \cos \varphi \\ ict &= b \sin \varphi \end{aligned} \qquad (14)$$

Since b is constant, differentiation gives:

$$\begin{aligned} dx &= - b \sin \varphi\, d\varphi \\ ic\, dt &= b \cos \varphi\, d\varphi \end{aligned} \qquad (15)$$

There follows

$$d\tau = \sqrt{(dt^2 - (dx/c)^2)} = bi/c.\, d\varphi \qquad (16)$$

Consequently φ increases uniformly with τ giving a picture of constant angular speed circular motion which Sommerfeld (1919) called 'cyclic motion'. It has a constant (purely imaginary) central acceleration in the x, ict plane caused by the applied force.

---

* *References*: See: Minkowski 'Raum und Zeit' 1908 (The translation given in the Dover reprint '*The Principle of Relativity*' is annotated on p.96 with Sommerfeld's 1923 notes from Sommerfeld *Atombau und Spektrallinien*.1919 etc.p.320)

## 3. Newton's Second Law for Rectilinear Motion

In 1906 Planck showed that Newton's second law of motion for rectilinear motion can be written in the relativistic form

$$F = \frac{dp}{dt} \qquad (1)$$

F is applied force and p is momentum given by

$$p = \frac{mv}{\sqrt{(1 - v^2/c^2)}} = mv\,\{1 + 1/6\,(v/c)^2 + ...\} \qquad (2)$$

m being what is usually called the rest mass. Planck's equation may easily be transformed to involve (hyperbolic) acceleration:

$$\frac{dp}{dt} = m\frac{d}{dt}\frac{v}{\sqrt{(1-v^2/c^2)}} = m\frac{1}{(1-v^2/c^2)^{3/2}}\frac{dv}{dt} = m\,\alpha \qquad (3)$$

This equation may also be written

$$\frac{dp}{dt} = m\frac{dV}{d\tau} \qquad (4)$$

Then Newton's law takes the familiar form

$$F = m\,\alpha = m\frac{dV}{d\tau} \qquad (5)$$

In words,

*Force = Rest mass x (hyperbolic) acceleration* (6)

This relation may be derived even more directly from the representation

$$p = mc\ \mathrm{sh}\ V/c \qquad (7)$$

Differentiation gives:

$$F = \frac{dp}{dt} = m\ \mathrm{ch}\ V/c\ \frac{dV}{dt} = m\frac{dV}{d\tau} \qquad (8)$$

For constant force the acceleration is also constant and the motion is hyperbolic.

---

* *References*: See Planck (1906a) and (1906b). The form of Newton's Law given above was proposed by the writer at the 2000 PIRT conference. It avoids the concept of velocity dependent mass, velocity dependence being in the acceleration not the mass.

## 4. Remarks on Newton's Law of Motion in Three Dimensions

In his 1905 paper Einstein wrote the relativistic 3 dimensional Newtonian equations of motion of a particle (an electron) in an electric field There were assumed as usual, inertial systems, here denoted S and S', with S for the observer and S' for the particle. S' moves with uniform velocity v relative to S along the x-axis with S and S' coincident at t=0. Newton's law in its classical form is assumed to apply in the frame S' instantaneously coincident with the particle so that relative to it

$$
\begin{aligned}
m\, d^2x'/ dt'^2 &= F'_x = e\, E'_x \\
m\, d^2y'/ dt'^2 &= F'_y = e\, E'_y \\
m\, d^2z'/ dt'^2 &= F'_z = e\, E'_z
\end{aligned}
\qquad (1)
$$

The accelerations in S' and in S are related by

$$
\begin{aligned}
d^2x'/ dt'^2 &= \gamma^3\, d^2x/ dt^2 \\
d^2y'/ dt'^2 &= \gamma^2\, d^2y/ dt^2 \\
d^2z'/ dt'^2 &= \gamma^2\, d^2z/ dt^2
\end{aligned}
\qquad (2)
$$

The electric fields are related by

$$
\begin{aligned}
E'_x &= E_x \\
E'_y &= \gamma\, (E_y - v/c\, H_z) \\
E'_x &= \gamma\, (E_z + v/c\, H_z)
\end{aligned}
\qquad (3)
$$

So from (1) it follows that to the observer in S the equations of motion are:

$$
\begin{aligned}
m\, \gamma^3\, d^2x/ dt^2 &= e\, E_x \\
m\, \gamma^2\, d^2y/ dt^2 &= e\, \gamma\, (E_y - v/c\, H_z) \\
m\, \gamma^2\, d^2z/ dt^2 &= e\, \gamma\, (E_z + v/c\, H_z)
\end{aligned}
\qquad (4)
$$

where on the right hand side are the components of the Lorentz force:

Einstein (1905) assumed acceleration has the same form in S' as in S and he interpreted the multiplying factors $m\gamma^3$ and $m\gamma^2$ as giving velocity-dependent longitudinal and lateral masses in the equations (4). In doing this Einstein was using, as he said, the interpretation then current though he was cautious in this interpretation and in 1907 used a three-dimensional form of Planck's 1906 formulation to rederive the equations of motion in a different way without using velocity-dependent masses. But even so, velocity dependent mass continued in use even until the present time.

---

* *References* The notion of the velocity-dependent mass of an electron was introduced by Lorentz in 1895 'Versuch…' and modified by Abraham-Föppl 1904 '*Lehrbuch der Elektrizität* , .'and Lorentz (1904c) to longitudinal and lateral masses.

## 5. Three-dimensional Particle Dynamics in Vector Form

The vector form of Planck's equation giving Newton's Law of Motion is

$$\mathbf{F} = \frac{d\mathbf{p}}{dt} \qquad (1)$$

where **F** is the applied force and **p** is the momentum vector

$$\mathbf{p} = \frac{1}{\sqrt{(1 - v^2/c^2)}} \,( mv_x, mv_y, mv_z) \qquad (2)$$

The following method of discussing the motion follows that used in electron optics (see Born & Wolf's "Optics"). The momentum is represented as

$$\mathbf{p} = p\,\mathbf{s} \qquad (3)$$

where p denotes the scalar value which is

$$p = \frac{mv}{\sqrt{(1 - v^2/c^2)}} \qquad (4)$$

and **s** is a unit vector in the direction of **v**. Differentiating the product Planck's equation is

$$\mathbf{F} = \frac{dp}{dt}\,\mathbf{s} + p\,\frac{d\mathbf{s}}{dt} \qquad (5)$$

where the first term on the right, the tangential component, is in the direction of **v** while the second term, the normal component, is perpendicular to **s.**

*(a) The tangential component*: This is

$$\mathbf{F.\,s} = \frac{dp}{dt} = \frac{m}{\{1 - v^2/c^2)^{3/2}}\,\frac{dv}{dt} \qquad (6)$$

On the right here is the same expression as in the rectilinear case. As a result there follows:

*tangential force = rest mass x tangential hyperbolic acceleration* (7)

---

*Reference*: The vector form of Planck's equation is due to Einsein (1907)

*(b) The normal component*: This becomes more explicit by introducing ds the differential arc length, ρ the radius of curvature and **n** the unit normal to the curve when we can put,

$$\frac{d\mathbf{n}}{dt} = \frac{d\mathbf{n}}{ds}\frac{ds}{dt} = \frac{1}{\rho}\,\mathbf{n}\,v \qquad (8)$$

Consequently the resolution of the force into tangential and normal components is

$$\mathbf{F} = \frac{dp}{dt}.\,\mathbf{s} + p\,\frac{v}{\rho}\,\mathbf{n} \qquad (9)$$

giving the central force as

$$p\,\frac{v}{\rho}\,\mathbf{n} = \gamma\,\frac{mv^2}{\rho}\,\mathbf{n} \qquad (10)$$

* *Work and energy*: Work is done solely by the tangential component of force. The rate of doing work is

$$\mathbf{F.v = (F.\,s)}\;v = \frac{dp}{dt}\,v = \frac{m\,v}{(1 - v^2/c^2)^{3/2}}\,\frac{dv}{dt} = \frac{dT}{dt} \qquad (11)$$

This is seen to be equal to the rate of change of kinetic energy T which is

$$T = \frac{m\,c^2}{\sqrt{(1 - v^2/c^2)}}. \qquad (12)$$

If a potential function V exists in the coordinate system used, then

$$\mathbf{F.dx} = -\,dV \qquad (13)$$

Integration of (11) then gives the equation for total energy E

$$V + T = E = \text{const.} \qquad (14)$$

# 6. Calculation of Central Force in Uniform Circular Motion

In two dimensional circular motion referred to the centre

$$\mathbf{p} = mc\ \mathrm{sh}\ (V/c)\ (-\sin\theta, \cos\theta) \tag{1}$$

$\theta$ being the polar angle. If angular velocity is uniform,

$$\frac{d\mathbf{p}}{dt} = mc\ \mathrm{sh}\ (V/c)\ (-\cos\theta, -\sin\theta)\ \frac{d\theta}{dt} \tag{2}$$

This gives acting force **F**. On transforming the time on the right hand side, there follows

$$\mathbf{F} = mc\ \mathrm{th}\ (V/c)\ (-\cos\theta, -\sin\theta)\ \frac{d\theta}{d\tau} = m\ v\ (-\cos\theta, -\sin\theta)\ \frac{d\theta}{d\tau} \tag{3}$$

which resembles the usual expression for the centrifugal force i.e. $mv\ d\theta/d\tau$ radially

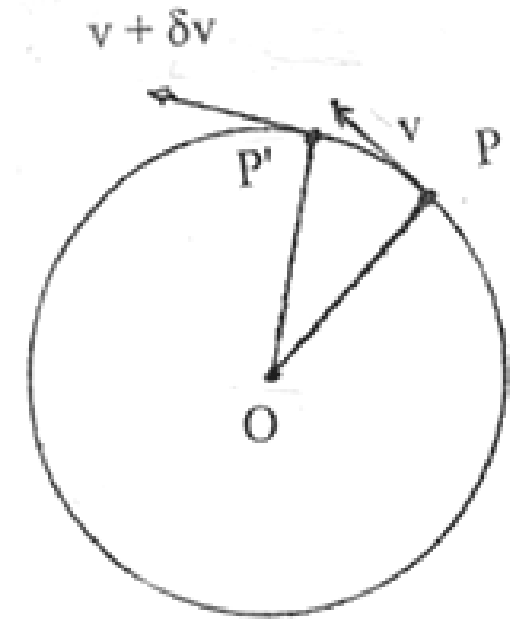

*Fig.1*: The space diagram

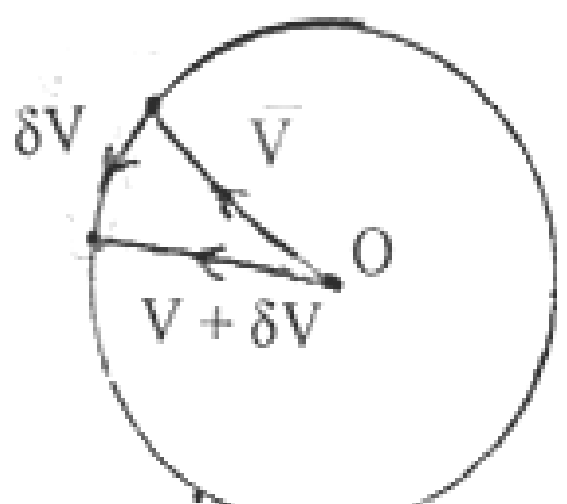

*Fig.2*: The velocity diagram (Hodograph)

The same calculation may be done by using the velocity diagram (hodograph) which in this case is a circle in the hyperbolic plane (fig.2). By the hyperbolic geometry of the circle the increment $\delta V$ is related to incremental angle $\delta\theta$ by

$$\delta V = c\ \mathrm{sh}\ (V/c)\ \delta\theta \tag{4}$$

Then the central force can be calculated as mass times vector momentum change:

$$F = m\ \frac{dV}{dt} = m\ \frac{d\theta}{dt.}\ c\ \mathrm{sh}\ (V/c) = m\ \frac{d\theta}{d\tau}\ c\ \mathrm{th}\ (V/c) = m\ \frac{d\theta}{d\tau}\ v \tag{5}$$

# 7. The Energy-Momentum Four-Vector

The energy-momentum four vector for a single particle of mass m and velocity v is defined here as

$$[E/c,\ p_x,\ p_y,\ p_z]^T \ = \ \left| \ \frac{mc}{\sqrt{(1 - v^2/c^2)}} \quad \frac{m\, v_x}{\sqrt{(1 - v^2/c^2)}} \quad \frac{mv_y}{\sqrt{(1 - v^2/c^2)}} \quad \frac{mv_z}{\sqrt{(1 - v^2/c^2)}} \ \right| \qquad (1)$$

Being a multiple of $[cdt,\ dx,\ dy,\ dz]^T$ it transforms the same way. Expressing the components in terms of the unit vector **n** in the direction of the velocity there is found

$$\begin{aligned} E/c &= mc \text{ ch } w &&= m \text{ ch } V/c \\ p_x &= mc \text{ sh } w \ n_1 &&= m \text{ sh } V/c \ n_1 = mc \text{ sh } V/c \sin\varphi \cos\theta \\ p_y &= mc \text{ sh } w \ n_2 &&= m \text{ sh } V/c \ n_2 = mc \text{ sh } V/c \sin\varphi \sin\theta \\ p_z &= mc \text{ sh } w \ n_3 &&= m \text{ sh } V/c \ n_3 = mc \text{ sh } V/c \cos\varphi \end{aligned} \qquad (2)$$

The components satisfy identically the equation defining a hyperbolic surface invariant under Lorentz transformation:

$$(E/c)^2 - (p_x^{\,2}+p_y^{\,2}+p_z^{\,2}) \ = \ (mc)^2 \qquad (3)$$

In consequence the dimensionality is effectively three not four. On taking differentials it follows that identically

$$(E/c)\ (dE/c) - (p_x\, dp_x+p_y\, dp_y+p_z\, dp_z) = 0 \qquad (4)$$

So that the differential vector

$$[dE/c,\ dp_x,\ dp_y,\ dp_z]^T \qquad (5)$$

is orthogonal to the energy-momentum vector and consequently space-like. The spherical parametrization of the surface (4) leads to the surface element squared of

$$dp_x^{\,2} + dp_y^{\,2} + dp_z^{\,2} - d\,(E/c)^2 \ = \ (mc)^2\ \{dw^2 + \text{sh}^2\, w\ (d\varphi^2 + \sin^2\varphi\ d\theta^2)\}$$

$$= (m)^2\ \{dV^2 + c^2\ \text{sh}^2\ (V/c)\ (d\varphi^2 + \sin^2\varphi\ d\theta^2)\} \qquad (6)$$

which is the standard Riemannian form for a hyperbolic space. The expression on the right corresponds to orthogonal incremental momentum components of m dV radially and of mc sh V/c dθ transversally.

## 8. Energy-Momentum of a System of Particles

The energy - momentum vector of a system of particles is defined by summation over all particles. The energy and momentum may be written

$$\begin{aligned} E/c &= \sum mc \operatorname{ch} w \\ \mathbf{p} &= \sum mc\, \mathbf{n} \operatorname{sh} w \end{aligned} \qquad (1)$$

Superimposition of equations results in the same transformation equations as for a single particle. Energy and momentum are connected by the identity:

$$(E/c)^2 - \mathbf{p.p} = (Mc)^2 \qquad (2)$$

With the help of suffices, M is seen to have the value

$$\begin{aligned} M^2 &= (\sum m_i \operatorname{ch} w_i)^2 - (\sum m_i \mathbf{n}_i \operatorname{sh} w_i) \mathbf{.} (\sum m_j \mathbf{n}_j \operatorname{sh} w_j) \\ &= \sum\sum m_i m_j \{\operatorname{ch} w_i \operatorname{ch} w_j - \mathbf{n}_i\mathbf{.n}_j \operatorname{sh} w_i \operatorname{sh} w_j\} \\ &= \sum\sum m_i m_j \operatorname{ch} w_{j/i} \end{aligned} \qquad (3)$$

Here M can be thought of as the total mass-energy of the particles. Associated with M a rapidity W can be defined by the equations

$$\begin{aligned} M \operatorname{ch} W &= \sum m \operatorname{ch} W \\ M\, \mathbf{n} \operatorname{sh} W &= \sum m\, \mathbf{n} \operatorname{sh} W \end{aligned} \qquad (4)$$

Division gives

$$\mathbf{n} \operatorname{th} W = (\sum m_i \mathbf{n}_i \operatorname{sh} w_i) / (\sum m_i \operatorname{ch} w_i) \qquad (5)$$

So that

$$\operatorname{th}^2 W = (\sum\sum m_i mj\, \mathbf{n}_i\mathbf{.n}_j \operatorname{sh} w_i \operatorname{sh} w_j)(\sum m_i \operatorname{ch} w_i)^2 \qquad (6)$$

The rapidity W here is similar to the classical velocity of the centre of mass.

On moving from S to S ' as expressed by w = W + w' the energy-momentum vector obeys the usual Lorentz transformation law, in matrix form

$$\begin{bmatrix} E/c \\ p \end{bmatrix} = \begin{bmatrix} \operatorname{ch} w & \operatorname{sh} w \\ \operatorname{sh} w & \operatorname{ch} w \end{bmatrix} \begin{bmatrix} E'/c \\ p' \end{bmatrix} \qquad \begin{bmatrix} E'/c \\ p' \end{bmatrix} = \begin{bmatrix} \operatorname{ch} w & -\operatorname{sh} w \\ -\operatorname{sh} w & \operatorname{ch} w \end{bmatrix} \begin{bmatrix} E/c \\ p \end{bmatrix} \qquad (6)$$

Equations (7) below transform to equations in system S ' by right matrix multiplication of the Lorentz transformation W.

## 9. Percussion

If the two particles have masses $m_1, m_2$ and velocities $v_1$, $v_2$ the basic equations are

$$\frac{m_1}{\sqrt{(1-v_1^{*2}/c^2)}} + \frac{m_2}{\sqrt{(1-v_2^{*2}/c^2)}} = \frac{m_1}{\sqrt{(1-v_1^2/c^2)}} + \frac{m_2}{\sqrt{(1-v_2^2/c^2)}}.$$

$$\frac{m_1 v_1^*}{\sqrt{(1-v_1^{*2}/c^2)}} + \frac{m_2 v_2^*}{\sqrt{(1-v_2^{*2}/c^2)}} = \frac{m_1 v_1}{\sqrt{(1-v_1^2/c^2)}} + \frac{m_2 v_2}{\sqrt{(1-v_2^2/c^2)}}. \qquad (1)$$

The asterisk denotes value after collision. The first equation is for the conservation of mass equivalent of energy and the second equation is for conservation of momentum.

These equations must hold independently of the frame of reference so if we take these equations to refer to inertial system S then they must also hold in a system S ' moving uniformly relatively to it. This property is most conveniently established using the hyperbolic representation when the conservation equations (1) become

$$m_1 \operatorname{ch} w_1^* + m_2 \operatorname{ch} w_2^* = m_1 \operatorname{ch} w_1 + m_2 \operatorname{ch} w_2$$
$$m_1 \operatorname{sh} w_1^* + m_2 \operatorname{sh} w_2^* = m_1 \operatorname{sh} w_1 + m_2 \operatorname{sh} w_2 \qquad (2)$$

In a system S ' moving with rapidity W relatively to S the equations will be

$$m_1 \operatorname{ch} (w_1^* - W) + m_2 \operatorname{ch} (w_2^* - W) = m_1 \operatorname{ch} (w_1 - W) + m_2 \operatorname{ch} (w_2 - W)$$
$$m_1 \operatorname{sh} (w_1^* - W) + m_2 \operatorname{sh} (w_2^* - W) = m_1 \operatorname{sh} (w_1 - W) + m_2 \operatorname{sh} (w_2 - W) \qquad (3)$$

On expanding the hyperbolic functions these equations are seen to be equivalent to the previous ones in (2). Further it can be seen that the validity of either of the equations (3) with respect to arbitrary translations W (i.e. the relativity principle) implies the validity of the other. It is convenient to write the relationships using the two-dimensional energy-momentum vector relating mass-energy E and momentum p:

$$[E/c \;\; p]^T = [mc \operatorname{ch} w \;\; mc \operatorname{sh} w]^T \qquad (4)$$

On moving from S to S ' as expressed by $w = W + w'$ the energy-momentum vector obeys the usual Lorentz transformation law, in matrix form

$$\begin{bmatrix} E/c \\ p \end{bmatrix} = \begin{bmatrix} \operatorname{ch} w & \operatorname{sh} w \\ \operatorname{sh} w & \operatorname{ch} w \end{bmatrix} \begin{bmatrix} E'/c \\ p' \end{bmatrix} \qquad \begin{bmatrix} E'/c \\ p' \end{bmatrix} = \begin{bmatrix} \operatorname{ch} w & -\operatorname{sh} w \\ -\operatorname{sh} w & \operatorname{ch} w \end{bmatrix} \begin{bmatrix} E/c \\ p \end{bmatrix} \qquad (5)$$

By right matrix multiplication of the Lorentz transformation w the equations then immediately transform to the corresponding form in system S '.

$$(E_1^*/c) + (E_2^*/c) = (E_1/c) + (E_2/c)$$
$$p_1^* + p_2^* = p_1 + p_2 \qquad (6)$$

* *Three-dimensional percussion*: equations for conservation of mass-energy in the usual form are:

$$\frac{m_1}{\sqrt{(1-v_1^{*\,2})}} + \frac{m_2}{\sqrt{(1-v_2^{*\,2})}} = \frac{m_1}{\sqrt{(1-v_1^2)}} + \frac{m_2}{\sqrt{(1-v_2^2)}} \tag{7}$$

For conservation of momentum components they are:

$$\frac{m_1 v_{x1}^*}{\sqrt{(1- v_1^{*\,2}/c^2)}} + \frac{m_2 v_{x2}^*}{\sqrt{(1- v_2^{*\,2}/c^2)}} = \frac{m_1 v_{x1}}{\sqrt{(1-v_1^2/c^2)}} + \frac{m\, v_{x2}}{\sqrt{(1-v_2^2/c^2)}}$$

$$\frac{m_1 v_{y1}^*}{\sqrt{(1- v_1^{*\,2}/c^2)}} + \frac{m_2 v_{y2}^*}{\sqrt{(1- v_2^{*\,2}/c^2)}} = \frac{m_1 v_{y1}}{\sqrt{(1-v_1^2/c^2)}} + \frac{m_2 v_{y2}}{\sqrt{(1-v_2^2/c^2)}}$$

$$\frac{m_1 v_{z1}^*}{\sqrt{(1- v_1^{*\,2}/c^2)}} + \frac{m_2 v_{z2}^*}{\sqrt{(1- v_2^{*\,2}/c^2)}} = \frac{m_1 v_{z1}}{\sqrt{(1-v_1^2/c^2)}} + \frac{m_2 v_{z2}}{\sqrt{(1-v_2^2/c^2)}} \tag{8}$$

These can be written concisely as

$$\begin{aligned} (E_1^*/c) + (E_2^*/c) &= (E_1/c) + (E_2/c) \\ p_1^* \mathbf{n}_1^* + p_2^* \mathbf{n}_2^* &= p_1 \mathbf{n}_1 + p_2 \mathbf{n}_2 \end{aligned} \tag{9}$$

Here p is scalar momentum and **n** is the unit vector in the direction of the velocity

---------------------------------------------------------------------------------------------------------

*Reference*: In the simple case of rectilinear collision the equations for percussion were established by Lewis & Tolman 1908.

# CHAPTER 7 – Differential Minkowski Space and Light Propagation

## 1. Differential Minkowski Space

The use of the differential form of the Lorentz transformation leads naturally to the concept of the differential Minkowski space of four-vectors (dt, dx, dy, dz) where all the ideas of normal Minkowski space apply to these differential vectors. Thus differential vectors are classified as time-like or space-like according to whether or not they lie in the differential Monge cone

$$c^2dt^2 = dx^2+dy^2+dz^2 \qquad (1)$$

Only time-like vectors have significance for physical motion because the requirement

$$(dx/dt)^2 + (dy/dt)^2 + (dz/dt)^2 < c^2 \qquad (2)$$

that the velocity should not exceed c, implies that

$$dx^2+dy^2+dz < c^2dt \qquad (3)$$

So the differential vector (dt, dx, dy, dz) lies within the light cone. The forward light-cone is usually of most interest being characterized by this last condition with $dt > 0$.

The light cone and the system of associated hyperbolic surfaces remain invariant relative to differential changes brought about by any homogeneous Lorentz trans-formation.

$$(cdt', dx', dy', dz') = L\,(cdt, dx, dy, dz) \qquad (4)$$

The use of the differential form here allows the variables t', x', y', z' to be related to t, x, y, z by non-homogeneous as well as homogeneous Lorentz transformations. This possibility is exactly in accord with Minkowski's original 1908 concept of the 'absolute world' as consisting of world vectors (t, x, y, z) invariant under the group of non-homogeneous Lorentz transformations. However the form of space he defined, i.e. the usual Minkowski space, is only invariant for homogeneous trans-formations corresponding to the chosen fixed origin. Differential Minkowski space gives more flexibility permitting invariance also with respect to translations. It is consequently more in accord with Minkowski's conception of the absolute world.

## 2. The Cayley-Klein Metric in Differential Minkowski Space

In his *Theory of Relativity* which first appeared in 1921, Pauli briefly observed that Varićak's results could be derived from the Cayley-Klein theory of projective geometry but apparently neither he nor others followed up this idea which will be described here in further detail

The space of differential vectors written in either of the two forms

$$(dt,\ dx,\ dy,\ dz),\ (c\ dt,\ dx,\ dy,\ dz) \qquad (1)$$

gives rise to a projective space with the differential Monge cone

$$c^2\, dt^2 - dx^2 - dy^2 - dz^2 = 0 \qquad (2)$$

as absolute. Since this locus is a real conic, the resulting projective space is hyperbolic. Vectors in this space lie within the Monge cone representing physically feasible motions satisfying the condition

$$dx^2 + dy^2 + dz^2 < c^2\, dt^2 \qquad (3)$$

The Cayley-Klein projective distance between two differential vectors with suffices 1, 2 is

$$\mathrm{ch}^{[-1]}\ \frac{(c^2\, dt_1\, dt_2 - dx_1\, dx_2 - dy_1\, dy_2 - dz_1\, dz_2)}{\sqrt{(c^2\, dt_1^2 - dx_1^2 - dy_1^2 - dz_1^2)}\ \sqrt{(c^2\, dt_2^2 - dx_2^2 - dy_2^2 - dz_2^2)}}. \qquad (4)$$

which is more conveniently written using vectors for the space part as

$$\mathrm{ch}^{[-1]}\ \frac{c^2\, dt_1\, dt_2 - d\mathbf{r}_1.d\mathbf{r}_2}{\sqrt{(c^2 dt_1^2 - d\mathbf{r}_1.d\mathbf{r}_1)}\ \sqrt{(c^2 dt_2^2 - d\mathbf{r}_2.d\mathbf{r}_2)}}. \qquad (5)$$

On dividing through by $dt_1\, dt_2$ and denoting velocities $d\mathbf{r}/dt$ by $\mathbf{v}$ and we see that the distance so defined is just the relative rapidity w of the two velocities $\mathbf{v}_1$, $\mathbf{v}_2$ given by

$$\mathrm{ch}\ w = \frac{c^2 - \mathbf{v}_1.\mathbf{v}_2}{\sqrt{(c^2 - \mathbf{v}_1.\mathbf{v}_1)}\ \sqrt{(c^2 - \mathbf{v}_2.\mathbf{v}_2)}}. \qquad (6)$$

so identifying the hyperbolic space as the Beltrami space of these velocity vectors.

In this space the Beltrami parametric representation

$$\begin{aligned} dx/dt &= v_x = c\ \mathrm{th}\ w\ n_1 = c\ \mathrm{th}\ w\ \sin\varphi\cos\theta \\ dy/dt &= v_y = c\ \mathrm{th}\ w\ n_2 = c\ \mathrm{th}\ w\ \sin\varphi\sin\theta \\ dz/dt &= v_z = c\ \mathrm{th}\ w\ n_3 = c\ \mathrm{th}\ w\ \cos\varphi \end{aligned} \qquad (7)$$

can be viewed in a corresponding homogeneous form arising from

$$(c\,dt,\ dx,\ dy,\ dz) = \text{const.}\ (\text{ch}\,w,\ \text{sh}\,w\ n_1,\ \text{sh}\,w\ n_2,\ \text{sh}\,w\ n_3) \qquad (8)$$

It implies

$$c^2\,dt^2 - dx^2 - dy^2 - dz^2 = (\text{const})^2 \qquad (9)$$

identifying the constant multiplier as $c\,d\tau$. Consequently

$$\begin{aligned}
c\,dt &= c\,d\tau\ \text{ch}\,w &&= c\,d\tau\ \text{ch}\,w \\
dx &= c\,d\tau\ \text{sh}\,w\ n_1 &&= c\,d\tau\ \text{sh}\,w \sin\varphi \cos\theta \\
dy &= c\,d\tau\ \text{sh}\,w\ n_2 &&= c\,d\tau\ \text{sh}\,w \sin\varphi \sin\theta \\
dz &= c\,d\tau\ \text{sh}\,w\ n_3 &&= c\,d\tau\ \text{sh}\,w \cos\varphi
\end{aligned} \qquad (10)$$

The parameterization (10) may also be written so as to give the components of the Minkowski four-velocity

$$\begin{aligned}
V_0 &= c\,dt/d\tau &&= c\ \text{ch}\,w &&= c\ \text{ch}\,w \\
V_1 &= dx/d\tau &&= c\ \text{sh}\,w\ n_1 &&= c\ \text{sh}\,w \sin\varphi \cos\theta \\
V_2 &= dy/d\tau &&= c\ \text{sh}\,w\ n_2 &&= c\ \text{sh}\,w \sin\varphi \sin\theta \\
V_3 &= dz/d\tau &&= c\ \text{sh}\,w\ n_3 &&= c\ \text{sh}\,w \cos\varphi
\end{aligned} \qquad (11)$$

Note that the space-time four-dimensionality arises from the use of homogeneous coordinates for three-dimensional motion.

---

* *Historical comment:* Because of its historical importance the remarks of Pauli are here quoted in full. In a footnote (*The Principle of Relativity,* p.74 in Engl. trans) he says:

> *" Varićak establishes a formal connexion between the Lorentz transformation, as well as the relativistic formulae for the Doppler effect, aberration of light, and reflection in a moving mirror, with the Bolyai-Lobachevski geometry ... This connexion with the Bolyai-Lobachevski geometry can be briefly described in the following way (this had not been noticed by Varićak): if one interprets $dx_1$, $dx_2$, $dx_3$, $dx_4$ as homogeneous coordinates in a three dimensional projective space, then the invariance of the equation $dx_1^2+dx_2^2+dx_3^2-dx_4^2 = 0$ amounts to introducing a Cayley system of measurement based on a real conic section. The rest follows from the well known arguments of Klein (Math.Ann. 4 1871 12)"*

Pauli only allows a 'formal connexion' and not a 'real connexion'! As regards the reference to Klein, it is remarkable that Klein himself, the authority on both projective non-Euclidean geometry and relativity, had managed to miss the precise relation between the two, even in his 1910 paper: 'Über die geometrische Grundlagen der Lorentzgruppe' devoted to exactly this question. The explanation appears to be that, as shown both in Klein's paper and in Pauli's comment above, at that time it was customary to think in terms of the pseudo-Euclidean space-like metric which does not lend itself to the Cayley metric. The only place where the writer has found Klein using the appropriate time-like form is in a footnote of his book *Die Entwicklung* …. (p.131, vol.II) where he quoted the Cayley-Klein distance formula in special relativity using the Minkowski imaginary angle The reversed Cauchy inequality, the justification for this formula, was not mentioned.

## 3 The Light Cone Condition and Coordinate Invariance

Suppose that a transformation of frames of reference is made resulting in the relation

$$[c\,dt',\, dx',\, dy',\, dz'\,]^T = \Lambda\,[c\,dt,\, dx,\, dy,\, dz\,]^T \qquad (1)$$

$\Lambda$ being a general Lorentz transformation. From this follows

$$c^2\,dt'^2 - dx'^2 - dy'^2 - dz'^2 = (c^2\,dt^2 - dx^2 - dy^2 - dz^2) \qquad (2)$$

So the Monge cones in the original and transformed spaces and their interiors correspond to each other (causality condition). More generally the same conclusion follows if there is a relation of the type

$$c^2\,dt'^2 - dx'^2 - dy'^2 - dz'^2 = \kappa^2\,(c^2\,dt^2 - dx^2 - dy^2 - dz^2) \qquad (3)$$

where $\kappa$ is any non-zero scalar multiplier. Such a relation may come about from the scalar multiplied form of the Lorentz transformation (e.g. the Voigt transformation) or from a nonlinear transformation of coordinates and time. Condition (3), fundamental to the theory; will be referred to as the *light-cone condition*. The following simple fact is basic.

*PROPOSITION*: The light-cone condition implies invariance of the Cayley-Klein metric.

*Proof:* Any linear combination of two time-like differential vectors:

$$(c\,dt,\, dx,\, dy,\, dz) = \lambda\,(c\,dt_1,\, dx_1,\, dy_1,\, dz_1) + \mu\,(c\,dt_2,\, dx_2,\, dy_2,\, dz_2) \qquad (4)$$

with $\lambda, \mu > 0$ lies on the segment joining the two points with suffices 1 and 2 and so from the convexity of the Monge cone is also time-like. Under linear transformation of differentials it transforms to the vector

$$(c\,dt',\, dx',\, dy',\, dz') = \lambda\,(c\,dt'_1,\, dx'_1,\, dy'_1,\, dz'_1) + \mu\,(c\,dt'_2,\, dx'_2,\, dy'_2,\, dz'_2) \qquad (5)$$

This similarly lies within the Monge cone in the transform space and is time-like. Applying the light-cone condition equating coefficients of $\lambda^2$, $\lambda\mu$ and $\mu^2$ we get

$$\begin{aligned}
c^2\,dt'^{\,2}_1 - dx'^{\,2}_1 - dy'^{\,2}_1 - dz'^{\,2}_1 &= \kappa^2\,(c^2\,dt_1^2 - dx_1^2 - dy_1^2 - dz_1^2)\\
c^2\,dt'_1 dt'_2 - dx'_1 dx'_2 - dy'_1 dy'_2 - dz'_1\,dz'_2 &= \kappa^2\,(c^2\,dt_1\,dt_2 - dx_1\,dx_2 - dy_1\,dy_2 - dz_1\,dz_2)\\
c^2\,dt'^{\,2}_2 - dx'^{\,2}_2 - dy'^{\,2}_2 - dz'^{\,2}_2 &= \kappa^2\,(c^2\,dt_2^2 - dx_2^2 - dy_2^2 - dz_2^2) \qquad (6)
\end{aligned}$$

From which follow invariance of the ratio giving Cayley-Klein metric and rapidity. The hyperbolic spaces in the original and transformed light-cones are correspond- ingly mapped on to one another isometrically. Rapidity and hyperbolic velocity are seen to be independent of the coordinate system under any differentiable coordinate transformation. Rapidity is an absolute invariant of the transformation and defines the hyperbolic geometry in Beltrami representation as described previously.

# 4. Normal Light Propagation

In 1924 Carathéodory published a general derivation of the light cone condition starting from an initial set of axioms relating to light propagation. The background to the paper is that in 1908, 1910 Bateman and Cunningham discovered that the equation of the wave surface as well as Maxwell's equations were also invariant under certain nonlinear transformations related to inversion which they called 'spherical wave transformations'. These papers thus raised the question of the exact nature of the group of transformations under which Maxwell's equations remain invariant and the relation of this group with Special Relativity This problem ascribed to Pauli by Carathéodory, was investigated in the paper.

From an initial axiomatic discussion Carathéodory established the existence of differentiable equations relating two systems of locally Euclidean coordinates and time of the form:

$$x' = X(x, y, z, t), \; y' = Y(x, y, z, t), \; z' = Z(x, y, z, t). \; t' = T(x, y, z, t) \qquad (1)$$

From these follow linear relations between corresponding differentials

$$[\, dx', dy', dz', dt' \,]^T = [\, J \,] \, [dx, dy, dz, dt \,]^T \qquad (2)$$

Here [ J ] is the Jacobian matrix. Using these linear relations we may express as a quadratic in terms of dx, dy, dz, dt the quantity

$$dx'^2 + dy'^2 + dz'^2 - c^2 \, dt'^2 \qquad (3)$$

For normal light propagation to be preserved the following condition must be satisfied:

$$dx'^{\,2} + dy'^{\,2} + dz'^{\,2} - c^2 \, dt'^{\,2} = \mu \, (x, y, z, t) \, (dx^2 + dy^2 + dz^2 - c^2 \, dt^2) \qquad (4)$$

This is a similar condition to that previously found by Bateman in connection with Maxwell's equations. It resembles the previous light cone condition but the multiplier $\mu$ (which should be positive) is a function of position and time instead of v. The condition is fundamental to Carathéodory's theory.

** Huyghens wavelets:* The condition (4) can be given a physical interpretation for it implies that the infinitesimal spherical wave (a Huyghens wavelet)

$$dx^2 + dy^2 + dz^2 = c^2 \, dt^2 \qquad (5)$$

transforms into a similar wavelet. These wavelets generate the wavefront which propagates according to equation of the wave-equation

$$(\partial V/\partial x)^2 + (\partial V/\partial y)^2 + (\partial V/\partial z)^2 - 1/c^2 \, (\partial V/\partial t)^2 = 0 \qquad (6)$$

The light cone condition implies that this equation is also invariant under the transformation considered. This is because the dual equation of the Monge cone, giving the condition that a hyper-plane with homogeneous coordinates ($n_0$, $n_1$, $n_2$, $n_3$) is tangential to the cone, is

$$n_1^2 + n_2^2 + n_3^2 - (n_0/c)^2 = 0 \qquad (7)$$

So the wave equation for V represents the condition that the vector

$$[\partial V/\partial x, \partial V/\partial y, \partial V/\partial z, \partial V/\partial t] \qquad (8)$$

is tangential to the li ne. On transforming to new coordinates, there will be a condition dual to the light-cone condition:

$$(n_1'^2 + n_2'^2 + n_3'^2 - (n_0'/c)^2) = \mu\,(x, y, z, t)^{-1}\,(n_1^2 + n_2^2 + n_3^2 - (n_0/c)^2) \qquad (9)$$

From this it immediately follows that the wave equation is satisfied in the new coordinates.

Carathéodory proved the invariance of the wavefront equation using the theory of characteristics for a partial differential equation and showed that any piecewise linear light path becomes transformed into another piecewise linear light path.

---

*References*: See Carathéodory:'On the axiomatics of special relativity …' (German) Preuss. Akad Wiss. 1924 reproduced in his Collected Works. There is also a shorter version in German in his collected works which is the translation of an encyclopaedia article published in Greek. The papers of Bateman and Cunningham are listed in the bibliography.

## 5. Conformal Transformation in Four Dimensions

With $l = ict$, $l' = ict'$ the light-cone condition can be written as

$$dx'^2 + dy'^2 + dz'^2 + dl'^2 = \mu\,(x, y, z, t)\,(dx^2 + dy^2 + dz^2 + dl^2) \qquad (1)$$

This is the condition for a conformal transformation in 4 dimensions. One such conformal transformation is the Lorentz transformation. Another is the inversion:

$$\begin{aligned} x' &= x / (x^2+y^2+z^2+l^2) \\ y' &= y / (x^2+y^2+z^2+l^2) \\ z' &= z / (x^2+y^2+z^2+l^2) \\ l' &= l / (x^2+y^2+z^2+l^2) \qquad (2) \end{aligned}$$

When translated back to x, y, z, t variables this coincides with one of the spherical wave transformations of Bateman & Cunningham. The importance of inversion is seen from the following result:

*THEOREM* (Liouville 1847): A conformal transformation in a space of more than 2 dimensions is represented by a sequence of similarity transformations and inversions.

The similarity transformations are those generated by translations and orthogonal transformations. A simplified proof was given by Carathéodory in his 1923 paper.

An important application of the spherical wave transformations to relativity, is the following example due to Bateman (1910)

*EXAMPLE* Consider the infinitesimal transformation

$$\begin{aligned} t' &= t\,(1 + f\,x) \\ x' &= x\,(1 + f\,x) + \tfrac{1}{2} f\,(-x^2 + c^2t^2 - y^2 - z^2) \\ y' &= y\,(1 + f\,x) \\ z' &= z\,(1 + f\,x) \end{aligned} \qquad (3)$$

f is here considered small so that only terms of the first order in f need be retained.
The conformal condition is satisfied with

$$\mu\,(x, y, z, t) = 1 + 2\,f\,x \qquad (4)$$

On substituting for t from the first equation into the second we find

$$x' = x + \tfrac{1}{2} f\,(x^2 + t'^2 - y^2 - z^2) \qquad (5)$$

With x, y, z fixed, the point with coordinate x' moves with constant acceleration f and as Bateman observed, the relation between t and t' agrees with that given by Einstein (1907) in his attempt to extend the Special Theory to an accelerated system.

This is an example of a non-Galilean transformation which nevertheless satisfies the condition of normal light propagation. Carathéodory considered that such transformations should be excluded from the invariance group of Special Relativity since he understood in his paper the Principle of Special Relativity to include the principle of mechanical relativity, . that all phenomena take place in an inertial frame. Consequently Carathéodory's conclusion was that the invariance group of Special Relativity is the non-homogeneous Lorentz group with scalar multiplication

However, since those nonlinear transformations obey the light cone condition, it follows from above that the Cayley-Klein metric defining an invariant rapidity can be introduced leading to the Beltrami interpretation together with its consequences e.g. all those which follow from the composition of velocities. The writer's opinion is that in view of this the status of the nonlinear transformations in the special theory must remain an open and important question.

---

*Reference*: Further comments were given in the writer's paper at the congress *'Constantin Carathéodory* ' Orestiada, Greece, Sept. 2000.

# *GENERAL BIBLIOGRAPHY*

---

*Note*: *PIRT Conference* refers to the conference *'Physical Interpretations of Relativity Theory',* held biannually from 1988 to 2006 at Imperial College, London and in 1907 and 1909 at Etvös University, Budapest. Proceedings for the London years were published by the Physics Department of Liverpool University but are no longer available. The papers for the London years 2000 to 2006 (conferences VII-X) are however available online at http://www.space-lab.ru/PIRT_VII-XII

# APPENDIX 1 – Some Historical Notes

## 1. Galilean Relativity and Newtonian Mechanics

The origin of the principle of relativity is nowadays usually credited to Galileo. Following the publication in 1632 of his great book 'Dialogues on the two World Systems - Ptolemaic and Copernican', Galileo was put on trial by the Inquisition and found guilty of teaching "that the Earth moves and is not the centre of the World". Aristotle had made a clear distinction between rest and motion Therefore it was natural to think if the Earth moved, either through rotation or by moving round the Sun, then this motion would be noticed. A falling body, for example, would not go straight down to Earth but fall askew. In the book the dialogue for the second day discusses at length the common observation that a person on a ship moving uniformly on a calm sea can be unaware of motion relative to the sea and can have the impression of being at rest. Further that, for example, a body dropped from the top of the mast falls straight down to the foot of the mast. This may be regarded as the first clear statement of the principle of relativity. It was used by Galileo in his later book 'Dialogues on two new Sciences' to show that the path of a projectile is a parabola.

A similar principle of relativity for uniform motion was used by Huyghens (1656) to establish the law governing the impact of colliding bodies. Knowing the law of collision for direct impact gives the law for other impacts on referring the motion to the moving centre of gravity of the two colliding bodies.

This principle of relativity was fundamental for the new mechanics which was then replacing the teaching of Aristotle. Aristotle, with his distinction between rest and motion, had actually stated that, if left alone, a body at rest will remain at rest and that a body in motion will remain in the same motion and his statements were apparently known to Newton. However in Newton's first law of motion of 1687- the law of inertia - uniform motion is seen as equivalent to being at rest although this is not apparent in the customary imprecise statement of the law which reads 'state of rest, or of uniform motion in a straight line' where the comma appears to separate distinct possibilities. As pointed out by Koyré, the more accurate translation of Newton's original Latin statement would read:

> *'A body perseveres in its state, at rest or in uniform motion in a straight line, unless compelled by an impressed force to change its state. '*

Here the dynamical state is understood to be 'at rest or in uniform motion in a straight line'.

As a deduction from his laws of motion, Newton stated the following which may be regarded as the first precise statement of the principle of mechanical relativity:

> *The motions of bodies included in a given space are the same among themselves, whether that space is at rest, or moves uniformly forward in a right line without any circular motion.*
>
> *Principia, Corollary V, (Motte translation)*

Newton defined also absolute space and time. The use of absolute space appears to be largely for the purpose of explaining rotational motion as illustrated by the well-known rotating bucket experiment, and he considered absolute space necessary to explain centrifugal force in rotating systems. So Newtonian absolute space came to replace Aristotelian 'rest'. Newton offered no discussion regarding absolute time. His use of absolute space, subsequently criticised, was defended by Euler (1748) as a necessary foundation for the analytical mechanics he initiated so successfully.Maxwell (1877) who needed absolute space for his aether theory of electromagnetic waves, stated that while Newtonian mechanics accepts that all motions must be defined relative to some origin and so are in this sense relative, the application of the second law of motion needs the concept of absolute motion so that in defining acceleration the motion of the origin does not have to be taken into account.

-------------------------------------------------------------------------------------------------------

*Notes*: Principal historical references are in the bibliography at the end of this appendix.
1) Galileo's dialogues of 1632 were withdrawn after publication, but were published in English translation by Salusbury in 1661, the year Newton went to Cambridge, and were generally known at that time. (cf Herivel: *Background to Newton's Principia*, Oxford 1965) Salusbury also translated Galileo's '*Two new sciences*' published in 1638 in Leyden which had more direct interest for mechanics with detailed derivations.
2) Huyghens' argument is described in Dugas: *History of Mechanics.*
3) Aristotle's statements appeared in *Physica* and *de Caelo*. They were subsequently used by d'Alembert in his *Encyclopedie* (cf. Compte *PIRT Conf.* 1998)
4) On Newton's statement of his laws see Koyré *Newtonian Studies,* London 1965 (Chapman & Hall) Chap. III, Appendix A The customary inaccurate statement of Newton's law is there traced to Motte's 1727 translation of Principia. A manuscript predating Principia stating clearly the equivalence of rest and uniform motion is quoted by Woodhouse: *Special Relativity,* Springer 2003.
5) Newton's Corollary V was stated as the Fourth Law of Motion two and a half years before Principia in an unpublished manuscript *De motu corporum in mediis regulariter cedentibus* (Woodhouse Special Relativity p.8)
6) On Euler's support for Newtonian absolute space and time in reply to Berkeley and Boscovich see Vasiliev. *Space, Time, Motion* London 1924 (Chatto & Windus)

## 2. Aberration

In 1827 Bradley discovered stellar aberration which causes an apparent displacement of a star in the direction of the Earths motion so that its observed position completes an ellipse round its true position in the course of a year.

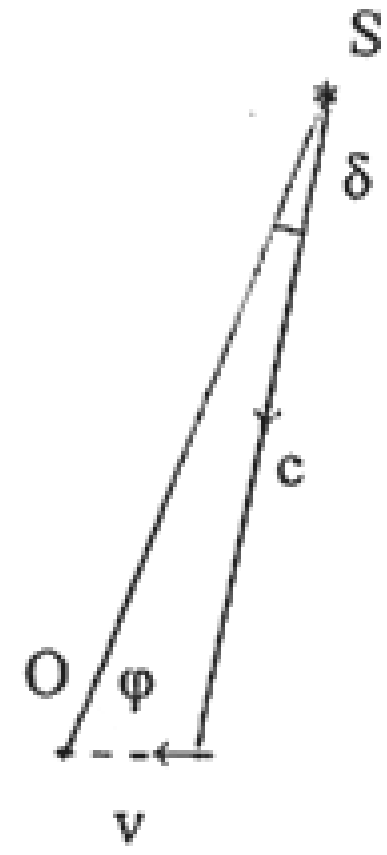

*Fig*. Aberration

Using Newton's corpuscular theory of light, Bradley explained aberration by vector addition of light velocity c with the forward velocity v of the observer O on the Earth so that the star S is observed at angle $\varphi$ as shown in the figure The sine rule gives approximately for the small aberration angle $\delta$

$$\delta = (v/c) \sin \varphi$$

From this the velocity of light c may be determined. Bradley found the speed of light to be the same for all stars examined and close to a value found by Römer in1769 from eclipses of Jupiter's satellites. Since in this derivation the angle of aberration $\delta$ depends on the speed of light, it might be expected to change if the light passed through a material of refractive index greater than one which would reduce its speed correspondingly. The test was made by Arago in 1810 with a negative result. Unable to find an explanation he asked Fresnel whether an explanation could be found using the wave theory of light which Young (1904) had shown could explain normal aberration. In reply Fresnel proposed his theory that the ether, at rest, is partially dragged by the Earth and as a result the velocity of the light in a medium moving with velocity v is given by the formula

$$c' = c/N + kv$$

Here N is the index of refraction and k is the dragging coefficient having the value.

$$k = 1 - 1/N^2$$

This formula was later experimentally investigated by Fizeau (1851) and found remarkably accurate. The result, strange at the time, subsequently found confirmation in the theory of relativity since the formula approximates addition of the velocities c/N and v by the composition rule when v/c is small (von Laue 1907, cf. Miller 1998)

* *The Doppler Effect*: Another optical phenomenon which came to be very important is the Doppler Effect which is by its nature a wave phenomenon. It was discussed in several papers from 1842 by Doppler who derived his formula and suggested it might be applied to the spectra of double stars. This was confirmed experimentally by Huggins and Miller in 1868

---

**Note*: Boscovich, already in 1766, had proposed using a telescope filled with water but this was not tried out until about a century later by Airy with a negative result.

## 3 The Doppler Effect:

Another optical phenomenon which came to be important is the Doppler Effect which is by its nature a wave phenomenon. It was discussed in several papers from 1842 by Doppler who derived his formula and suggested it might be applied to the spectra of double stars. This was confirmed experimentally by Huggins and Miller in 1868

In the elementary theory of the classical Doppler Effect for one-dimensional motion the medium is assumed stationary and motions of the observer O and source S are related to it as in the figure.

$O \rightarrow v_O$ $S \rightarrow v_S$

--------o-------------------o--------

*Fig.* Doppler Effect

Frequency f of the wave relative to the medium is found from wave speed c and wave-length $\lambda$ by:

$$f = c/\lambda \tag{1}$$

Motion of the source with velocity $v_s$ changes wave-length $\lambda$ to a wave-length $\lambda'$:

$$\lambda' = \lambda\,(1 + v_S/c) \tag{2}$$

while motion $v_o$ of the observer changes wave speed relative to the observer to:

$$c' = c + v_O = c\,(1 + v_O/c) \tag{3}$$

The observed frequency f ' for motion both of source and observer is then

$$f' = \frac{c'}{\lambda'} = f\,\frac{(1 + v_O/c)}{(1 + v_S/c)} \tag{4}$$

For velocities considerably less than c, binomial approximation results in the linear Doppler formula depending only on relative velocity $v_O - v_S$ of the observer to source.

$$f' = f\,\{1 + (v_O - v_S)/c\} \tag{5}$$

The formula is exact when and only when the source is at rest relative to the medium of propagation, otherwise the formula neglects a second-order effect for motion of source relative to the medium, a limitation emphasized by Drude 1900.

---

*References*: Doppler's papers are conveniently consulted in his collected papers: *Abhandlungen* Leipzig 1907 reprinted as Ostwald's Klassiker No. 161, pp. 1-24. The initial application to stellar spectra was made by Huggins & Miller *Proc.Roy.Soc.* 1868.

## 4 . Maxwell's equations and aether theories

As remarked above, Maxwell believed that the concept of absolute space is necessary for the foundations of Newtonian mechanics and he had assumed a material aether absolutely at rest when he established his electromagnetic theory of light (1873). He even published physical constants for the aether.

* *The early work of Lorentzon aberration and Doppler Effect:* Stokes in 1845 had proposed a theory of aberration which assumed that the Earth dragged the aether in such a way that the relative velocity between them became zero on the surface. This theory led to results at variance with Fresnel's hypothesis. Lorentz's early work aimed at reconciling the ideas using Maxwell's equations. In connection with this he developed a theory in 1886 covering both aberration and the Doppler effect using what he called the *principle of invariance of the phase* which states that the quantity

$$\Phi \ = \ 2\pi f\,(t - (lx + my + nz))$$

remains unaltered with change from rest to motion. Here f denotes frequency and (l m n) the direction cosines of the wave-normal. This principle was later used by Einstein.

* *The transformation of Voigt* (1887) Investigating the Doppler Effect using aether theory, Voigt showed that the 3-dimensional wave equation is unaltered by the change of variables

$$x' \ = \ x - vt, \quad y' \ = \ y\,\surd(1- v^2/c^2), \quad z' \ = \ z\,\surd(1- v^2/c^2), \quad t' \ = \ t - (v/c^2).x$$

This differs only by a scalar multiple from the Lorentz transformation later used. Voigt's contribution was however isolated and not followed up.

* *The contraction hypothesis*: The Michelson-Morley experiments of 1887, 1888 established that there is not any second order effect which enables a moving system to be distinguished from one that is at rest. The difficulty of explaining this fact resulted in the *contraction hypothesis* proposed independently by FitzGerald (1889) and Lorentz (1892). According to this hypothesis, motion of the electrons relative to the ether causes a length contraction in the direction of motion by the factor $\surd(1-v^2/c^2)$, the electrons then becoming ellipsoidal.

**The Maxwell-Herz equations and the theory of electrons* : In 1890 Herz published two papers, on Maxwwell's equations, one for systems at rest and one for systems in mortion. They were republished in his 1892 book. With the appearance of the Hertz form and discovery of the electron (1891) Lorentz developed his *theory of electrons* to study electrical phenomena in moving media. This theory explained electrical phenomena in terms of a free flow of electrons in the aether, the aether exerting a force on the electrons due to its electric and magnetic fields, a force which became known as the *Lorentz force*.

# 5 Lorentz's development of his transformation

Lorentz's derivation of the transformation originated in his attempt (1895) to find what transformation of variables is necessary for Maxwell's equations to keep their same form in a coordinate system S at rest in the aether and a systerm S' moving uniformly relative to it.

He first made a Galilean transformation from coordinates in S to moving ('relative') coordinates coinciding with S' by

$$x_r = x - vt, \quad y_r = y, \quad z_r = z,$$

Also, he changed time to a mathematically convenient variable he called 'local time':

$$t_r = t - (v/c^2).x$$

He then found that, to the first order in v/c, the behaviour of light in the moving system S' can, be deduced from an appropriate set of transformed electromagnetic quantities and by making this transformation of variables to a corresponding solution for the stationary system S.
He called this the Principle of Corresponding States.

Later in 1899 and 1904 Lorentz arrived at a transformation not limited to first order effects. He found it necessary, after the initial Galilean transformation, to make a further transformation

$$x' = l \beta x_r, \quad y' = l y_r, \quad z' = l z_r, \quad t' = l t/\beta - l \beta v x_r /c^2$$

together with a transformation relating electrical and magnetic fields and charge density. Here l is a positive multiplier (possibly a function of v) and β the constant, nowadays denoted by γ, set initially at an approximating value

$$\beta = 1 + v^2/2c^2$$

but later at the exact value

$$\beta = 1/\sqrt{(1 - v^2/c^2)}$$

Lorentz's revised theory of 1904 demonstrated the complete invariance of Maxwell's equations under this transformation.

From a different point of view, the same transformation with l=1 had also been made by Larmor in his 1900 book 'Aether and Matter'

-----------------------------------------------------------------------------------------------------

## 6. Poincaré on the Lorentz Transformation and the Principle of Relativity

After the negative result of the Michelson-Morley experiments of 1887, 1888, Poincaré had taken a close interest in work of Lorentz commenting critically on it (see remarks in the bibliography), In 1899 he speculated that no difference would be found between the Maxwell-Hertz equations for fixed and moving systems even for higher order approximations since electrical, like mechanical systems, depend only on relative motions (see his 1901 book), This view found verification from the revised theory of Lorentz (1904). described above. But instead of using the two-stage transformation of Lorentz, Poincaré (1905) substituted from the two sets of equations and obtained the direct transformation which he named the Lorentz transformation.

$$x' = l\,\beta\,(x - vt), \quad y' = l\,y, \quad z' = l\,z, \quad t' = l\,\beta\,(t - (v/c^2)x)$$

Poincaré observed that when $l=1$ the combination and the inverse of such transformations have the same form so giving rise to a one dimensional group, which he called the *Lorentz group* which he subsequently in 1906 extended to three dimensions. This form of this transformation with the scalar multiplier unity includes that of Voigt (1887) as Lorentz (1909) later acknowledged saying that Voigt's paper had escaped him all those years and adding:

> *'The idea of the transformation might therefore have been borrowed from Voigt, and the proof that it does not alter the form of the equations for the free ether is contained in his paper'* (Theory of Electrons 1909)

*The Principle of Relativity*: In a lecture at St. Louis in 1904 Poincaré proposed several principles applying to all physical phenomena one of which was:

> *The principle of relativity, according to which the laws of physical phenomena must be the same for a stationary observer as for an observer carried along in a uniform motion of translation so that we have not and can not have any means of discerning whether or not we are carried along in such a motion*

In his 1905 paper, which appeared shortly before that of Einstein, Poincaré questioned the existence of absolute motion for linear motions although he had in his previous writings been careful to point out the exceptional nature of rotating systems.

---

1) Poincaré H: L'état actuel et l'avenir de la physique mathématique, Congress of Arts and Sciences, St. Louis, USA Sept 1904, *Bull. Sci. Math* 28 1904 302– ; Engl tr. (G.B. Halstead) *Monist* Jan 1905 rpr Poincaré: *Value of Science*'

2) The usual form of the Lorentz transformation originated in correspondence between Poincaré and Loremtz between late 1904 and early 1905 (see Miller1981)

## 7. Einstein’s 1905 paper

The Theory of Relativity is most frequently credited to Einstein. This is justified in the sense that, although many of the significant ideas in the theory had already been noticed by Poincaré, it was Einstein who first presented them as a coherent theory. Einstein’s interest in the theory of relativity dates from his student days in Zürich, the initial idea (1901) being based on relative motion. This finds an echo in the introduction to his 1905 paper saying

> *'Maxwell's electrodynamics – as usually understood at the present time – when applied to moving bodies, leads to asymmetries which do not appear to be inherent in the phenomena. Take, for example, the reciprocal electrodynamic action of a magnet and a conductor. The observable phenomenon here depends only on the relative motion of the conductor and the magnet, whereas the customary view draws a sharp distinction between the two cases in which either the one or the other of these bodies is in motion.'*

The non-existence of the aether is also affirmed in this introduction. However the subsequent theory of the 1905 paper was not based on relative motions but followed the Lorentz theory which Einstein had studied together with Poincaré’s writings and other works.

Einstein’s version of the Principle of Relativity denied absolutes both for space and, most remarkably, for time. All inertial systems in relative uniform motion were considered equivalent for the description of physical laws and the velocity of light the same in all frames of reference Each observer keeps his own time measured by a personal clock. The status of rotating systems was left undefined and in fact, ignored. The 1905 paper was followed by a restatement with modifications in the 1907 paper in 'Jahrbuch der Physik'

General recognition of the approach of the 1905 paper depended on experimental verification After initial unclear results and an unsuccessful attempt to verify the theory using the transverse Doppler Effect, final experimental proof was reported by Bücherer by measurements on fast moving electrons at the same time as Minkowski's famous 1908 lecture and established a general acceptance of the relativity principle as Einstein had given it.

At the time, Einstein's contribution was regarded as a variation of the Lorentz theory and was referred to as the Lorentz-Einstein theory but later Lorentz himself acknowledged the immense simplification of Einstein's relativity hypothesis

> *‘If I had to write the last chapter now, I should certainly have given a more prominent place to Einstein's theory of relativity by which the theory of electromagnetic phenomena in moving systems gains a simplicity that I had not been able to attain. The chief cause of my failure was my clinging to the idea that the variable t only can be considered as the true time and that my local time t' must be regarded as no more than an auxiliary mathematical quantity. In Einstein's theory, on the contrary, t' plays the same part as t; if we want to describe phenomena in terms of x', y', z', t' we must work with these variables exactly as we could do with x, y, z, t ’*
>
> Lorentz: 1915 note added on p.321 to *'Theory of Electrons'* (1906).

*Notes on Einstein etc*:
1) Detailed biographic comments for Einstein including his indebtedness to others and use of texts such as Föppl and Drude may be found in his Collected Works, vol. II edited by Stachel. Many of these comments are reproduced in Stachel's '*Einstein's Miraculous Year*'.They would refer to the period about 1903 when Einstein, with friends he called the Olympia Academy, was studying Maxwell's equations, works by Lorentz, Poincaré as well as philosophical works of Hume and Mach
2) Early work of Einstein on relative motion is recorded by two letters of 1901 to his future wife Mileva Marić, the first of which, dated March 27, said "how happy and proud I shall be when our work on relative motions is brought to a victorious conclusion".This letter has led to controversial speculations on the contribution of Marić to the theory of relativity and its philosophy. Einstein's 1901 reference to 'our' work shows her initial participation although it is known that soon after she had to give up scientific work for personal and family reasons. It is interesting that she, like Varićak, came from the region very much influenced by ideas originating from Boscovich who had denied absolute motion in favour of relative motion. Varićak made an extensive study of Boscovich's works and wrote many scholarly articles on them (references in Kurepa 1965). He brought to notice Boscovich's article of 1755 which contained, according to Silberstein (1914, p. 38 footnote) 'many clear and radical ideas regarding the relativity of space, time and motion'
3) Bücherer A.H: Messungen an Becquerel strahlen. Die experimentalle Bestätigung der Lorentz-Einsteinsche Theorie, *Phys Z,* 9 1908 755-762. It is also in *Ann Phys* 1909

## 8. Bibliography on the History of the Theory of Relativity

These are in chronological order up to the appearance of the Poincaré-Einstein theory.

## 9. Secondary Literature and Collected Works

# APPENDIX 2 - Mathematical Notes

## 1. Spherical Trigonometry

A spherical triangle is formed by three great-circle arcs on the surface of a sphere. The vertices and the angles at these vertices are usually denoted by A, B, C and the lengths of the sides by a, b, c. The sides may alternatively be characterized by the angles α, β, γ they subtend at the centre of the sphere which are related to a, b, c by

$$\alpha = a/R,\ \beta = b/R,\ \gamma = c/R$$

R is here the radius. The principal formulae are then, relative to vertex A:

(a) *The cosine formulae*:

$$\cos\alpha = \cos\beta\cos\gamma + \sin\beta\sin\gamma\cos A \quad \text{etc.}$$
$$\cos a/R = \cos b/R \cos c/R + \sin b/R \sin c/R \cos A \quad \text{etc.}$$

(b) *The sine formula*:

$$\sin A/\sin\alpha = \sin B/\sin\beta = \sin C/\sin\gamma$$

(c) *The polar cosine formulae*:

$$\cos A = -\cos B\cos C + \sin B\sin C\cos\alpha \qquad \text{etc.}$$

* *Remark*: The sine rule may be deduced from the cosine formula by writing

$$\sin^2 A = 1 - \cos^2 A = \frac{(\sin\beta\sin\gamma)^2 - (\cos\alpha - \cos\beta\cos\gamma)^2}{(\sin\beta\sin\gamma)^2}$$

$$\begin{aligned} f^2 &= (\sin\beta\sin\gamma)^2 - (\cos\alpha - \cos\beta\cos\gamma) \quad (f>0) \\ &= (1 - \cos^2\beta)(1 - \cos^2\gamma) - \cos^2\alpha + 2\cos\alpha\cos\beta\cos\gamma - \cos^2\beta\cos^2\gamma \\ &= 1 - \cos^2\alpha - \cos^2\beta - \cos^2\gamma + 2\cos\alpha\cos\beta\cos\gamma \end{aligned}$$

$$\frac{\sin A}{\sin\alpha} = \frac{f}{\sin\alpha\sin\beta\sin\gamma}.$$

The sine rule follows from the symmetry of this formula.

(b) *The polar cosine formula*: Any point on a sphere has an equatorial great circle with this point as a pole. An orientation of this great circle may be determined by the right handed corkscrew rule.Then the three vertices of any triangle have three associated great circles forming the *polar triangle*. The relation between triangle and polar triangle is reciprocal.Angles α, β, γ, A, B, C of either triangle correspond to π-A, π-B, π-C, π-α, π-β, π-γ of the other. The cosine rule for the polar triangle gives polar cosine formulae

* *Right-angled triangles*:

Taking B as the right angle as shown, the formulae for a right angled triangle are set out below following the arrangement in Todhunter's: *Spherical Trigonometry*

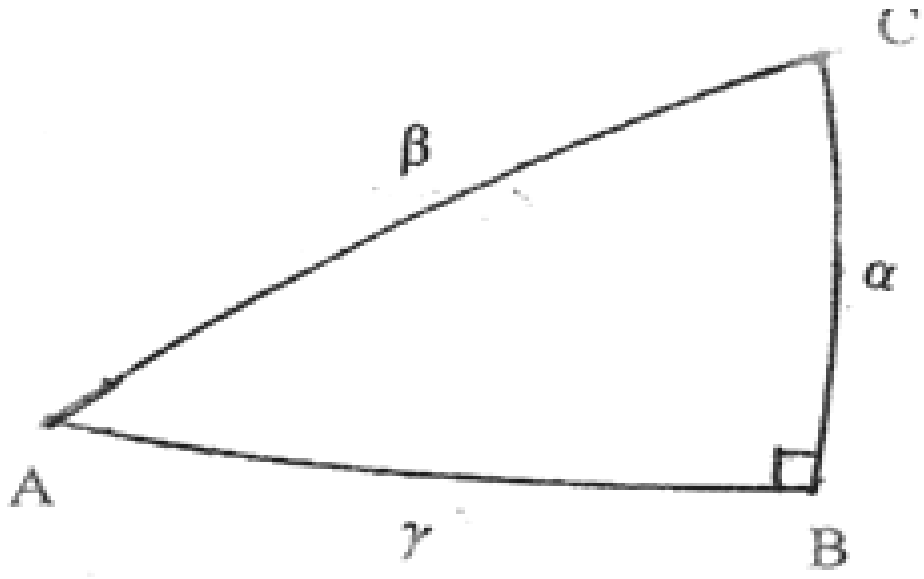

*Fig.* A right-angled spherical triangle

The formulae for a right angled triangle are the following (with B as right angle)

(a) *The cosine rule (Pythagoras' theorem)*

$\cos\beta = \cos\alpha \cos\gamma$

(b) *The cotangent rule*

$\cos\beta = \cot A \cot C$

(c) *The sine rule*

$\sin\alpha = \sin A \sin\beta$
$\sin\gamma = \sin C \sin\beta$

(d) *The adjacent side rule*:

$\tan\gamma = \tan\beta \cos A$
$\tan\alpha = \tan\beta \cos C$

(e) *The tangent rule*:

$\tan\alpha = \tan A \sin\gamma$
$\tan\gamma = \tan C \sin\alpha$

(f) *The adjacent angle rule*:

$\cos A = \cos\alpha \sin C$
$\cos C = \cos\gamma \sin A$

Some may be deduced geometrically by projecting the triangle on to a tangent plane as described in the text while others are simple deductions from these. These 10 formulae cover all possible combinations of the unknowns A, B, a, b, c. They may all be deduced from a diagram called *Napier's Rule*. For further details on these matters see Todhunter & Leathem's book.

The formulae may be conveniently stated in words (Todhunter 1886 p. 18)

(a) Cos hyp = product of cosines of sides
(b) Cos hyp = product of cotangents of angles
(c) Sine side sin opposite angle × sin hyp
(d) Tan side = tan hyp × cos included angle
(e) Tan side = tan opposite angle × sin other side
(f) Cos angle = cos opposite side × sin other angle

---

* *Reference*: Todhunter & Leathem: *Spherical Trigonometry* London 1914 etc. An excellent textbook originally written for schools The above formulae are as given in this book but for the case when B is a right angle.

* *Spherical excess*: From the polar cosine formula it may be deduced that the sum of the angles in radians of a spherical triangle is greater than π.

$$\cos A = -\cos B \cos C + \sin B \sin C \cos a/R$$
$$< -\cos B \cos C + \sin B \sin C = -\cos (B+C) = \cos (\pi\text{-}B\text{-}C)$$

from which

$$A < \pi - B - C, \quad \text{i.e.} \quad A+B+C < \pi$$

The *spherical excess* is the positive difference in radians

$$E = (A+B+C) - \pi$$

It gives the area of a spherical triangle by *Giraud's formula*

$$\text{area} = E\, R^2$$

* *Lagrange's formula for excess*:Lagrange gave several formulae for finding the excess. The most appropriate for this book is that of Lagrange (1799) which is

$$\cot (E/2) = \frac{\cos \beta/2 \cos \gamma/2 + \sin \beta/2 \sin \gamma/2 \cos A}{\sin \beta/2 \sin \gamma/2 \sin A} = \frac{\cot \beta/2 \cot \gamma/2 + \cos A}{\sin A}$$

Lagrange's derivation was the following

$$\cot (E/2) = -\tan (A + B + C)/2 = -\frac{\tan A/2 + \tan (B + C)}{1 - \tan A/2 \ \tan (B + C)/2}$$

Now use is made of the identity

$$\frac{\tan (B + C)/2}{\cot A/2} = \frac{\cos (\beta - \gamma)/2}{\cos (\beta + \gamma)/2}$$

which is one of 'Napier's analogies' (See Todhunter & Leathem) Substituting for tan (B + C)/2, leads to the required expression

$$\cot (E/2) = -\frac{\tan A/2 \cos (\beta + \gamma)/2 + \cot A/2 \cos (\beta - \gamma)/2}{\sin \beta/2 \sin \gamma/2 \sin A}$$
$$= \frac{\cos \beta/2 \cos \gamma/2 + \sin \beta/2 \sin \gamma/2 \cos A}{\sin \beta/2 \sin \gamma/2 \sin A}$$

Lagrange showed this may be converted to a symmetric form as in the analogous hyperbolic case below.

---

*Reference*: Lagrange's 1799 paper: 'Solution de quelques problems ...' may conveniently be read in his Collected Works: *Oeuvres* VII 331-359 (see especially 339)

## 2. HyperbolicTrigonometry

We pass from the formulae for a sphere of constant Gaussian curvature $R^2$ to the corresponding formulae for a two-dimensional hyperbolic space of constant Gaussian curvature $- R^2$ by changing R into iR (Taurinus 1826).This is equivalent to changing angles from $\alpha$, $\beta$, $\gamma$ to $\alpha/i$, $\beta/i$, $\gamma/i$ i.e. to $-i\alpha$, $-i\beta$, $-i\gamma$. Then reinterpreting sine, cosine, tangent of imaginary angles as sinh, cosh, tanh (here abbreviated sh, ch, th) there are found the corresponding hyperbolic formulae:

(a) *The hyperbolic cosine formula*:

$$\mathrm{ch}\,\alpha = \mathrm{ch}\,\beta\,\mathrm{ch}\,\gamma - \mathrm{sh}\,\beta\,\mathrm{sh}\,\gamma\cos A \quad \text{etc.}$$
$$\mathrm{ch}(a/R) = \mathrm{ch}(b/R)\,\mathrm{ch}(c/R) - \mathrm{sh}(b/R).\mathrm{sh}(c/R).\cos A \quad \text{etc.}$$

(b*) The hyperbolic sine formulae*:

$$\sin A/ \mathrm{sh}\,\alpha = \sin B/ \mathrm{sh}\,\beta = \sin C/ \mathrm{sh}\,\gamma$$

(c) *The hyperbolic polar cosine formula*:

$$\cos A = - \cos B \cos C + \sin B \sin C\,\mathrm{ch}\,\alpha \quad \text{etc.}$$

* *Remark*: The hyperbolic sine rule may be deduced in the same way as in the spherical case from the hyperbolic cosine formula by writing

$$\sin^2 A = 1 - \cos^2 A = 1 - \frac{(\mathrm{ch}\,\beta\,\mathrm{ch}\,\gamma - \mathrm{ch}\,\alpha)^2}{(\mathrm{sh}\,\beta\,\mathrm{sh}\,\gamma)^2}$$

From this the sine rule follows from the symmetrical right hand side of

$$\frac{\sin A}{\mathrm{sh}\,\alpha} = \frac{\sqrt{(1 - \mathrm{ch}^2\alpha - \mathrm{ch}^2\beta - \mathrm{ch}^2\gamma + 2\,\mathrm{ch}\,\alpha\,\mathrm{ch}\,\beta\,\mathrm{ch}\,\gamma)}}{\mathrm{sh}\,\alpha\,\mathrm{sh}\,\beta\,\mathrm{sh}\,\gamma}$$

---

* *References on hyperbolic geometry*: The literature is considerable. Standard older texts are those of Bonola and Sommerville as well as the books of Coxeter and Rosenfeld  Bonola includes a translation of Bolyai's work.  Not much of the original work of Lobachevski is easily available in translation  Apart from a small introductory volume originally published in German, there is available one article in French, 'Géometrie imaginaire' and  an English translation of a later work 'Pangeometry' both listed at the end of this appendix.

*Right-angled hyperbolic triangles*: With B as right angle the formulae corresponding to the spherical case are: the following covering all possibilities

(a) *The hyperbolic cosine rule* (*Pythagoras' theorem*)

$\text{ch}\,\beta = \text{ch}\,\alpha\;\text{ch}\,\gamma$

(b) *The hyperbolic cotangent rule:*

$\text{ch}\,\beta = \cot A \cot C$

(c) *The hyperbolic sine rule*

$\sin A = \text{sh}\,\alpha/\text{sh}\,\beta$
$\sin C = \text{sh}\,\gamma/\text{sh}\,\beta$

(d) *The hyperbolic adjacent side rule:*

$\cos A = \text{th}\,\gamma/\text{th}\,\beta$
$\cos C = \text{th}\,\alpha/\,\text{th}\,\beta$

(e) *The hyperbolic tangent rule*:

$\text{th}\,\alpha = \tan A\;\text{sh}\,\gamma$
$\text{th}\,\gamma = \tan C\;\text{sh}\,\alpha$

(f) *The hyperbolic adjacent angle rule*:

$\cos A = \sin C\;\text{ch}\,\alpha$
$\cos C = \sin A\;\text{ch}\,\gamma$

The tangent and adjacent side rules may be related to tangential projection as

$\tan A = (\text{th}\,\alpha/\text{th}\,\gamma)\;\text{sech}\,\gamma$ $\qquad$ $\cos A = \text{th}\,\gamma/\text{th}\,\beta$
$\tan C = (\text{th}\,\gamma/\text{th}\,\alpha)\;\text{sech}\,\alpha$ $\qquad$ $\cos C = \text{th}\,\alpha/\text{th}\,\beta$

As in the spherical case, these formulae may be set out using a hyperbolic form of Napier's Rule (see Sommerville's: Non-Euclidean Geometry, pp 68,74)

* *Infinite* triangles In hyperbolic geometry it is possible to have infinite triangles with one or more vertex angles zero. One such triangle is shown below formed by an asymptotic parallel to a straight line CA from a point B not on CA and a perpendicular from B to CA.

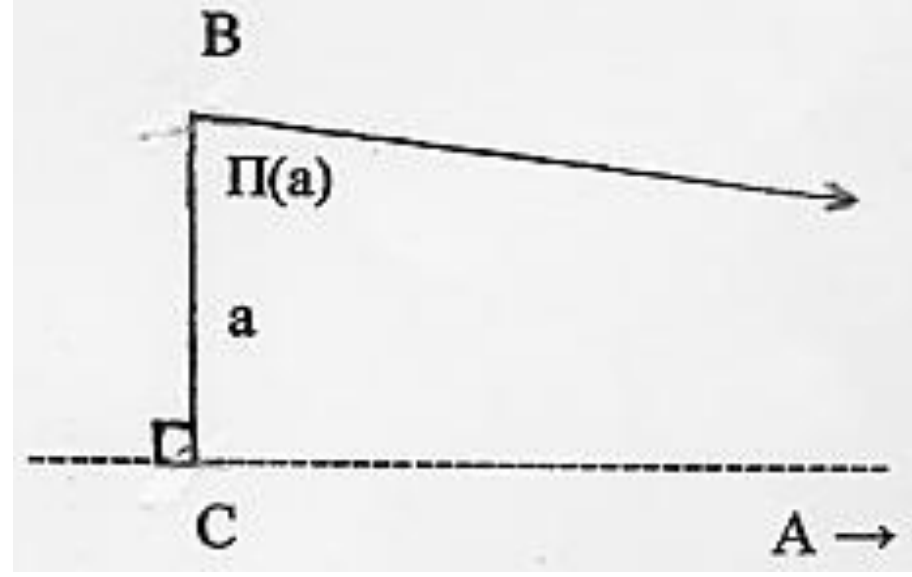

*Fig*. The parallel angle Π of Lobachevsky

The angle at B is the *parallel angle of Lobachevsky* denoted by him by $\Pi(\alpha)$ showing its dependence on the length ratio $\alpha$ (= a/R) of the perpendicular BC. The right-angled triangle ABC has one zero angle and two infinite sides. On putting $A = 0$, $B = \Pi(\alpha)$, $C = \pi/2$, $b = \infty$, $c = \infty$ in the above polar cosine formula for cos A follows,

$$1 = \sin \Pi(\alpha)\;\text{ch}\,\alpha$$

$$\sin \Pi(\alpha) = 1/\text{ch}\,\alpha, \qquad \cos \Pi(\alpha) = \text{th}\,\alpha, \qquad \tan \Pi(\alpha) = 1/\text{sh}\,\alpha$$

These may all be deduced using tan half-angle formulae from Lobachevsky's formula

$$\tan \{\tfrac{1}{2}\,\Pi(\alpha)\} = \exp(-\,\alpha)$$

* *Hyperbolic defect*: Following a similar method to the spherical case we have from the polar cosine formula

$$
\begin{aligned}
\cos A &= -\cos B \cos C + \sin B \sin C \ \mathrm{ch}\, a/R \\
&> -\cos B \cos C + \sin B \sin C \\
&= -\cos (B+C) \\
&= \cos (\pi-B-C)
\end{aligned}
$$

From which follows

$$A < \pi-B-C \quad \text{i.e. } A+B+C < \pi$$

The *hyperbolic defect* is defined as the positive difference

$$D = \pi - (A+B+C)$$

It gives the area of a hyperbolic triangle by a formula due to Gauss

$$\text{area} = D\, R^2$$

* *The Hyperbolic Form of Lagrange's Formula for Excess (Area)*:

Lagrange's formula becomes

$$\cot (D/2) = \frac{\mathrm{ch}\, \beta/2\ \mathrm{ch}\, \gamma/2 + \mathrm{sh}\, \beta/2\ \mathrm{sh}\, \gamma/2 \cos A}{\mathrm{sh}\, \beta/2\ \mathrm{sh}\, \gamma/2 \sin A}$$

*LEMMA*: In any hyperbolic triangle,

$$\tan (B + C)/2 = \frac{\mathrm{ch}\, (\beta-\gamma)/2}{\mathrm{ch}\, (\beta+\gamma)/2} \cot \frac{A}{2}$$

This is the first of Napier's analogies for a hyperbolic triangle. Its proof is similar to the spherical case. From it the Lagrange formula follows easily:

$$\cot (D/2) = \tan (A+B+C)/2 = \frac{\tan A/2 + \tan(B+C)/2}{1 + \tan A/2 \tan(B+C)/2}$$

By the lemma the right hand side is

$$\text{rhs} = \frac{\tan(A/2)\ \mathrm{ch}(\beta+\gamma)/2 - \cot(A/2)\ \mathrm{ch}(\beta-\gamma)/2}{\mathrm{ch}(\beta+\gamma)/2 - \mathrm{ch}\, (\beta-\gamma)/2}$$

$$= \frac{\mathrm{ch}\, \beta/2\ \mathrm{ch}\, \gamma/2 + \mathrm{sh}\, \beta/2\ \mathrm{sh}\, \gamma/2 \cos A}{\mathrm{sh}\, \beta/2\ \mathrm{sh}\, \gamma/2 \sin A}$$

---

*Reference*: Sommerville (1914) gives Gauss' simple proof of his formula for area.

## 3. Cayley-Klein Projective Metric

Cayley (1869) showed how the concept of distance could be introduced into projective geometry relative to an absolute conic (or simply 'absolute') having equation

$$a(x, x) = 0$$

Here a(x, y) is a bilinear form in the vectors x, y of a projective space.

$$a(x, y) = \sum_{i, j = 1} \sum a_{ij}\, x_{i.} y_j$$

Projective distance ρ is defined relative to the absolute by the formula

$$\rho = \cos^{[-1]} \frac{a(x,y)}{\sqrt{\{a(x,x)\, a(y,y)\}}}.$$

In three dimensions setting

$$a(x, x) \;=\; x_1^2 + x_2^2 + x_3^2$$

makes the absolute an imaginary locus and the Cayley definition gives the angle between the vectors x and y or, if these lie on a sphere with centre at the origin, the arc length on the surface of the sphere.

Klein (1871) considerably developed Cayley's idea by a method based on cross-ratios leading to a unified treatment by projective geometry of Euclidean, spherical and Lobachevskian geometries according to the nature of the absolute conic. He introduced the terms parabolic, elliptic or hyperbolic respectively to denote the three basic geometries.

In the hyperbolic case the quadratic a(x, x) is semi-definite so the absolute becomes a real ellipse and in view of the reversed Cauchy inequality the Cayley formula must be written as

$$\rho = \mathrm{ch}^{[-1]} \frac{a(x,y)}{\sqrt{\{a(x,x)\, a(y,y)\}}}.$$

Ordinary Euclidean geometrical ideas apply in the interior of this ellipse but straight lines are considered to meet only if the meeting point lies in its interior. So lines do not necessarily meet and are then considered parallel. Lines meeting on the ellipse are defined as asymptotically parallel, the bounding conic thus representing infinity. With these conventions it is found that all Euclidean axioms are satisfied except those relating to parallels

---

-* *References*: The original reference to Cayley's work is the last part of his 'Sixth Memoir on Quantics': *Phil. Trans*. 1859. Kleins' work was published in his elegantly written *Nicht Euklidische Geometrie* reprinted by Chelsea 1927. See also Veblen & Young': *Projective Geometry* (Ginn) for a fine exposition.

* *Relation to the Riemannian form*: It is easy to show an equivalent formula for ρ is

$$\mathrm{sh}\,\rho = \frac{\sqrt{\{a(x,y)^2 - a(x,x).a(y,y)\}}}{\sqrt{\{a(x,x)\,a(y,y)\}}}$$

On putting $y = x + dx$ there is found for the infinitesimal metric element squared:

$$ds^2 = \frac{a(x,dx)^2 - a(x,x)\,a(dx,dx)}{a(x,x)^2} = \frac{a(x,dx)^2}{a(x,x)^2} - \frac{a(dx,dx)}{a(x,x)}$$

It relates Cayley-Klein and Riemannian metrics. This formula shows that tangential elements to the surface $a(x, x) = \text{const}\ (> 0)$ satisfying the constraint that $a(x,dx) = 0$ have metric element squared

$$ds^2 = -\,a(dx, dx)/\text{const.}$$

Note the unexpected minus sign.

## 4 The Beltrami-Klein Representation:

Plane hyperbolic geometry is represented by the geometry of line segments within a circle as in fig.1. Two line segments are considered to be asymptotically parallel if they mwwt on the circle as in fig.2.

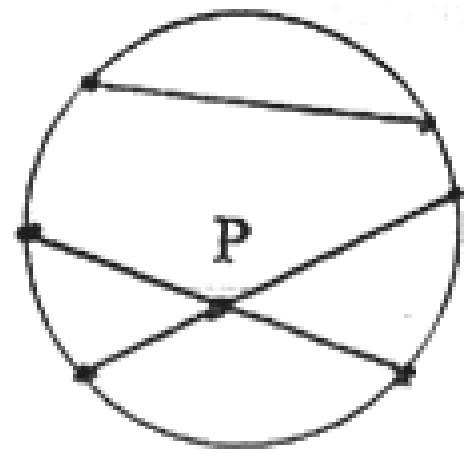

*Fig.1* Intersecting and non-intersecting lines.

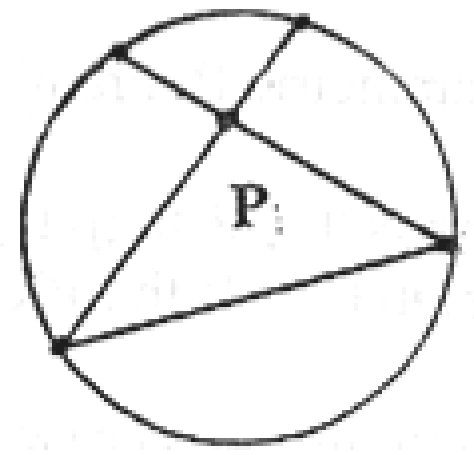

*Fig*.2: Asymptotic parallels from a point P

*The cross-ratio metric* A point at radial distance r from the centre of a circle of radius R has hyperbolic distance ρ

$$\rho = \frac{R}{2} \ln \left\{\frac{R + r}{R - r}\right\} = R\,\mathrm{th}^{[-1]}\,(r/R),$$

so

$$r = R\,\mathrm{th}\,(\rho/R)$$

*Further formulae forthe Cayley-Klein metric of the Beltrami disc*: with the equation of the circle and its parametrization as

$$X^2 + Y^2 = R^2$$

$$X = r\cos\theta = R\,\mathrm{th}\,(\rho/R)\cos\theta,\quad Y = r\sin\theta = R\,\mathrm{th}\,(\rho/R)\sin\theta$$

The Cayley-Klein metric is found from the corresponding homogeneous form (the absolute) by writing x/z, y/z for X and Y giving the equation and it homogeneous parametrization aa

$$x^2 + y^2 - R^2 z^2 = 0$$

$$x = R \,\mathrm{sh}\,(\rho/R) \cos\theta, \quad y = R \,\mathrm{sh}\,(\rho/R) \sin\theta, \quad z = \quad \mathrm{ch}\,(\rho/R)$$

The absolute in projective coordinates (x, y, z) gives the Cayley-Klein distance ρ as

$$\mathrm{ch}\,(\rho/R) = \frac{(R^2 z_1 z_2 - x_1 x_2 - y_1 y_2)}{\sqrt{\{(R^2 z_1^2 - x_1^2 - y_1^2)(R^2 z_2^2 - x_2^2 - y_2^2)\}}}.$$

$$= \frac{(R^2 - X_1 X_2 - Y_1 Y_2)}{\sqrt{\{(R^2 - X_1^2 - Y_1^2)(R^2 - X_2^2 - Y_2^2)\}}}.$$

From this is found sh and ch and then th as sh/ch

$$\mathrm{sh}\,(\rho/R) = \frac{R\sqrt{\{(X_1 - X_2)^2 + (Y_1 - Y_2)^2 - (X_1 Y_2 - X_2 Y_1)\}}}{\sqrt{\{(R^2 - X_1^2 - Y_1^2)(R^2 - X_2^2 - Y_2^2)\}}}$$

$$\mathrm{ch}\,(\rho/R) = \frac{(R^2 - X_1 X_2 - Y_1 Y_2)}{\sqrt{\{(R^2 - X_1^2 - Y_1^2)(R^2 - X_2^2 - Y_2^2)\}}}.$$

*The differential metric element:* The squared value is found from sh (ρ/R) as

$$\frac{R^2\{(R^2 - Y^2)\, dX^2 + 2XY\, dX\, dY + (R^2 - X^2)\, dY^2\}}{(R^2 - X^2 - Y^2)^2}$$

$$= \frac{R^2 (dX^2 + dY^2)}{(R^2 - X^2 - Y^2)} + \frac{R^2 (XdY - YdX)^2}{(R^2 - X^2 - Y^2)^2}$$

Changing back to polar coordinates gives metric element squared as

$$\frac{dr^2}{1 - r^2/R^2} + \frac{r^2\, d\theta^2}{(1 - r^2/R^2)^2} = d\rho^2 + R^2 \,\mathrm{sh}^2 (\rho/R)\, d\theta$$

*The 3-dimensional case*: The region within the circle becomes the *Klein-Beltrami ball*

$$X^2 + Y^2 + Z^2 < R^2$$

The radial distance ρ from the origin (0, 0, 0) to the point **r** = (X, Y, Z) is as before

$$\mathrm{th}\,(\rho/R) = r/R$$

So the parametric form of the Beltrami representation for direction unit vector (l, m, n) is

$$X = R \,\mathrm{th}\,(\rho/R)\, l, \quad Y = R \,\mathrm{th}\,(\rho/R)\, m, \quad Z = R \,\mathrm{th}\,(\rho/R)\, n$$

The corresponding homogeneous form (x, y, z, t) in Weierstrass coordinates is

$$x = R \,\mathrm{sh}\,(\rho/R)\, l, \quad y = R \,\mathrm{sh}\,(\rho/R)\, m, \quad z = R \,\mathrm{sh}\,(\rho/R)\, n, \quad t = \mathrm{ch}\,(\rho/R)$$

$$-x^2 - y^2 - z^2 + R^2 t^2 = R^2$$

*The Cayley metric*: Using notation **r** to denote vectors (X, Y, Z), equivalent forms for the non-homogeneous Cayley metric giving distance from $\mathbf{r}_1$ to $\mathbf{r}_2$ are

$$\mathrm{ch}\,\rho = \frac{R^2 - \mathbf{r}_1.\mathbf{r}_2}{\sqrt{\{(R^2 - \mathbf{r}_1.\mathbf{r}_1)(R^2 - \mathbf{r}_2.\mathbf{r}_2)\}}}$$

$$\mathrm{sh}\,\rho = \frac{\sqrt{\{R^2\,(\mathbf{r}_2 - \mathbf{r}_1).(\mathbf{r}_2 - \mathbf{r}_1) - [(\mathbf{r}_1.\mathbf{r}_1)\,(\mathbf{r}_2.\mathbf{r}_2) - (\mathbf{r}_1.\mathbf{r}_2)^2]\}}}{\sqrt{\{(R^2 - \mathbf{r}_1.\mathbf{r}_1)(R^2 - \mathbf{r}_2.\mathbf{r}_2)\}}}$$

$$= \frac{\sqrt{\{R^2(\mathbf{r}_2 - \mathbf{r}_1).(\mathbf{r}_2 - \mathbf{r}_1) - (\mathbf{r}_1 \times \mathbf{r}_2).(\mathbf{r}_1 \times \mathbf{r}_2)\}}}{\sqrt{\{(R^2 - \mathbf{r}_1.\mathbf{r}_1)(R^2 - \mathbf{r}_2.\mathbf{r}_2)\}}}$$

The last equality follows from the Lagrange identity

$$(\mathbf{r}_1.\mathbf{r}_1)(\mathbf{r}_2.\mathbf{r}_2) - (\mathbf{r}_1.\mathbf{r}_2)^2 = (\mathbf{r}_1 \times \mathbf{r}_2).(\mathbf{r}_1 \times \mathbf{r}_2)$$

The formula for th ρ may be found by division.

* *Riemannian metric:* Putting $\mathbf{r}_1 = \mathbf{r}$ and $\mathbf{r}_2 = \mathbf{r} + d\mathbf{r}$ in the formula for sh ρ or th ρ the non-dimensional metric element squared is found as:

$$\frac{R^2}{(R^2 - \mathbf{r}.\mathbf{r})^2}\,\{(R^2 - \mathbf{r}.\mathbf{r})(d\mathbf{r}.d\mathbf{r}) + (\mathbf{r}.d\mathbf{r})^2\}$$

Now in spherical coordinates,

$$d\mathbf{r}.d\mathbf{r} = (dr)^2 + r^2\,(d\varphi^2 + \sin^2\varphi\, d\theta^2), \quad \mathbf{r}.d\mathbf{r} = x dx + y dy + z dz = \tfrac{1}{2}\, d(r^2) = r\, dr$$

$$(R^2 - \mathbf{r}.\mathbf{r})(d\mathbf{r}.d\mathbf{r}) + (\mathbf{r}.d\mathbf{r})^2 = R^2\,(dr)^2 + (R^2 - r^2)\, r^2\,(d\theta^2 + \sin^2\theta\, d\varphi^2)$$

So

$$\frac{R^2}{(R^2 - r^2)}\{dr^2 + r^2\,(d\varphi^2 + \sin^2\varphi\, d\theta^2)\} + \frac{R^2}{(R^2 - r^2)^2}\,(\mathbf{r}.d\mathbf{r})^2$$

$$= \frac{(dr/R)^2}{(1 - r^2/R^2)^2} + \frac{1}{(1 - r^2/R^2)}\,(r/R)^2\{d\varphi^2 + \sin^2\varphi\, d\theta^2\}$$

Further transformation to Beltrami coordinates gives the metric element squared as.

$$d\rho^2 + R^2\, \mathrm{sh}^2\,(\rho/R)\,\{d\varphi^2 + \sin^2\varphi\, d\theta^2\}$$

## *5. Historical references for the mathematics appendix*